\UseRawInputEncoding

\documentclass[trackchanges, twocolumn,twocolappendix]{aastex7}

\usepackage{comment}
\usepackage{enumitem}

\newcommand {\msun}{M$_{\odot}$}

\usepackage{color}
\usepackage{graphics}
\usepackage{amsmath}
\usepackage{multirow}
\usepackage{booktabs}
\usepackage{lipsum}
\graphicspath{{./}{./}}

\usepackage{amsmath}

\newcommand{\HI}{\ion{H}{1}$\,$}
\newcommand{\HIcm}{\ion{H}{1}~21-cm}

\newcommand{\LHI}{$\mathcal{L}_{HI}$}

\newcommand{\LessActive}{\textit{Less populated}}
\newcommand{\MoreActive}{\textit{More populated}}

\def\rev#1{\textcolor{black}{#1}}
\def\sRev#1{\textcolor{black}{#1}}
\def\fRev#1{\textcolor{black}{#1}}

\defcitealias{peeples19}{FOGGIE~I}
\defcitealias{corlies20}{FOGGIE~II}
\defcitealias{zheng20}{FOGGIE~III}
\defcitealias{simons20}{FOGGIE~IV}
\defcitealias{lochhaas21}{FOGGIE~V}
\defcitealias{lochhaas23}{FOGGIE~VI}
\defcitealias{wright24}{FOGGIE~VII}
\defcitealias{Acharyya24}{FOGGIE~VIII}

\defcitealias{trapp25}{FOGGIE~XII}

\usepackage{ragged2e}

\begin{document}

\title{Figuring Out Gas and Galaxies In Enzo (FOGGIE). XIII. On the Observability of Extended \HI\ Disks and Warps}

\shorttitle{Observability of \HI\ Disks in FOGGIE}

\author[orcid=0000-0001-7813-0268,sname='North America']{Cameron W. Trapp}
\affiliation{Center for Astrophysical Sciences, William H.\ Miller III Department of Physics \& Astronomy, Johns Hopkins University, 3400 N.\ Charles Street, Baltimore, MD 21218}
\email[show]{ctrapp2@jhu.edu}

\author[0000-0003-1455-8788]{Molly S.\ Peeples}
\affiliation{Space Telescope Science Institute, 3700 San Martin Dr., Baltimore, MD 21218}
\affiliation{Center for Astrophysical Sciences, William H.\ Miller III Department of Physics \& Astronomy, Johns Hopkins University, 3400 N.\ Charles Street, Baltimore, MD 21218}
\email{molly@stsci.edu}

\author[0000-0002-7982-412X]{Jason Tumlinson}
\affiliation{Space Telescope Science Institute, 3700 San Martin Dr., Baltimore, MD 21218}
\affiliation{Center for Astrophysical Sciences, William H.\ Miller III Department of Physics \& Astronomy, Johns Hopkins University, 3400 N.\ Charles Street, Baltimore, MD 21218}
\email{tumlinson@stsci.edu}

\author[0000-0002-2786-0348]{Brian W. O'Shea}
\affiliation{Department of Computational Mathematics, Science, and Engineering, Michigan State University, East Lansing, MI, US}
\affiliation{Department of Physics and Astronomy, Michigan State University, East Lansing, MI, US}
\affiliation{Facility for Rare Isotope Beams, Michigan State University, East Lansing, MI 48824, USA}
\affiliation{Institute for Cyber-Enabled Research, 567 Wilson Road, Michigan State University, East Lansing, MI 48824}
\email{bwoshea@msu.edu}

\author[0000-0002-1685-5818]{Anna C.\ Wright}
\affiliation{Center for Computational Astrophysics, Flatiron Institute, 162 Fifth Avenue, New York, NY 10010}
\email{awright@flatironinstitute.org}

\author[0000-0003-4804-7142]{Ayan Acharyya}
\affiliation{INAF - Astronomical Observatory of Padova, vicolo dell’Osservatorio 5, IT-35122 Padova, Italy}
\email{ayan.acharyya@inaf.it}

\author[0000-0002-6804-630X]{Britton D.\ Smith}
\affiliation{Institute for Astronomy, University of Edinburgh, Royal Observatory, EH9 3HJ, UK}
\email{Britton.Smith@ed.ac.uk}

\author[orcid=0009-0000-7559-7962,sname='Saeedzadeh']{Vida Saeedzadeh}
\affiliation{Center for Astrophysical Sciences, William H.\ Miller III Department of Physics \& Astronomy, Johns Hopkins University, 3400 N.\ Charles Street, Baltimore, MD 21218}
\email{vsaeedz1@jh.edu}

\author[0000-0001-7472-3824]{Ramona Augustin}
\affiliation{Leibniz-Institut f{\"u}r Astrophysik Potsdam (AIP), An der Sternwarte 16, 14482 Potsdam, Germany}
\email{raugustin@aip.de}


\begin{abstract}
Atomic Hydrogen (\HI) is a useful tracer of gas in and around galaxies, and can be found in extended disk-like structures well beyond a system's optical extent. Here, we investigate the properties of extended \HI\ disks that emerge in six Milky Way-mass galaxies using cosmological zoom-in simulations from the Figuring Out Gas \& Galaxies in Enzo (FOGGIE) suite. This paper focuses on the observability of the extended \HI\ in these systems. We find overall agreement with observational constraints on the \HI\ size-mass relation. To facilitate direct comparisons with observations, we present synthetic \HIcm\ emission cubes. By spatially filtering our synthetic cubes to characterize the absence of short baselines in interferometric maps, we find that such observations \rev{at 20 Mpc retain $\sim 96\%-99\%$ of total \HI\ emission on average, but can miss up to $\sim15\%$ of \HI\ signal outside the central disk due to missing short spacings. This effect is small for more isolated systems, but more significant for more strongly interacting systems, as there is more diffuse signal. This preferentially removes low column density, low velocity dispersion gas in the circumgalactic medium (CGM)}. The amount of observable material depends strongly on its distribution, \rev{distance}, and the system's observed orientation, preventing the formulation of a simple correction factor. Therefore, to fully characterize extended disks, their CGMs, and the interfaces between them, including data from large single-dish radio telescopes is likely necessary. 


\end{abstract}


\section{Introduction} \label{sec:introduction}

Disks are one of the most common features of galaxies in the modern Universe. To describe a galaxy often amounts to describing its disk, and yet disks vary widely in their size, mass, thickness, kinematics, gas content, and star formation rates. Disks follow a number of well-behaved scaling relations, such as the Tully-Fisher relation between mass and rotational velocity \citep{1977A&A....54..661T, mcGaugh00-TullyFisher}, the Kennicutt-Schmidt relation between gas surface density and star formation rate \citep{kennicutt98}, and the \HI\ Size-Mass relation \citep{wang16-HISizeMass}. Because disks are conspicuous, large, and lit up by stars, they are among the best-understood components of the cosmic baryon cycle. 

Even though the bulk properties of disks are well-studied, many questions remain, and there is ample room for continued discovery. Of particular interest is the question of how galactic star formation is sustained. Gas depletion times are short ($\sim 1$ Gyr) for Milky-Way like disks \citep{2012ApJ...758...73S}, so in order for observed star formation rates to exist, there must be robust mechanisms that feed disks from the Circumgalactic Medium (CGM). One way to observe this material is through absorption. Astronomers seeking to explain how extragalactic gas accretes to form disks have examined the correlations between \HI\ disks and the CGM \citep{2015ApJ...813...46B} and pursued CGM absorption observations well inside $\sim 50$ kpc. In order to fully characterize the distribution and dynamics of this material, \HI\ radio emission studies are necessary. The question of how to study this previously undetected material outside the inner disk is a difficult one. Studies utilizing large single dish radio telescopes \cite[e.g.,][]{thilker04,pisano2014:GBT-NGC2997,deBlok2014:GBT-NGC2403,Qian2025:FAST-NGC2683} can probe low column density gas, but are lacking in spatial resolution. Interferometric emission studies \cite[e.g.,][]{fraternali02,Walter_2008-THINGS,heald2011:HALOGAS,deBlok24-Mhongoose} provide both high spatial and spectral resolution datacubes, as well as high sensitivity, allowing for the distribution of \HI\ to be more fully characterized.

\rev{While offering exquisite spatial resolution, interferometric studies sparsely sample the spatial frequencies within the UV plane when collecting data \citep{clark99,thompson99}, and are therefore not sensitive to all spatial scales. In particular, synthesized images from interferometers lack data from missing short baselines, known as the ``zero spacing problem" \citep{braun85}, meaning they are not sensitive to diffuse signals above a certain spatial scale. In the context of nearby \sRev{($\lesssim10-20$ Mpc)}, Milky Way-mass galaxies, this corresponds to missing spatially diffuse signals at scales greater than $\sim50-100$ kpc, depending on the target distance and array configuration. This could correspond to missing out on a diffuse component of \HI\ that overlays the disk or large-scale, diffuse features in the CGM.}

There has been much work done to address this problem. Radio observations with a large, single-dish antenna do not suffer from the zero spacing problem, and are frequently used to check the integrated flux of an interferometric image. Galactic \HI\ surveys see little difference between the total observed fluxes of single-dish and interferometric studies \cite[e.g.,][]{Walter_2008-THINGS,reynolds22-Wallaby,deBlok24-Mhongoose}, with \HI\ below column densities of $10^{19}~\rm cm^{-2}$ making up a small fraction ($\sim2\%$) of the total \HI\ mass around star forming systems \citep{pingel18}. Studies have also historically sought to directly include single-dish data during the image deconvolution process through a process known as joint-deconvolution, filling in the information from missing short-baselines \citep{braun85,deul87,stanimirovic99,faridani18-JointDecovolution,pingel18,plunkett23}. Recent galactic HI studies and surveys have incorporated data from both single dish observations and interferometers \cite[e.g.,][]{eibensteiner2023,wang2024-FEASTS_I,koch2025-GBT_VLA_DualConv}. The significance of this zero spacing problem is not well constrained in the context of \HI\ in the CGM and extended disks, \sRev{as interferometer sensitivities \citep{deBlok24-Mhongoose} have only recently gotten down to column density limits expected for \HI\ in the CGM from simulations ($\lesssim10^{18} \, \rm{cm}^{-2}$, e.g. \citealt{trapp22}).}

Cosmological, hydrodynamical simulations are another means by which we can attempt to understand the formation, evolution, and observability of gaseous disks. This paper is the second in a series of papers in which we study the extended disks, polar rings, and warps that emerge naturally for the six Milky-Way-like galaxies in the FOGGIE suite of simulations \citep[Figuring Out Gas \& Galaxies in Enzo; ][]{peeples19}. The FOGGIE halos were selected to yield galaxies at $z = 0$ that resemble the Milky Way, but they were not tuned to produce extended \HI\ disks or larger disk-like structures. Even so, all of them have extended \HI\ at various points in their evolution. This paper focuses on the observability of these extended \HI\ disks. To that end, we create synthetic \HIcm\ datacubes in order to draw direct comparisons with recent and upcoming observational studies. To \rev{isolate and} investigate the effects that the missing short baselines problem has on the observability of extended \HI\ emission, we spatially filter these datacubes using a \rev{Gaussian high pass filter to mimic various interferometric studies}. We additionally consider the effects that viewing angle has on both the \rev{observable distribution of gas in interferometric studies} as well as observed dynamics.

The first paper in this series \citep{trapp25} investigates the morphology and origin of the extended and misaligned \HI\ disks in these systems, and will hereafter be referred to as \citetalias{trapp25}. 

Briefly, \citetalias{trapp25} found that all FOGGIE \HI\ disks have misaligned features (i.e., warps or polar rings) at some point during their evolution; however, their strengths, lifetimes, and origins vary significantly. We also found a correlation between the relative abundance of \HI\ in the CGM and the morphology of the disk. While the six FOGGIE halos lie along a continuum of relative \HI\ abundance, we subdivide our sample in two. The halos with relatively lower amounts of CGM \HI\ are classified as \LessActive\, while the halos with higher amounts of CGM \HI\ are classified as \MoreActive. The \LessActive\ systems form \rev{thin, extended, coherently rotating disks} and have hot inner CGMs near their virial temperature, while the \MoreActive\ systems do not. \rev{Additionally, systems that are \MoreActive\ have more diffuse \HI\ in their system.} The observational differences between these classifications will be discussed in this paper.

In Section~\ref{sec:simulations} we briefly describe the FOGGIE simulations. In Section~\ref{sec:obs_comp} we describe our disks and discuss how they broadly compare with observations. In Section~\ref{sec:synthetic_obs_methods} we detail the methodology used to create the synthetic \HI\ datacubes. In Section~\ref{sec:synthetic_obs_results}, we discuss the effects of filtering out diffuse components on the observability of material. Finally, Section~\ref{sec:viewing_angle_effects} discusses the effects that viewing angle has on what material is detectable in these systems and the observed dynamics.

\begin{deluxetable*}{lccccccc}
\tablecaption{Key properties and brief description of the FOGGIE halos at $z=0$ \label{tab:halo_masses}}
\tablehead{
\colhead{\textbf{Halo}} &
\colhead{$\mathbf{M_{\rm 200}}$} &
\colhead{$\mathbf{M_{*,disk}}$} &
\colhead{$\mathbf{M_{\rm HI,200}}$} &
\colhead{$\mathbf{M_{\rm HI,disk}}$} &
\colhead{$\mathbf{R_{\rm disk}}$} &
\colhead{$\mathbf{H_{\rm disk}}$} &
\colhead{\textbf{Description}} \\
\colhead{} &
\colhead{[10$^{12}$ M$_{\odot}$]} &
\colhead{[10$^{10}$ M$_{\odot}$]} &
\colhead{[10$^{10}$ M$_{\odot}$]} &
\colhead{[10$^{10}$ M$_{\odot}$]} &
\colhead{[kpc]} &
\colhead{[kpc]} &
\colhead{}
}
\startdata
\textbf{\textit{Less Populated}} \\
Tempest   & 0.50 & 4.02 & 1.28 & 1.27 & 21.8 & 1.25 & \textit{Strong warp} \\
Maelstrom & 1.01 & 7.24 & 2.72 & 2.36 & 24.2 & 0.69 & \textit{Thin disk} \\
Blizzard  & 1.14 & 9.02 & 2.79 & 2.33 & 22.0 & 0.69 & \textit{Strong warp} \\
\hline
\textbf{\textit{More Populated}} \\
Cyclone   & 1.69  & 17.41   & 3.88  & 3.17  & 38.2  & 2.36  & \textit{Polar ring} \\
Hurricane & 1.71 & 17.20 & 3.73 & 2.94 & 33.3 & 0.69 & \textit{Two polar rings} \\
Squall    & 0.80 & 10.48 & 0.73 & 0.50 & 16.8 & 1.81 & \textit{Misaligned inner disk} \\
\hline
\hline
\enddata
\tablecomments{The masses shown are either the mass within $R_{200}$ (e.g. $M_{\rm HI,200}$), or the mass within our disk definition (e.g. $M_{\rm HI,disk}$). The difference between these two corresponds to the \HI\ mass in the CGM. $R_{200}$ is the radius enclosing an average density of 200$\times$ the critical density of the Universe at $z=0$. $R_{\rm disk}$ is the radius at which the mean \HI column density falls below $1.25\times10^{20}$ cm$^{-2}$, a commonly used observational contour \citep[e.g.,][]{wang16-HISizeMass,bluebird20:chilesVI}. $H_{\rm disk}$ is the \HI\ scale height. We divide our sample in two based on the relative abundance of \HI\ in their CGMs. \LessActive\ systems have little \HI\ in their CGMs relative to the central disk and, therefore, a lower amount of diffuse signal compared to the \MoreActive\ systems. \LessActive\ systems form thin, coherently rotating, extended disks and have an inner CGM near their virial temperature; \MoreActive\ systems do not.}
\end{deluxetable*}


\section{FOGGIE Simulations} \label{sec:simulations}

In this study, we compare six zoom-in galaxy formation simulations of Milky Way-mass halos from the Figuring Out Gas \& Galaxies in Enzo (FOGGIE) suite to the latest \HI\ 21-cm observations. The FOGGIE simulations were run using the open-source adaptive mesh refinement code Enzo \citep{bryan14,brummel-smith19}.
As discussed in \citetalias{trapp25}, these simulations resolve the CGM through a novel ``forced refinement'' scheme, forcing high spatial resolution in diffuse gas \citep{peeples19} in combination with density and cooling refinement \citep{simons20}. See Appendix A of \citetalias{trapp25} for more details of the relative contributions of the refinement criteria.

\rev{The FOGGIE simulations include density and metallicity-dependent cooling and a metagalactic background \citep{haardt12} using the {\sc Grackle} chemistry and cooling library \citep{smith17}, including self-shielding of gas owing to \HI\ opacity at $z\leq15$ \citep{emerick19}. The code solves a non-equilibrium six species chemical reaction network at run time, tracking \HI, \ion{H}{2}, \ion{D}{1}, \ion{D}{2}, \ion{He}{1}, \ion{He}{2}, \ion{He}{3}, and e$^{-}$. All metal species are grouped in a single field, allowing for metallicity-dependent cooling assuming ionization equilibrium and solar abundances.}

\rev{Simulations are run with Enzo's native star formation and supernovae thermal feedback schemes \citep{cen_ostriker_06}. This feedback is underpowered, in part leading to these systems having overmassive and overly-concentrated stellar profiles. Star formation occurs in dense gas with a converging flow. Gas is turned into star particles in proportion to the local gas mass, with a minimum star-particle mass of 1000 M$_{\odot}$ at high redshift, ramping up to 10,000 M$_{\odot}$ between $z=2$ and $z=1$ (see \citealt{wright24} for a more thorough discussion of stellar feedback and star formation prescriptions). Dark Matter particle mass is $1.39\times10^{6}~\rm{M}_{\odot}$.}

These simulations have a comoving box size of 100$h^{-1}$ Mpc with a 256$^{3}$ root grid. The FOGGIE simulations use a forced-refinement box with a length of 288 kpc comoving on a side. Every gas cell within this box centered on and moving with the main galaxy, is forced to a minimum comoving spatial resolution (1100 pc comoving). At each timestep of the simulations, cells within this box are allowed to further refine based on density. This generation of FOGGIE simulations implemented a ``cooling refinement'' criterion to better resolve thermally unstable gas. This criterion refines cells such that their size is smaller than the cooling length (sound speed $\times$ cooling time). The density and cooling refinement criteria are allowed to resolve up to 274 pc comoving.

This forced refinement scheme is of particular importance in the context of extended \HI\ emission, as the spatial scales of low column density gas outside the main disk are only now starting to be constrained. Many of these observational studies are conducted with interferometers, which are only sensitive to spatial scales up to a maximum angular size limit, determined by their shortest baseline. Importantly for this study, the largest permitted cell size in the forced refinement region (1.1 kpc comoving) is much smaller than the largest relevant angular scale for these interferometers, for typical distances (e.g., $\sim$60 kpc for the 21-cm line at 10 Mpc, \rev{or $\sim$120 kpc at 20 Mpc}).

The FOGGIE simulations have predicted direct observables for a range of CGM and galaxy topics. These include:
absorption \citep{peeples19} and emission lines \citep{corlies20, lochhaas25,saeedzadeh25}, absorption-line observations of the CGM of the Milky Way \citep{zheng20}, and the interface between galaxies and the intergalactic medium \citep{lehner22}.

In this study, we analyze all six halos in the second generation of FOGGIE simulations near redshift $z=0$. All halos were selected to be Milky Way-like at $z=0$ in terms of their mass and merger history (see \citealt{wright24}). After $z=2$, there are no major mergers with the exception of Squall, which experiences a 2:1 merger at $z=0.7$. For a more detailed analysis of the evolution of these halos' \HI\ profiles over time, see \citetalias{trapp25}.

\section{Disk Properties} \label{sec:obs_comp}

\begin{figure}[ht!]
\begin{centering}
\includegraphics[width=0.45\textwidth]{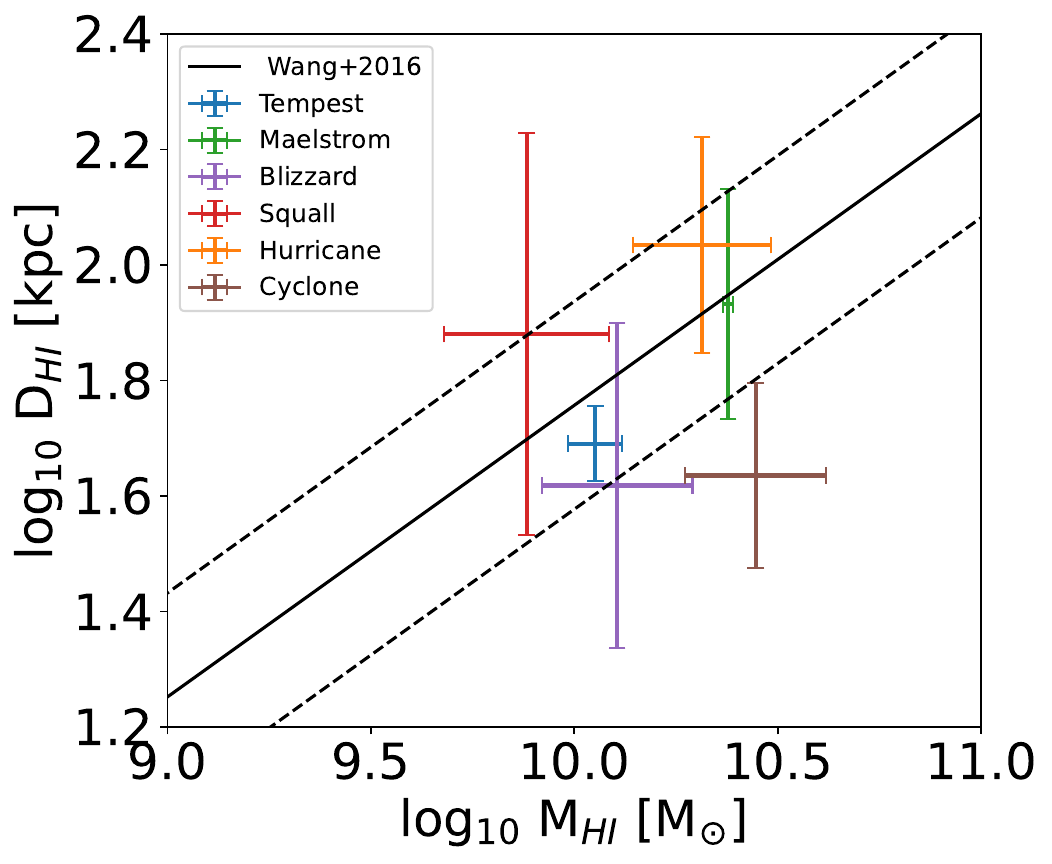}
\caption{
Scaling relations for the six \HI\ disks presented in this study. The points correspond to the mean values for a given system between $z=0$ and $z=0.5$, with bars showing the standard deviation. \textit{Top:} The \HI\ Size-Mass relation \citep{broeils97-HISizeMass} for these systems. $M_{HI}$ is calculated as the total disk \HI\ mass. $D_{HI}$ is twice the radius at which the mean column density drops below $1.25\times10^{20}~\rm cm^{-2}$. The solid (dashed) line corresponds to the observational correlation (and 3$\sigma$ scatter) found in \citet{wang16-HISizeMass}.
\label{fig:disk_scaling}}
\end{centering}
\end{figure}

Table~\ref{tab:halo_masses} presents basic properties of the sizes and masses of these systems.
In order to calculate disk masses, we use our disk definition introduced in \citetalias{trapp25}. In brief, we define the disk as a contiguous region with \ion{H}{1} number density above $\sim 6 \times 10^{-3}$ cm$^{-3}$ at $z=0$. To avoid missing low-density regions in the disk associated with stellar feedback or dynamical effects, we fill both fully enclosed holes and ``donut'' holes that completely pierce the disk.

As previously mentioned, we divide the halos into two broad categories based on the amount of \HI\ in their CGM relative to the central disk \sRev{(see the ``ideal" panels of Fig.~\ref{fig:interferometric_projections} and Fig.~\ref{fig:interferometric_projections_2} for a visualization of these two categories)}. The \LessActive\ systems (Tempest, Maelstrom, and Blizzard) have a relatively small amount of \HI\ in their CGMs, form thin, extended, and coherently rotating disks, and have hot inner CGMs near their virial temperature at $z=0$. 
Tempest is the smallest system by mass, and shows little \HI\ outside its central disk ($\sim10^{8}$ \msun). It shows a strong warp (inclination of $\sim20$\textdegree) at redshift $z=0$. Maelstrom has more \HI\ mass in its CGM ($\sim3.6\times10^{9}$ \msun), but is less populated in comparison to other systems. It does not have a significant warp at $z=0$. Blizzard has a more significant amount of \HI\ in its CGM ($\sim4.6\times10^{9}$ \msun) as well as a strong warp at redshift $z=0$ (inclination of $\sim20$\textdegree). 

The \MoreActive\ systems (Cyclone, Hurricane, and Squall) have relatively more \HI\ in their CGMs, have thicker disks, and have cooler inner CGMs at $z=0$. Despite having a large stellar mass, Squall has relatively little \HI\ in its disk, although there is a relatively large amount in its CGM ($\sim2.3\times10^{9}$ \msun). At $z=0$ it has a small ($\sim2$ kpc), misaligned inner disk. Hurricane is the most active system and has a relatively large amount of \HI\ outside its central disk ($\sim7.9\times10^{9}$ \msun). Additionally, it has two distinct polar rings at $z = 0$. Cyclone has a nearly perfectly orthogonal polar ring, as well as a large amount of CGM material. For a more detailed analysis of the properties, origins, and evolution of these misaligned features, as well as how this categorization relates to disk morphology and thin disk formation, see \citetalias{trapp25}.

Observations of \HI\ disks find a tight correlation between the observed \HI\ mass and the spatial size of the \HI\ disk \citep[e.g.,][]{broeils97-HISizeMass,wang16-HISizeMass,bluebird20:chilesVI}. A commonly used metric is the correlation between the \HI\ mass of the galaxy and the diameter of the $N_{HI}=1.25\times 10^{20} ~\rm{cm}^{-2}$ contour. The top panel of Fig.~\ref{fig:disk_scaling} shows this relation for the six halos considered in this study from redshift $z=0-0.5$. This redshift range was chosen to \sRev{highlight the evolution of individual systems} and match the redshift range of the CHILES survey \citep{bluebird20:chilesVI}, which has galaxies that closely follow this relation with $D_{HI}$ ranging from a few kpc to $\sim$60 kpc. The points represent the mean values between these redshifts, with the bars showing the standard deviation. The \HI\ mass, $M_{HI}$, is the total within the defined disk region. The disk diameter, $D_{HI}$, is calculated as 2 times the radius at which the mean face-on column density drops below $1.25\times 10^{20}~\rm{cm}^{-2}$. All the \LessActive\ systems (Tempest, Maelstrom, and Hurricane) fall within the observed scatter from \citet{wang16-HISizeMass}. Squall and Blizzard are on the upper and lower edges, respectively. Cyclone is overmassive for its size. This is likely due to its recent interaction history with satellite galaxies, leading to a relatively large amount of material in the CGM and a relatively compact disk. \sRev{Some systems evolve quite significantly during this redshift range. The two least populated systems, Tempest and Maelstrom, show relatively little evolution.}

\section{Synthetic \HIcm\ Observations} \label{sec:synthetic_obs_methods}

\begin{deluxetable}{l|c|c|c|c|c}
\tablecaption{\rev{Survey parameters varied for the four survey analogs considered in this study. Parameters for SKA-MID were estimated based on the performance of MHONGOOSE following \citet{deBlok24-Mhongoose}.}  \label{tab:survey_parameters} }
\tablehead{
\colhead{Survey} \vline & $\sigma$ & $b_{\rm min}$  & $N_{\rm{HI}}$ & $\Theta_{\rm{PB}}$  & $\Delta v$ \\
 & ["] & [m] & [cm$^{-2}$] & [\textdegree] & [km/s]
}
\startdata
MHONGOOSE-LR  & 65 & 29 & $1\times10^{18}$ & 1 & 1.4 \\
MHONGOOSE-HR  & 22 & 29 & $5\times10^{18}$ & 1 & 1.4 \\
SKA-MID-LR  & 65 & 29  & $2.8\times10^{17}$ & 1 & 1.4 \\
SKA-MID-HR  & 22 & 29  & $1.4\times10^{18}$ & 1 & 1.4 \\
\enddata
\tablecomments{$\sigma$ is the effective angular resolution, $b_{\rm{min}}$ is the minimum baseline length simulated, $N_{\rm{HI}}$ is the \HI\ column density sensitivity,  $\Theta_{\rm{PB}}$ is the primary beam FWHM, and $\Delta v$ is the velocity resolution.}
\end{deluxetable}

In this section, we investigate the observability of the \HI\ distributions presented in this study via the creation of synthetic \HIcm\ emission cubes. Importantly, when the effects of missing short baselines are considered, the observed distribution of \HI\ changes based on the survey parameters and target galaxy.


In order to draw comparisons with multiple \HIcm\ surveys, we created synthetic datacubes for all six of our halos with a variety of instrumental parameters. The parameters considered for the synthetic surveys were: spatial resolution, spectral resolution, primary beam full-width half-max (FWHM), minimum baseline length, and sensitivity. For our fiducial suite of synthetic images, we considered all systems at an observed inclination of 40\textdegree~and a distance of \rev{20 Mpc.} Both distance and inclination can have effects on the observability of specific \HI\ components; however, \sRev{we chose these values to match the typical distances and inclinations of galaxies with comparable \HI\ mass in the MHONGOOSE sample \citep{deBlok24-Mhongoose}.}

The dimensions of the final image were selected such that the point spread function of the synthetic survey is oversampled by a factor of four and the field-of-view (FOV) \rev{is large enough to ensure that the low spatial frequency data is similarly resolved in Fourier space (350 kpc). Note, we only consider gas within the forced refinement region (288 kpc comoving on a side), as gas is well resolved in this region}. In order to capture the high rotational velocities in the central regions of the FOGGIE disks, we increased the bandwidth beyond survey parameters as needed for each system.

To create our synthetic datacubes, we considered all grid cells containing non-zero \HI\ gas mass within the forced refinement region of the simulation. Gas outside this region was not considered, due to its poorer spatial and mass resolution. The Voigt profiles, including Gaussian core and damping wings, were then calculated for each cell. The flux densities were normalized by the cells' column density and projected onto the final image as follows:


\begin{equation}
    \frac{F_{\nu,cell}}{\rm cm^{-2}~s} = \bigg{(}\frac{M_{\rm{HI},\it{cell}}}{m_{\rm HI}} \bigg{)} \bigg{(} \frac{l_{cell}^{-2}}{\rm cm^{-2}} \bigg{)} \bigg{(}\frac{\phi_{\nu,cell}^{Voigt}}{\rm s} \bigg{)}
\end{equation}
where $M_{\rm{HI},\it{cell}}/m_{\rm HI}$ is the number of \HI\ atoms, $l_{cell}$ is the cell size, and $\phi_{\nu,cell}^{Voigt}$ is the normalized Voigt profile for the cell at its Doppler velocity. We assume the gas is optically thin \citep{zwaan1997:optically-thin-justification,Walter_2008-THINGS}, so the \HI\ flux and mass differ only by a constant factor. While this assumption may break down in the interstellar medium, it holds in the CGM. We therefore left our datacubes in units of $\rm cm^{-2}~s$ for simplicity, such that the integration over all channels gives the \HI\ column density.

\begin{figure*}[ht!]
\plotone{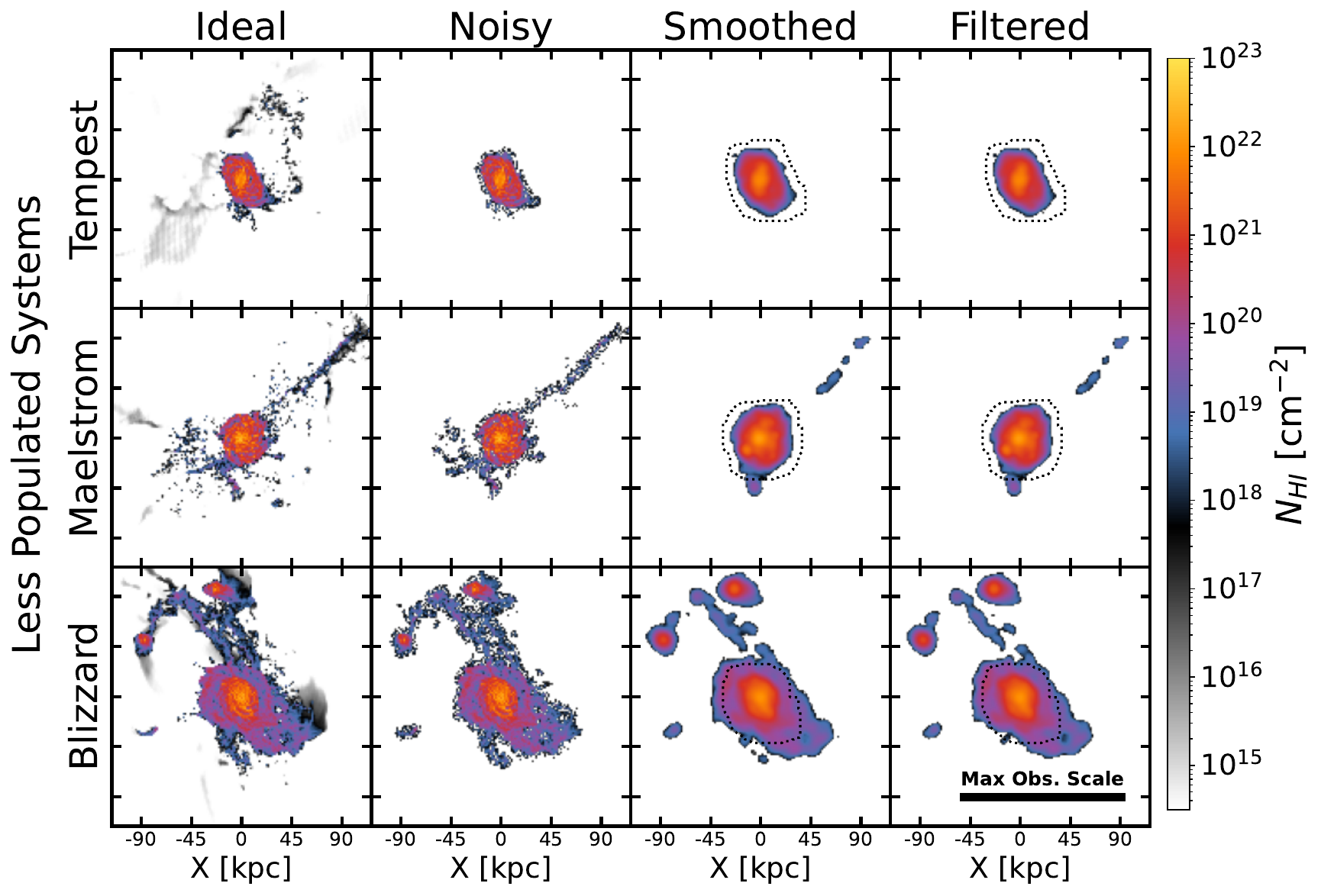}
\caption{
Effects of various steps of the synthetic \HI\ imaging pipeline for the three \LessActive\ galaxies at 20 Mpc at an inclination of 40\textdegree. \sRev{Plots show the MHONGOOSE-LR survey parameters. See Appendix~\ref{sec:appendix_survey_vis} for the other survey realizations.} The scale bar on the \sRev{bottom} right shows the maximum observable angle at this minimum baseline. The first column (\textbf{Ideal}) shows the ideal column density projections \sRev{(vertical striping arises from projecting lower resolution grid cells along non-grid axes, see Sec.~\ref{sec:disc-on_resolution}).} The second column (\textbf{Noisy}) adds Gaussian noise with a standard deviation equal to 0.2 times the sensitivity limit. For this column and the final two, only sources identified as significant are plotted. The third column (\textbf{Smoothed}) shows the effects of \rev{primary beam} and the Gaussian smoothing kernel, which \rev{limits the field of view} and removes signals with small spatial scales, respectively. The final column (\textbf{Filtered}) shows the effect of spatially filtering the data. The three galaxies presented respond to this step in distinct ways, related to how much diffuse gas is present around the disk. This has little effect on Tempest and Maelstrom, but has a larger effect on Blizzard due to the presence of more diffuse signal. \sRev{The dashed lines in the right two panels show the dilated mask used to separate disk and CGM \HI.}
\label{fig:interferometric_projections}}
\end{figure*}

\begin{figure*}[ht!]
\plotone{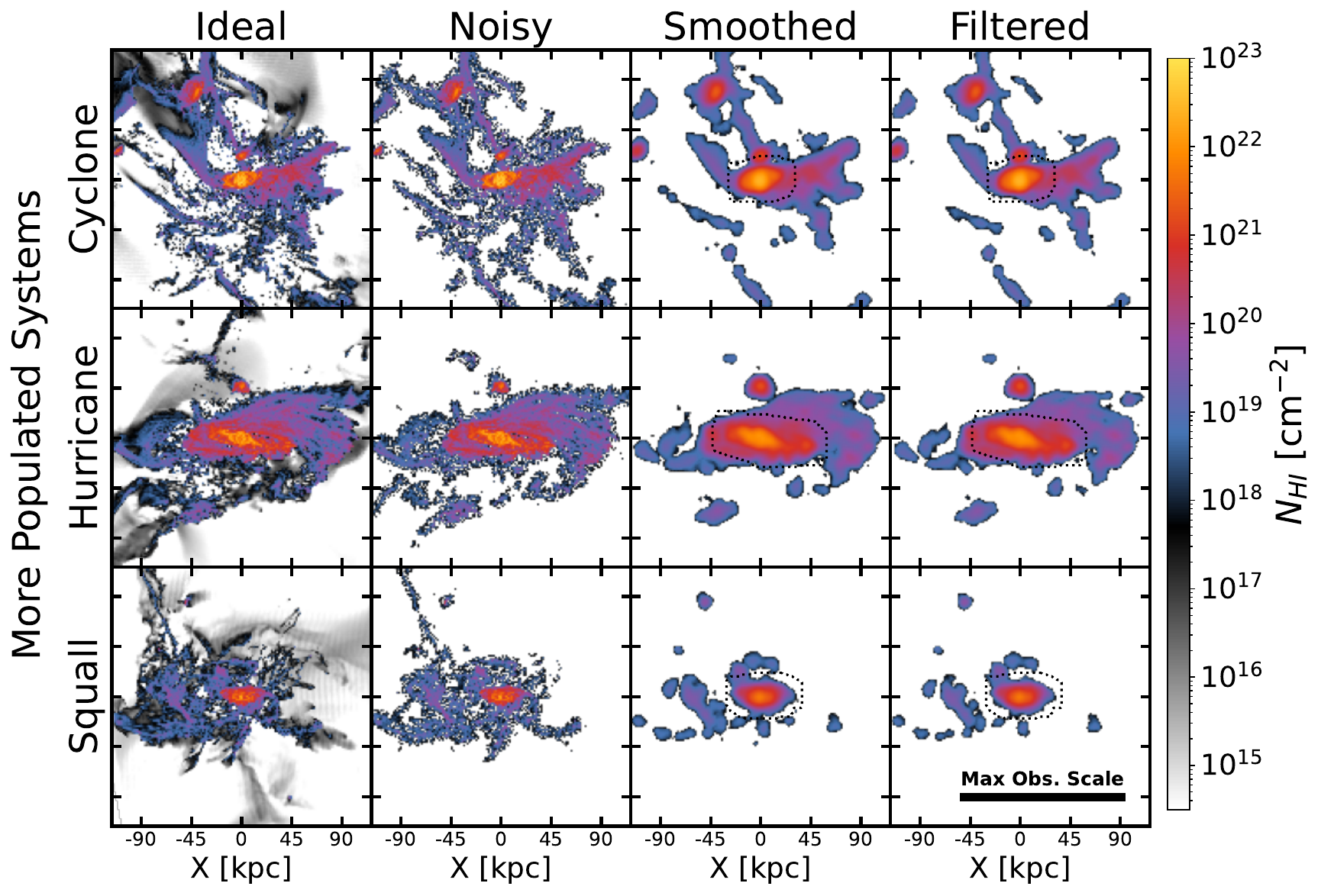}
\caption{
Same as Fig.~\ref{fig:interferometric_projections}, but for the three \MoreActive\ systems at 20 Mpc. These systems have a much larger amount of both small-scale and diffuse material in the CGM that is lost in the smoothing and filtering steps, respectively.
\label{fig:interferometric_projections_2}}
\end{figure*}

This gives us the ``ideal'' datacube. In order to make full mock spectra, we further degrade the images in a few steps: adding the effect of the primary beam, smoothing spatially, adding noise, and filtering out low spatial frequencies. The effects of these steps are visualized in Fig.~\ref{fig:interferometric_projections} and Fig.~\ref{fig:interferometric_projections_2}.

We model the primary beam (which effectively sets the field of view) as a Gaussian with a FWHM equal to 1\textdegree\ to match MeerKAT and \rev{potential Square Kilometer Array (SKA) realizations}. This lowers the signal at the edge of the field of view.

The effective column density sensitivity of the study is determined by the strength of the noise we add. To create our ``noisy'' datacube, we model the noise as a Gaussian noise distribution with a standard deviation equal to 0.2 times the synthetic survey column density limit, such that 5$\sigma$ detections correspond to the sensitivity limit. Under the assumption that the noise is largely uncorrelated with the signal \citep{veronese2025}, we created this noise cube separately and added it to the ideal datacube.

In order to smooth the data, we convolve the ideal datacube and noise cube with a Gaussian beam. The FWHM was set to the effective spatial resolution of the synthetic survey, and the noise cube was rescaled as necessary to match the target sensitivity. This gave us our ``smoothed'' datacube.

\begin{deluxetable}{l|c|c|c|c}
\tablecaption{\rev{Ratio of total (disk, satellites, and CGM) integrated \HI\ fluxes from the filtered to the ideal datacubes for the four survey analogs. Spatial filtering has little effect on the total \HI\ signal, as it preferentially removes low column density gas in the CGM.}  \label{tab:total_flux_ratios} }
\tablehead{
\colhead{Halo} \vline & \multicolumn{4}{c}{Integrated \HI\ Flux Ratios (Filtered/Ideal)} \\
\cline{2-5}
 & \colhead{MHONGOOSE-LR} \vline & \colhead{MHONGOOSE-HR}  \vline & \colhead{SKA-MID-LR} \vline & \colhead{SKA-MID-HR}
}
\startdata
Tempest  & 0.988 & 0.992 & 0.990 & 0.996  \\
Maelstrom  & 0.986 & 0.990 & 0.989 & 0.995  \\
Blizzard  & 0.973 & 0.965 & 0.984 & 0.985  \\
Cyclone  & 0.970 & 0.962 & 0.980 & 0.982  \\
Hurricane  & 0.970 & 0.955 & 0.983 & 0.980  \\
Squall  & 0.939 & 0.914 & 0.969 & 0.954  \\
\enddata
\end{deluxetable}

\begin{deluxetable*}{l|cc|cc|cc|cc|cc}
\setlength{\tabcolsep}{3pt}   
\tablecaption{ Ratio of observable \HI\ mass in the CGM to actual \sRev{(ideal)} \HI\ mass for different survey analogs at 20 Mpc\label{tab:observed_masses}}
\tablehead{
\colhead{Halo} \vline &
\multicolumn{2}{c|}{MHONGOOSE-LR} &
\multicolumn{2}{c|}{MHONGOOSE-HR} &
\multicolumn{2}{c|}{SKA-MID-LR} &
\multicolumn{2}{c}{SKA-MID-HR} \\
\cline{2-11}
\colhead{} \vline &  \colhead{Smoothed}  & \colhead{Filtered} \vline & \colhead{Smoothed} & \colhead{Filtered} \vline &
\colhead{Smoothed} & \colhead{Filtered} \vline & \colhead{Smoothed} & \colhead{Filtered} 
}
\startdata
\textbf{\textit{Less Populated}} & & & & & & & & & &\\
Tempest   & 0.000 & 0.000 & 0.002 & 0.000 & 0.000 & 0.000 & 0.023 & 0.014 \\
Maelstrom & 0.497 & 0.466 & 0.391 & 0.379 & 0.754 & 0.715 & 0.641 & 0.612 \\
Blizzard  & 0.847 & 0.719 & 0.708 & 0.639 & 1.000 & .935 & 0.898 & 0.839 \\
\hline
\textbf{\textit{More Populated}}& & & & & & & & & &\\
Cyclone   & 0.868 & 0.804 & 0.716 & 0.666 & 0.969 & 0.901 & 0.889 & 0.843 \\
Hurricane & 0.887 & 0.740 & 0.717 & 0.641 & 0.984 & 0.902 & 0.911 & 0.843 \\
Squall    & 0.507 & 0.399 & 0.162 & 0.143 & 0.828 & 0.721 & 0.588 & 0.520 \\
\hline
\hline
\enddata

\tablecomments{
All surveys can observe almost all \HI\ gas in the total datacubes (within 1--3\% on average, see \rev{Table~\ref{tab:total_flux_ratios}}). Depending on the specific system, however, survey parameters perform drastically differently in recovering the CGM gas mass. \sRev{While the spatial filtering has a lesser effect on \LessActive\ systems, it can reduce recoverable CGM emission by $\sim$15\%, as defined by the difference between the Filtered and Smoothed ratios.}
The last two columns, analogous to what may be visible with SKA-MID, perform the best for each system. Relative survey performance for individual systems depends on the abundance of diffuse and compact components in the CGM. Values are calculated without satellite contribution.}

\end{deluxetable*}

\begin{figure}[ht!]
\begin{centering}
\includegraphics[width=0.49\textwidth]{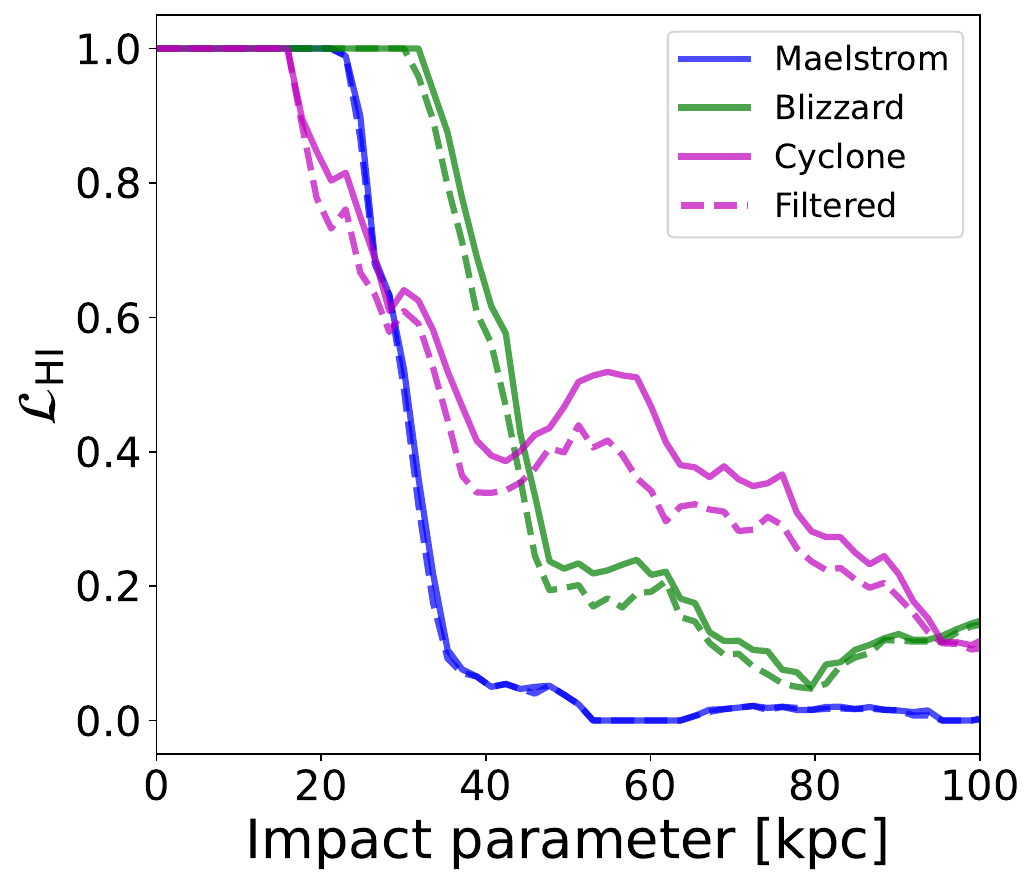}
\caption{ \HI\ covering fraction (\LHI) as a function of impact parameter for the ``smoothed'' (solid) and ``filtered'' (dashed) moment-0 maps shown in Fig.~\ref{fig:interferometric_projections} and Fig.~\ref{fig:interferometric_projections_2} at 20 Mpc. As expected, the curve for Maelstrom does not change significantly, as there is little diffuse material that is filtered out. For both Blizzard and Cyclone, the filtered curve is shifted to the left by a few kpc due to diffuse material being filtered out. Covering fractions are calculated in 2 kpc thick annuli centered on the given impact parameter.} 
\label{fig:interferometric_LHI}
\end{centering}
\end{figure}

\begin{figure*}[ht!]
\begin{centering}
\includegraphics[width=0.8\textwidth]{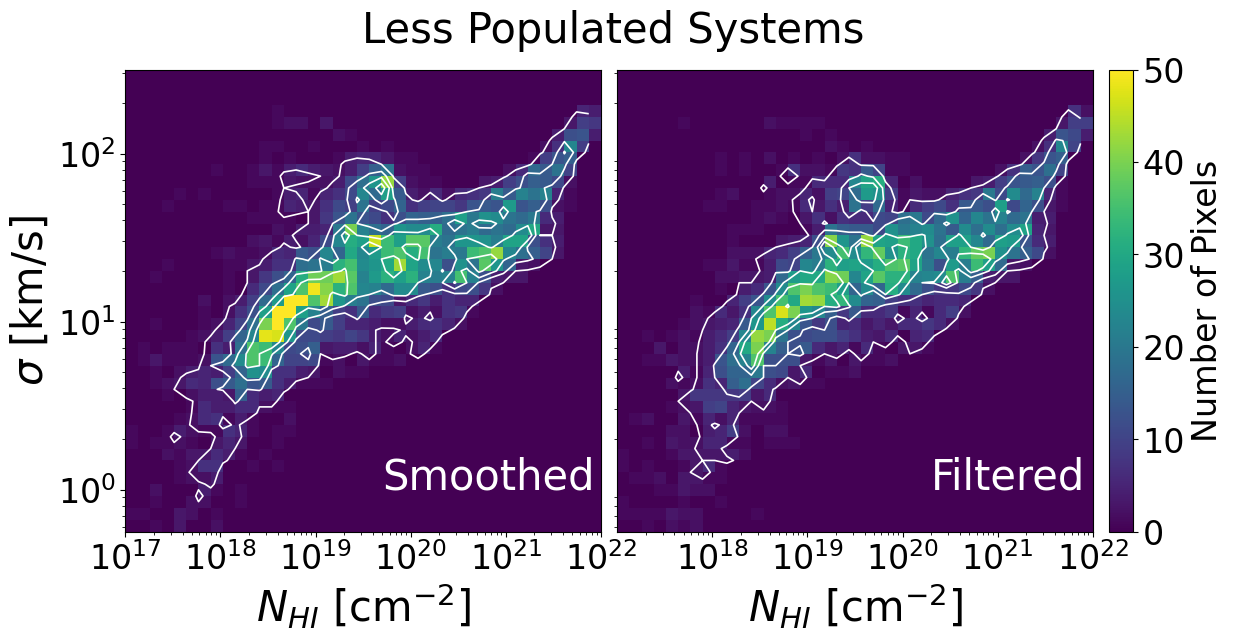}
\caption{
Log histogram of \HI\ column density ($N_{HI}$) versus velocity dispersion ($\sigma$) summed for the three \LessActive\ systems (Tempest, Maelstrom, and Blizzard) at 20 Mpc. $N_{HI}$ was corrected for inclination (i) by a factor of $\cos{i}$. The contours show the 25th, 50th, 75th, and 95th percentile curves. The synthetic survey parameters are analogous to the MHONGOOSE survey's low-resolution case (beam FWHM $\sim 65$'', sensitivity $\sim10^{18}~\rm{cm}^{-2}$, minimum baseline $= 29$ m). The plot on the left shows the distribution for the smoothed datacube (without filtering). There are two main peaks in the distribution: high column density and high velocity dispersion gas in the top right, and low column density and lower velocity dispersion gas in the bottom left. The plot on the right shows the same distribution for the filtered datacubes. The high column density peak is largely unaffected; however, the low column density peak is reduced by the filtering step. This implies that this gas is primarily diffuse and may not be visible to interferometric observations. 
\label{fig:interferometric_scatterplot}}
\end{centering}
\end{figure*}

\begin{figure*}[ht!]
\begin{centering}
\includegraphics[width=0.8\textwidth]{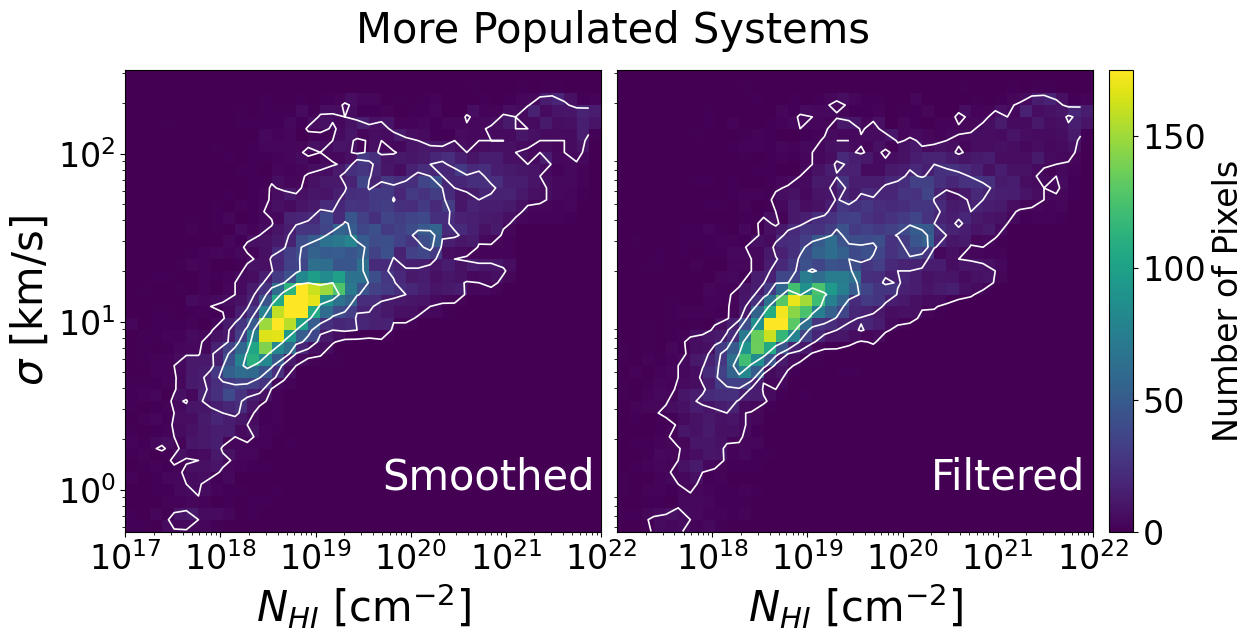}
\caption{
Same as Fig.~\ref{fig:interferometric_scatterplot}, but for the \MoreActive\ systems (Cyclone, Hurricane, and Squall) at 20 Mpc. The low column density peak in these systems is much more dominant. While this peak is reduced during the filtering steps, the effect is less dramatic than in the \LessActive\ systems.
\label{fig:interferometric_scatterplot2}
}
\end{centering}
\end{figure*}

\begin{figure}[ht!]
\begin{centering}
\includegraphics[width=0.475\textwidth]{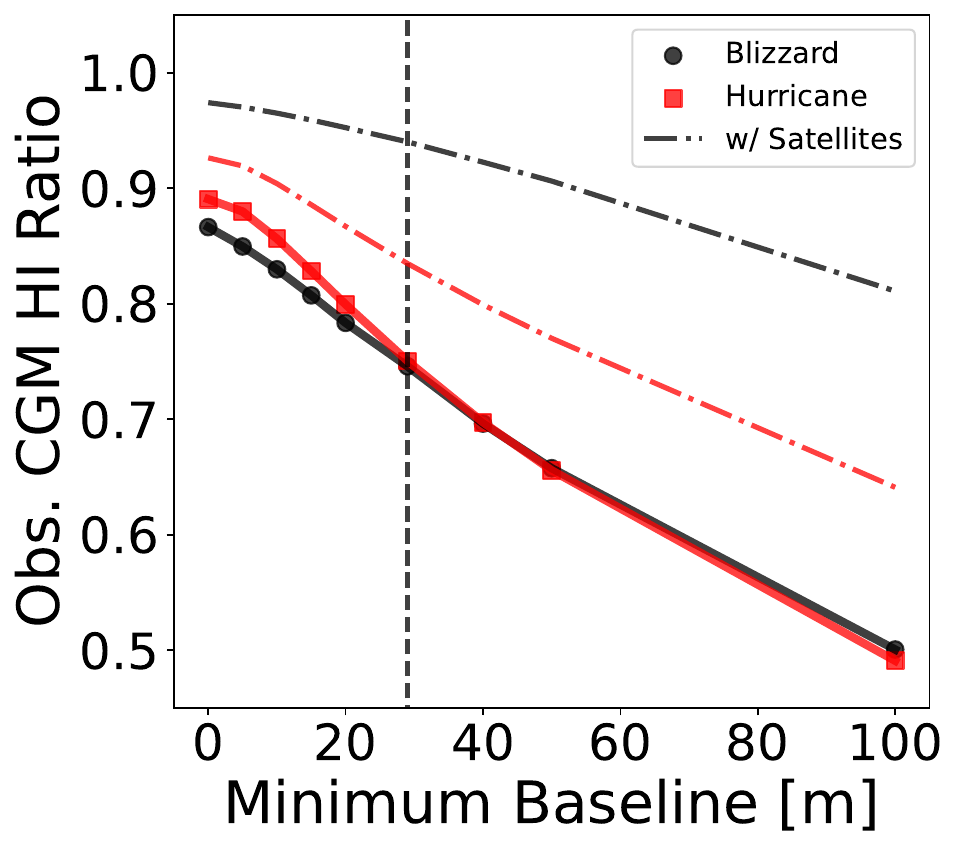}
\caption{
Evolution of how much \HI\ emission in the CGM is observable as a function of minimum baseline at 20 Mpc. The rest of the survey parameters are analogous to the MHONGOOSE survey's low-resolution case (beam FWHM $\sim$ 65'', sensitivity $\sim10^{18}~\rm{cm}^{-2}$).
Solid lines show the CGM ratio with satellites removed, while the dash-dotted lines show the ratio with satellites present.
The vertical dashed line shows the MeerKAT minimum baseline ($b_{min}=29~\rm m$). With satellites removed, observing $\sim$90\% of the CGM signal would require minimum baselines of a few meters, likely requiring single dish data.
\label{fig:minimum_baseline_dependence}}
\end{centering}
\end{figure}

Finally, we consider the effect that missing short baselines has on filtering out diffuse signals. Each baseline in an interferometric survey samples a specific location in spatial frequency space (the UV plane). Longer baselines sample high spatial frequencies, while shorter baselines sample lower spatial frequencies. Interferometers are limited by their lack of zero-distance baselines, meaning they are missing out on the most diffuse signals. In order to fully investigate the theoretical observability of the gas in our halos, we simulate this effect in a conservative way.

To start, we apply a Fast-Fourier Transform (FFT) to each spectral slice and then apply a Gaussian high-pass filter to remove the low spatial frequency components. The \rev{half-width-half-maximum} of the high-pass filter is set equal to the minimum observable spatial frequency for a given minimum baseline \rev{at the wavelength of the given slice ($b_{\rm min}/\lambda$)}. The selection of the Gaussian high-pass filter as opposed to a sharper filter was made to minimize any ringing artifacts in the resulting image. \rev{We do not consider sparse sampling of the rest of the UV plane, experimentally isolating the effects of the missing short baselines} 

\rev{Although we use a Gaussian high-pass filter, significant filtering artifacts were still introduced due to removing the low-frequency components.} To extract all significant signal, we transformed the filtered images back into image space and deconvolved the resulting dirty images with the H{\"o}gbom CLEAN algorithm \citep{hogbom74-CLEAN}.
\rev{The dirty beam is known from the spatial smoothing and filtering steps, and is uniform across the image in this case. To match the deep cleaning done in the full-depth MHONGOOSE images \citep{deBlok24-Mhongoose}, each slice was cleaned individually down to 0.5$\sigma$, where $\sigma$ is the noise level in the dirty image. We utilize a clean mask generated from the ideal image, only cleaning real components above 0.5$\sigma$. This is primarily done for computational efficiency and to avoid cleaning noise. Cleaning using the auto-masking strategy utilized in the MHONGHOOSE survey \citep{deBlok24-Mhongoose}, in which a clean mask is iteratively developed by cleaning to successively deeper noise levels, gives similar results.} The sum of the CLEAN components and the residuals\footnote{\rev{We do not correct the residuals, as the correction factor should be $\sim$1 in this case.}} was used to create our ``filtered'' datacubes. \rev{We note that in the case of small/zero-width Gaussian high-pass filters, this process recreates the smoothed image.} 

The final step in our synthetic datacube pipeline is the creation of source masks to identify spaxels with statistically significant signal. This was done using the {\sc SoFia-2} software package \citep{westmeier21-sofia2}. Parameters were chosen to mimic the corresponding observational surveys \citep{deBlok24-Mhongoose}. In brief, we use the SoFiA-2 ``smooth and clip (S+C)'' method using spatial kernels of 0 and 4 pixels, velocity kernels of 0, 9, and 25 channels, and a threshold of 5$\sigma$. \rev{This was run on the noisy, smoothed, and filtered datasets.}
 
We chose four different combinations of parameters to target four different synthetic surveys. \rev{These parameters are summarized in Table~\ref{tab:survey_parameters}. This parameterization simplifies the differences between various surveys, isolating the effects of sensitivity, spatial resolution, and the minimum baseline length. The first two are analogous to the MHONGOOSE survey at two different resolution targets: low-resolution (MHONGOOSE-LR) and high-resolution (MHONGOOSE-HR) \citep{deBlok24-Mhongoose}. The final \rev{two surveys are} analogous to what a future SKA may achieve at the same resolution targets. Their survey parameters were selected to be similar to MHONGOOSE's performance. Sensitivity was estimated following \citet{deBlok24-Mhongoose}, wherein it is scaled up proportionally to the SKA-MID baseline design collecting area relative to MeerKAT's collecting area. As the authors note, this neglects different baseline distributions, dish designs, and antenna temperatures.} \sRev{The diameter of SKA antennae will be different than MeerKAT, resulting in a slightly smaller primary beam; however, we do not account for this difference.}

\rev{To check that our filtering/cleaning steps recovered a realistic amount of \HI\, we considered the change in integrated flux between the ideal, smoothed, and filtered datacubes. We show the ratio of the integrated flux in the total (disk, satellites, and CGM) filtered datacube to the total ideal datacube in Table~\ref{tab:total_flux_ratios}. The smoothing and filtering steps reduce the overall \HI\ signal by only $\sim1-3\%$ on average. The most extreme case (Squall) loses $\sim6-9\%$ of its signal, similar to the average difference between the MHONGOOSE and GBT observations \citep{deBlok24-Mhongoose}. Note, when comparing the total filtered cubes to the smoothed datacubes, the filtered datacubes have 0-2\% less flux.} 
 
\section{The effects of filtering out diffuse components} \label{sec:synthetic_obs_results}

Fig.~\ref{fig:interferometric_projections} and Fig.~\ref{fig:interferometric_projections_2} show the effects that each step of our synthetic imaging pipeline has on the column density projections (moment 0 map). These figures mimic the MHONGOOSE survey's low-resolution case (beam FWHM = 65'', sensitivity~$\sim10^{18}~\rm{cm}^{-2}$, primary beam FWHM  = 1\textdegree, \citealt{deBlok24-Mhongoose}). Of particular interest is the difference between how diffuse and compact components respond to various steps.

The first column (``ideal'') shows the ideal datacube at the target pixel resolution.
The second column (``noisy'') shows the effect of the column density cutoff imposed by the noise parameter. This step cuts out most of the diffuse gas.
The third column (``smoothed'') shows the effect of the Gaussian smoothing. Aside from the obvious degradation of the spatial resolution, this also has the effect of reducing some small spatial scale signals below the column density sensitivity.
The fourth column (``filtered'') shows the effects of the spatial filtering and cleaning. Diffuse components on the scale of the maximum observable size are filtered out. The disk itself is largely unaffected; however, the emission at the disk edge and in the CGM is reduced. For Maelstrom and Tempest, which do not have much detectable diffuse gas at this sensitivity, there is little effect. For the other halos, there are more significant effects due to the presence of more diffuse, high column density gas in these systems.

 Fig.~\ref{fig:interferometric_LHI} shows the covering fraction of the smoothed and filtered images as a function of impact parameter for three representative halos with the MHONGOOSE-LR survey parameters. For Maelstrom, there is little difference between the two curves. For Blizzard and Cyclone, however, the filtered curve is shifted down slightly. This implies that for systems with more diffuse extended emission, such as is seen in the \MoreActive\ systems and Blizzard due to its more recent interactions, interferometric studies \rev{may underestimate disk sizes and covering fractions slightly}. This reduction will only occur when there is a significant amount of spatially diffuse signal at the disk edge above the column density limit, and is therefore highly galaxy-dependent.

The quantitative effects of the smoothing and filtering steps \rev{for the CGM} are shown in Table~\ref{tab:observed_masses}. This table shows the observable mass fractions (ratio of the mass in the smoothed/filtered images versus the ideal images). Each row shows only the mass ratio for the CGM, with the contributions from the disk and satellites removed.
The detectable \HI\ emission from the disk (and therefore the system as a whole) is largely unaffected by both the smoothing and spatial filtering steps, with each survey able to detect $\sim$98-99\% of the HI. The disk is masked out using our disk definition (see \citetalias{trapp25}) projected onto the image and smoothed at the target survey resolution.
The disk removal is based on the full 3D gas distribution in the simulation and does not depend on the chosen line of sight.
To mask out the satellites, we initially identify subhalos via {\sc ROCKSTAR}, and then mask them out using the same method as was used for the central disk.

\rev{For systems with more extended \HI\ emission, the filtering step reduces the observable CGM \HI\ mass by up to $\sim$15\%.} The first two columns are analogous to the MHONGOOSE survey at two different resolution targets \citep{deBlok24-Mhongoose}. The two surveys on the right are analogs to what may be possible with SKA-MID. The relative performance between surveys varies system by system and depends on the spatial scales of the CGM gas. \sRev{In general, the higher resolution parameterizations see less of an effect from the short-spacing problem than their counterparts. This is because the spatially diffuse material that gets filtered out is preferentially low column density, and therefore lost to noise, similar to what is seen in observations \cite[e.g.][]{wang2024-FEASTS_I}.}

\fRev{Depending on the system and survey parameters, much of the \HI\ outside the central disk can be missed due to having too low of a column density. Tempest is a unique case, as most of its CGM \HI\ is in small-scale, low column density clumps that are not detectable at the physical resolutions and sensitivities considered here. Other systems/survey parameterizations can lose anywhere from $\sim$0-70\% of the \HI\ solely due to sensitivity limitations. See Appendix~\ref{sec:appendix_survey_vis} for a quantification of each step in this pipeline.}

Fig.~\ref{fig:interferometric_scatterplot} shows 2-D histograms of the column density versus velocity dispersion of each pixel in the smoothed and filtered images for the three \LessActive\ halos in our sample (Tempest, Maelstrom, and Blizzard) for the MHONGOOSE-LR survey parameters. The histograms for the smoothed datacubes show two peaks at high and low column densities \rev{to facilitate direct comparisons with MHONGOOSE survey results \citep{deBlok24-Mhongoose}.} The low-column-density peak \sRev{($\sim5\times 10^{18}\,\rm{cm}^{-2}$)} is not seen in observations \rev{of galaxies in this mass range}, despite being present in other simulations \citep{marasco25:mhongoose-sim-comp,lin2025-FeastsCompWithSimulations}. The histograms for the filtered datacubes show a reduced peak, implying that this gas may not be seen in purely interferometric studies. \rev{At this resolution and distance, there is a population of high column density gas at high projected dispersion ($\sim$ 100 km/s), which is not seen in observations. This reduces the concentration of the high column density peak and is likely a result of the overconcentrated star formation \citep{wright24}.}

Fig.~\ref{fig:interferometric_scatterplot2} shows the same 2-D histograms for the \MoreActive\ systems, which show relatively lower column density, low velocity dispersion gas in both cases. The overall filtering effect of reducing this peak is the same, although the low column density peak is still dominant. The diffuse gas in these systems, despite having relatively little mass compared to the central disks, occupies large volumes and is, therefore, represented by a large number of cells in the simulation and spaxels in the resulting \HI\ datacubes. This causes the low column density peak to remain dominant over the higher column density peak. The removal of satellites had little effect on these plots.

\rev{This implies that the \MoreActive\ systems have an overabundance of this low column density \HI\ compared with the observed galaxies in the MHONGOOSE survey. Given that these systems host a relatively large number of satellites compared with the \LessActive\ systems, this could be in part a selection effect or point to an underlying issue with the \HI\ distribution in the simulation. We note that, unlike other simulation comparisons \cite[e.g.][]{marasco25:mhongoose-sim-comp,lin2025-FeastsCompWithSimulations}, the FOGGIE simulations devote more computational power to resolving the CGM, where most of this discrepancy lies. Explicit treatment of radiation transfer may, in part, alleviate this tension \cite[e.g.][]{lucchini26}.}

Finally, Fig.~\ref{fig:minimum_baseline_dependence} shows how the observable CGM \HI\ ratio depends on the minimum baseline length for Blizzard and Hurricane. All other survey parameters match the MHONGOOSE-LR survey. \rev{The observable CGM ratio begins to flatten out around a minimum baseline of 5 m.} The points at 0 m represent what can be observed at the given sensitivity limit for the given smoothing, \rev{and are equivalent to the smoothed images, implying the deconvolutional process is reconstructing the bulk of the signal.} Given that dish sizes for the Very Large Array (VLA) and MeerKAT are 25 m and 13.5 m, respectively, such small baselines are unrealistic. Joint deconvolutions with single dish data \rev{would likely be needed} to recover this signal.

\section{Effects of Viewing Angle}\label{sec:viewing_angle_effects}

\begin{figure*}[ht!]
\begin{centering}
\includegraphics[width=0.75\textwidth]{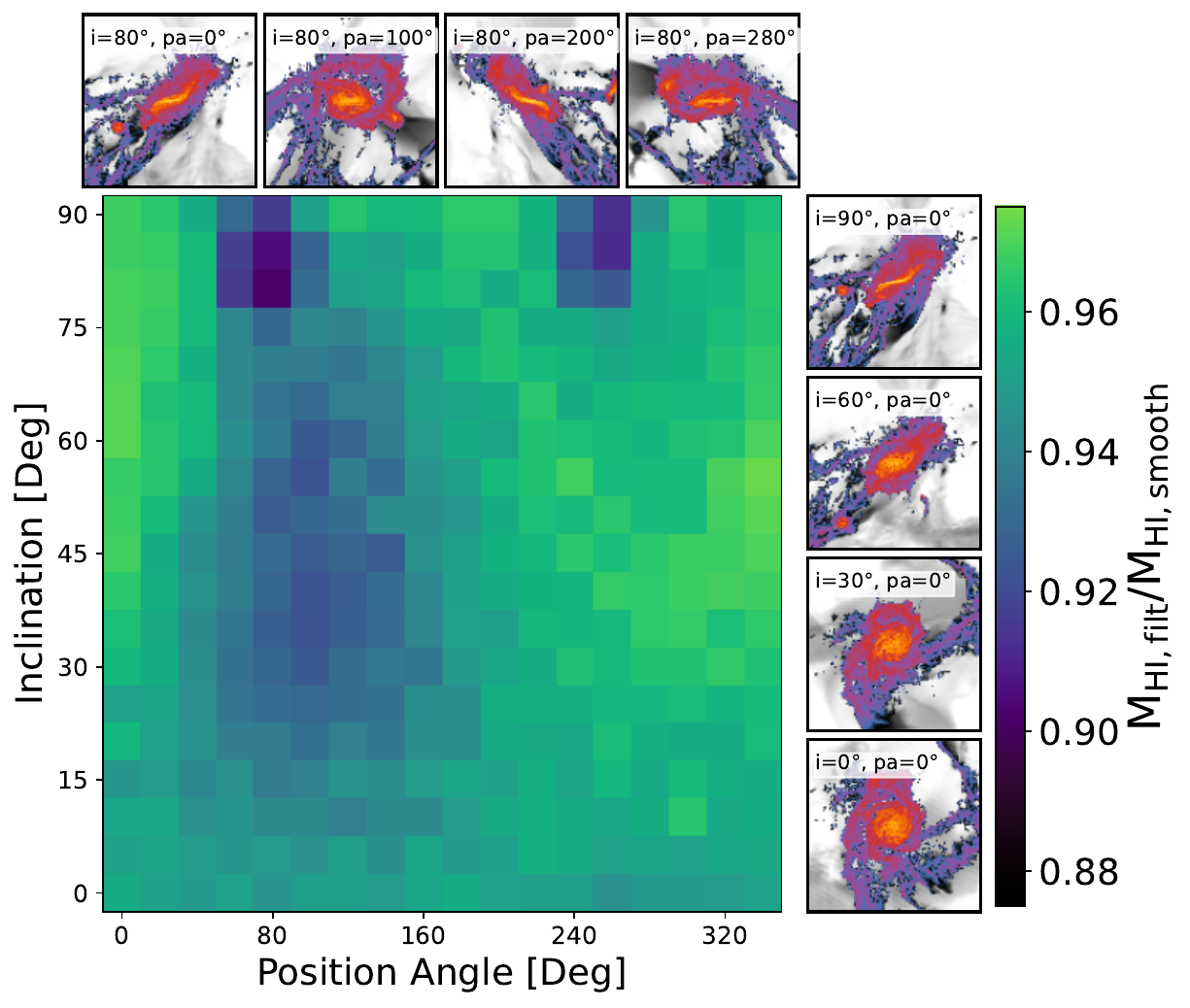}
\caption{
Ratio of observable \HI\ in the filtered image to observable \HI\ in the smoothed image for Blizzard at various observed inclinations and position angles (at $z=0$, D=20 Mpc). This isolates the effect of missing short baselines. Imaging parameters correspond to MHONGOOSE-LR. Projections on the top and right show the ideal \HI\ column density maps at various orientations. There is a general increase in this ratio at higher inclination, as edge-on galaxies appear more compact on the sky. At lower inclinations, there is little dependence on position angle; however, there is a clear pattern at higher inclinations. When the galaxy is more edge-on, there are two regions of lower observability at position angles of $\sim$100\textdegree~and $\sim$280\textdegree~corresponding to when the warped component of the disk is face-on with the observer, causing dips in the observable CGM ratio.
\label{fig:angular_dependence}}
\end{centering}
\end{figure*}

\begin{figure*}[ht!]
\plotone{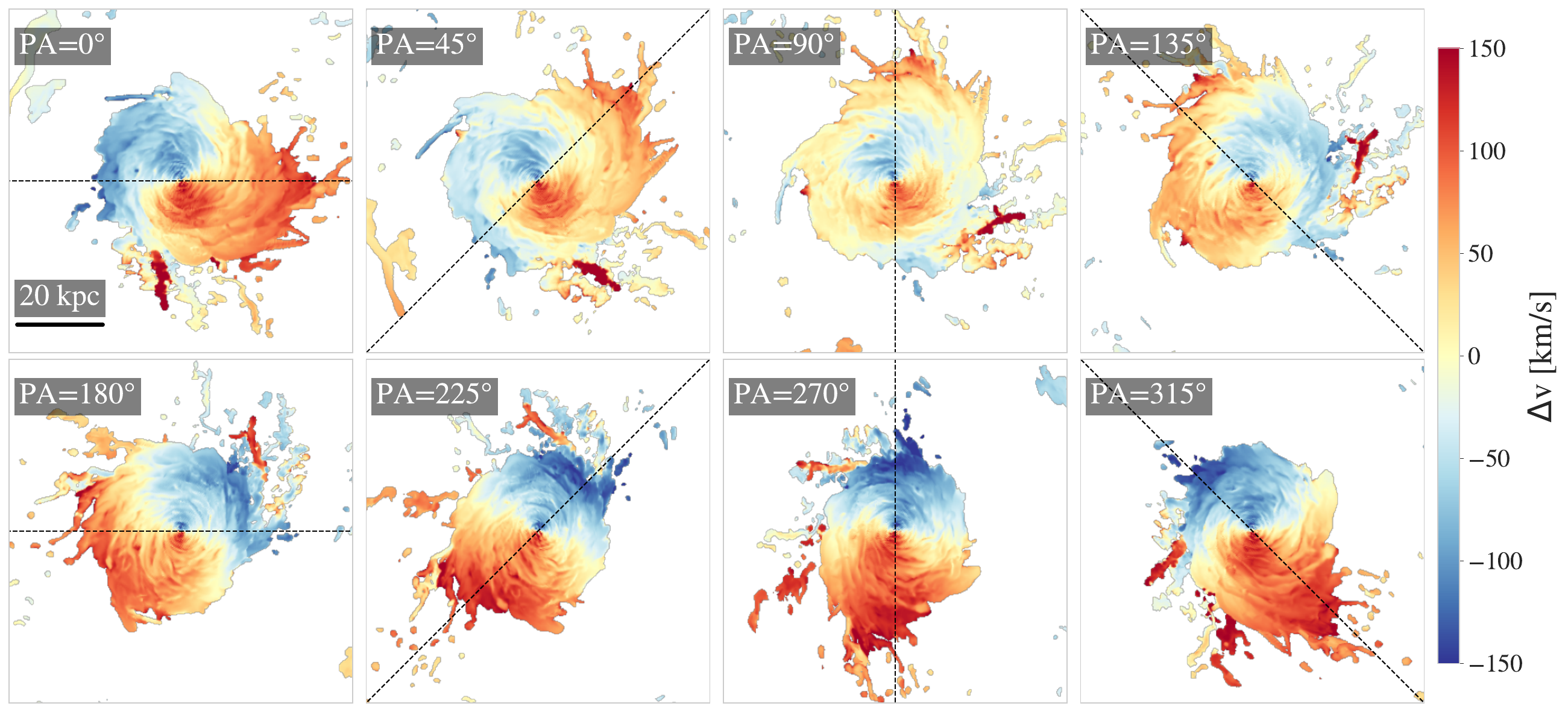}
\caption{
Ideal first moment maps (projected velocity component) for Tempest at various position angles. Only pixels above $N_{HI}>10^{16}$ cm$^{-2}$ are shown for visual clarity. All images are projected at an inclination of 20\textdegree~at redshift $z=0$. The black dashed line shows the direction of the warp, fit by eye. The warp causes what appear to be velocity inversions when it is anti-aligned with the observed disk major axis (PA=90\textdegree).
\label{fig:tempest_projected_velocity}}
\end{figure*}

\begin{figure*}[ht!]
\plotone{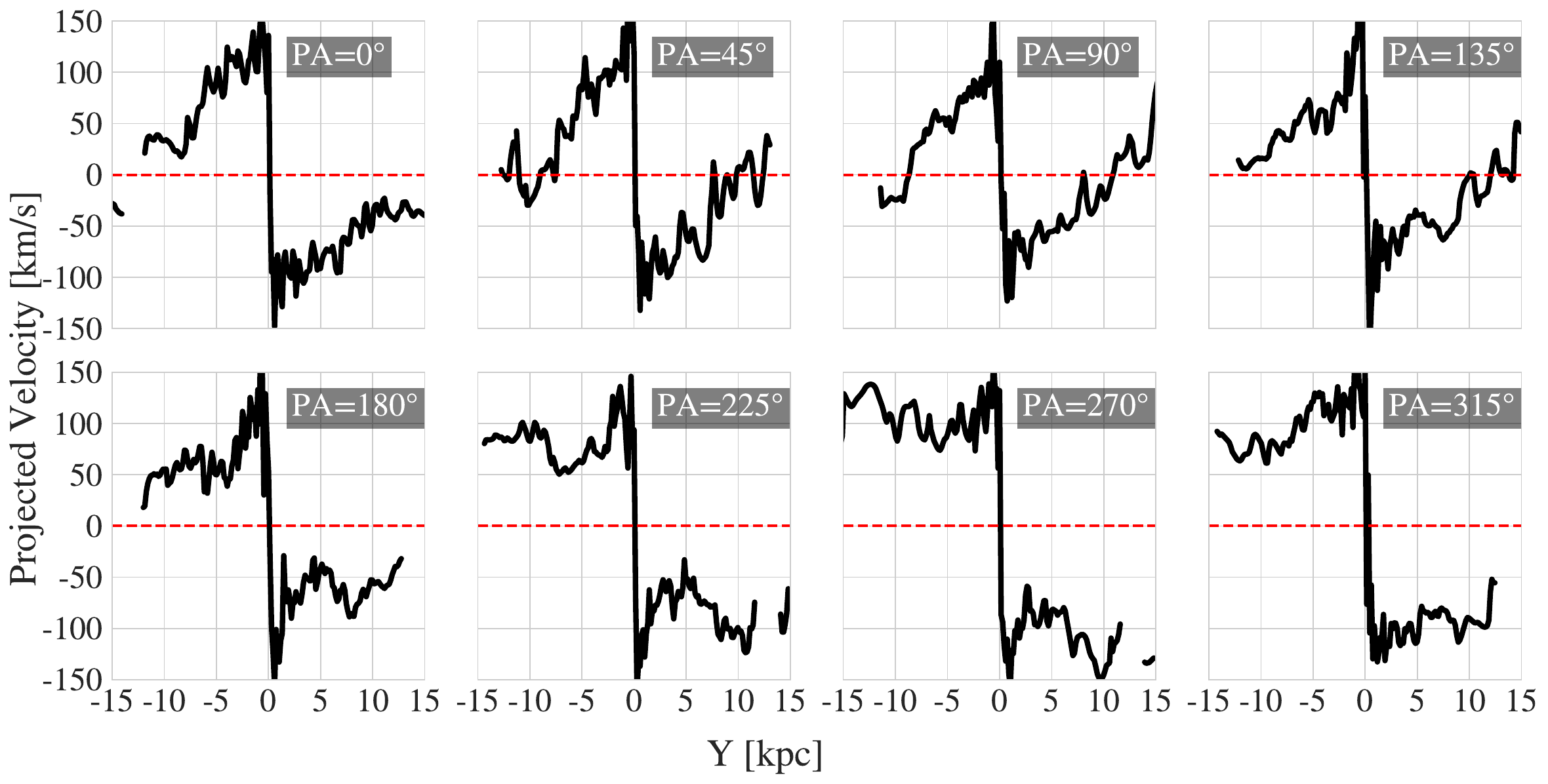}
\caption{
Corresponding Position-Velocity plots for the ideal first moment maps shown in Fig.~\ref{fig:tempest_projected_velocity}. Velocities are taken from the disk minor axis. Velocity inversions caused by the warp can be seen in $PA=45$\textdegree$-135$\textdegree. The rotation curves appear much flatter in $PA=225$\textdegree$-315$\textdegree, despite being strongly centrally peaked. 
\label{fig:tempest_ideal_pv_diagrams}}
\end{figure*}

The observed viewing angle can have significant effects on what information can be reliably inferred from observation. The most salient effect is the inclination of the system towards or away from the observer. \rev{This effect is well known in how it can affect the} observable structure of the galaxy (i.e., edge-on versus face-on) \cite[e.g.][]{burstein91} as well as \rev{line of sight velocity contributions} of various velocity components (e.g., rotational velocity is easier to measure in highly inclined disks, while vertical velocities are easier to measure in less inclined disks) \citep{rogstad74,wong04,teodoro15-3dbarolo}. The observed position angle versus the direction of the misalignment is of particular importance when considering misaligned gas, as is seen in the outer disks of many of our systems, as well as their CGMs. In this section, we discuss specifically how these effects relate to the observed spatial scales on the sky and their consequences for observability in interferometric studies, as well as how these misaligned features manifest dynamically at a variety of viewing angles with Blizzard and Tempest as case studies.

For this section, we discuss two viewing angles: inclination (i) and position angle (PA). The inclination angle describes the tilt of the disk towards or away from the observer, with a value of $i=90$\textdegree~corresponding to an edge-on view, and a value of $i=0$\textdegree~corresponding to a face-on view. In a coordinate system with the z-axis defined by the rotational axis of the disk, this inclination angle is equivalent to the pitch of the system. The position angle describes the rotation around the z-axis, equivalent to the yaw of the system. In a cylindrically symmetric system, this position angle would have little to no effect. In systems with misaligned extended disks and CGMs, the position angle can affect the projected spatial scale and observed dynamics, as discussed below. These angles can be visualized in the projections on the top and right of Fig.~\ref{fig:angular_dependence}.

The projected spatial scale of the \HI\ on the sky is highly relevant to the study of extended \HI\ distribution with radio interferometers. For a well-defined, disky system, an edge-on (high inclination) view will have relatively small spatial scales. A face-on view (low inclination), on the other hand, will take up a larger portion of the sky. Fig.~\ref{fig:angular_dependence} shows how the observable \HI\ ratio in the CGM varies with both inclination and position angle for Blizzard. As expected, the observable ratio generally increases with increasing inclination, as the edge-on projection is more compact and therefore there is less spatially diffuse signal that is lost to the missing short baseline problem. This is complicated by the significant warp in the outer disk of Blizzard. This results in a significant evolution with position angle at intermediate and high inclinations. At low inclinations ($\sim$0\textdegree--15\textdegree), there is little evolution with position angle. What evolution there is is likely caused primarily by field of view \rev{and noise effects}. At intermediate inclinations ($\sim$15\textdegree--75\textdegree), there is a region of lower observability around PA=100\textdegree. At high inclinations ($\sim$75\textdegree--90\textdegree), a second region of lower observability appears around PA=280\textdegree. These position angles correspond to where the warp of Blizzard is more face-on, causing it and kinematically aligned CGM gas to take up larger spatial scales in the sky. Further misalignment of the CGM gas exaggerates this effect at PA=100\textdegree, while limiting it at PA=280\textdegree\, in this case. This trend with inclination is present in the rest of the systems, and the trend with position angle is similar in other systems with misaligned features. We specifically do not remove satellites from this plot to avoid confusion in masking out slightly different regions in different projections. Overall trends remain the same.

\rev{As discussed in \citetalias{trapp25}, Tempest shows a significant warp in its outer disk. Of interest is how this appears in projection, particularly when viewed at lower inclinations, where it will not be as readily apparent.} Fig.~\ref{fig:tempest_projected_velocity} shows how the ideal first moment maps for Tempest vary as a function of observed position angle at a fixed inclination ($20$\textdegree). \rev{These plots show line-of-sight velocities, which relate to the 3D velocities of the gas as:}
\begin{equation}
\begin{aligned}
V_{\mathrm{los}}(R) =\; & V_{\mathrm{sys}} \\
&+ V_{\mathrm{rot}}(R)\cos\theta\,\sin i \\
&+ V_{\mathrm{rad}}(R)\sin\theta\,\sin i \\
&+ V_{z}(R)\cos i
\end{aligned}
\end{equation}
\rev{wherein $V_{\rm{los}}(R)$ is the line of sight velocity as a function of radius, $V_{\rm{sys}}$ is the systematic line of sight velocity offset, $\theta$ is the azimuthal angle in the disk plane measured from the projected major axis, and $i$ is the observed inclination. $V_{\rm rot}$, $V_{\rm rad}$, and $V_{z}$ are the three velocity components in cylindrical space defined by a disk or ring at a given inclination. Modeling these three velocity components can be difficult, as they can often be degenerate depending on inclination and misaligned structure \cite[e.g.][]{wong04}.}

The choice of position angle=0\textdegree~is arbitrary and was chosen such that the warp orientation (dashed line) was left to right at PA=0\textdegree. The large vertical velocities associated with the orbits of gas in the warp appear as a deviation from the dipole signature associated with disk rotation. The strength and sign of this deviation depend primarily on the angle between the warp axis and the observed kinematic axes of the galaxy. When the warp is aligned with the disk major axis (PA=270\textdegree), there is no qualitatively noticeable effect, although the projected velocities increase at the warp. When the warp is more aligned with the disk minor axis (e.g., PA=0\textdegree, 180\textdegree), there are noticeable deviations from rotation. When the warp is anti-aligned with the observed disk major axis (PA=90\textdegree), a velocity inversion is present, where the sign of the projected velocity switches near the disk edge.

Such velocity inversions can be observed in \HI\ observations of galaxies \rev{and are often signs of warps} \citep{lin2025-FEASTS-M51, Healy_2024-WarpedMhongooseDisk} (see Sec.~\ref{sec:disc-comparison_between_sim_and_obs}).  This is further visualized in Fig.~\ref{fig:tempest_ideal_pv_diagrams}, which shows the corresponding Position-Velocity diagrams for these plots. Velocities were taken from the central pixel along a vertical line in each orientation. Notably, when the warp axis is perfectly aligned with the observed disk major axis (PA=270\textdegree), the \rev{projected velocities} appear relatively flat despite having a strong central peak. 

\rev{Multiple modeling techniques exist to extract some or all of these velocity components, including global fits to 2D moment maps \cite[e.g.][]{sellwood15-Diskfit}, 2D tilted ring models \cite[e.g.][]{rogstad74,oh18-2DBAT}, 3D tilted ring models \cite[e.g.][]{jozsa07-Tirific,teodoro15-3dbarolo}, and other fits to 3D datacubes, such as {\sc Rotcurve} \citep{shachar26-RotCurve}, {\sc GalPak3D} \citep{Bouche15-Galpak3d}, and {\sc DysmalPy} \citep{price21-DysmalPy}. The projection of various velocity components can become degenerate at lower inclinations ($\lesssim25$\textdegree), and can become increasingly degenerate if there is significant warping \citep{jozsa07-Tirific,kamphuis15}.} \sRev{Note, we only visualize projections in Fig.~\ref{fig:tempest_projected_velocity}, while many of these techniques fit the full 3-D datacube.}

\section{Discussion of Implications For Observations} \label{sec:discussion}

While the FOGGIE simulations broadly agree with the observed \HI\ Size-Mass relation, the details of how much \HI\ there exists outside the inner disk, as well as the observable orientation and dynamics of the \HI\ in the extended disk, depend heavily on the observational parameters and observed viewing angle. In order to interpret observations properly and make robust comparisons to simulations, these effects should be taken into account. In the following subsections, we will discuss the feasibility of observing the extended \HI\ emission outside of the disk in real galaxies (Sec.~\ref{sec:disc-how_to_observe}), how to compare between simulations and observations (Sec.~\ref{sec:disc-comparison_between_sim_and_obs}), and numerical resolution effects important in this study (Sec.~\ref{sec:disc-on_resolution}).

\subsection{Observing extended \HI\ emission} \label{sec:disc-how_to_observe}

Detecting low column density gas in the CGM has been a longstanding problem of interest in astronomy. Historically, the detection of this gas was largely constrained to absorption studies \cite[e.g.,][]{chen12,nielsen13,werk14,2015ApJ...813...46B,kacprzak15,berg2019,peroux19,zahedy19,lehner25}, as column densities were too low to be observed in emission. Simulation studies have predicted a variety of morphologies, and often show sharp drop-offs in \HI\ column density at the disk edge \cite[e.g.,][]{mina21,trapp22,rahul23-TNG50_CGM,piacitelli25}. Many of these simulation suites focus their resolution elements on high-density regions, which preferentially targets the inner star-forming disk. The simulations presented here instead focus computational power on better resolving the material in the CGM, allowing for increasingly complex structures to emerge that are not seen in other suites. This further complicates the picture, as it allows for both spatially diffuse and spatially compact structures to be resolved within the CGM of our simulations (see \citetalias{trapp25} for more details). To fully characterize the \HI\ outside the disk, studies must be sensitive to \emph{both} of these spatial scales.

Even with only the six galaxies presented here, we see a large variance in the amount of detectable CGM material in different interferometry studies (Fig.~\ref{fig:interferometric_projections},~\ref{fig:interferometric_projections_2}). In systems like Tempest, for instance, small spatial scale, clumpy material dominates the CGM. This material is almost entirely missing from \rev{the synthetic} observations due to sensitivity and resolution effects for most instruments, and the effects of missing short baselines are largely negligible. \rev{In Maelstrom, the small-scale clumpy material is largely detectable in the smoothed dataset. Filtering has little effect on these small-scale features, reducing the observable CGM fraction by a negligible amount.}

Systems like Blizzard or Hurricane, however, show much more diffuse signal. While signal is still lost to resolution effects, much is also lost due to the missing short baseline problem. Furthermore, the effects of this spatial filtering can be highly dependent on the observed viewing angle of the system in combination with its morphology. This makes the formulation of a simple correction to account for this missing signal impossible, or at the very least, subject to large uncertainties. While a loss of up to \rev{$\sim$15\%} of observable CGM \HI\ due to interferometric effects may seem acceptable, it is important to keep in mind that this effect is strongly biased towards missing diffuse components of gas. As it is currently unclear how much of the actual baryon content in the CGMs of real galaxies is in the diffuse component versus clumpy structures, \rev{and likely varies significantly based on the local environment, interaction history, and relative \HI\ abundance of a system, it is difficult to know how much may be missing from a given observation.}

\rev{The dense sampling of short baselines in SKA precursors is important in making sure as much of this signal is retained as possible, allowing these surveys to detect much of the observable (and total) \HI\ (Table~\ref{tab:total_flux_ratios}).} The next generation of interferometers will \rev{even deeper column densities to be probed}. As seen in Table~\ref{tab:observed_masses}, the enhanced spatial resolution and sensitivity that an instrument similar to SKA-MID will provide will allow for much more of this material to be detected. Systems still suffer from the missing short baseline problem to some extent. To alleviate this, these spatial frequencies should be sampled \rev{or reliably inferred}. As seen in Fig.~\ref{fig:minimum_baseline_dependence}, depending on the system, baselines of down to 5 meters \rev{need to be sampled to characterize all observable material at the target sensitivity.}
\rev{As introduced previously, one known way to directly address this problem is joint deconvolution of interferometric data with large single-dish radio data.}
There have recently been an increasing number of studies utilizing data from the Five-hundred-meter Aperture Spherical radio Telescope (FAST) \citep{wang2024-FEASTS_I} as well as the Green Bank Telescope (GBT) \citep{eibensteiner2023,koch2025-GBT_VLA_DualConv}. The \rev{continued} combination of large single dishes with high-sensitivity interferometry surveys such as MHONGOOSE \citep{deBlok24-Mhongoose,wang26-FeastsAndMhongoose} and future surveys with SKA-MID will be of particular interest.

\subsection{On comparisons between observation and simulation} \label{sec:disc-comparison_between_sim_and_obs}

As instruments become able to probe deeper \HI\ column densities and cosmological simulations continue to advance in both resolution and complexity of physics, robust comparisons between the two are vital for both validating simulations as well as understanding and interpreting observations. We highlight two broad effects in this study. As emphasized in Fig.~\ref{fig:interferometric_scatterplot}, the consideration of missing short baselines alters the observed distribution of gas.
With this consideration, the three \LessActive\ systems in our sample track recent observations much more closely \citep{marasco25:mhongoose-sim-comp,lin2025-FeastsCompWithSimulations}, although the \MoreActive\ systems still show an overabundance of low column density \HI. The importance of robust forward modeling applies to many different topics in astrophysics, including HI-based rotation curves \cite[e.g.,][]{maccio16, brooks17,ruan25}, metallicity gradients \cite[e.g.,][]{hemler21:TNG50-Zgradients, Acharyya24,graf24:FIRE-MetallicityGradients}, simulation-based inference \cite[e.g.,][]{hahn2022}, and more.

A second important effect is the significant and non-intuitive effects that the viewing angle has on these systems. The effect of inclination and the observed position of the disk relative to the major axis is well known on measured kinematics. As shown in Fig.~\ref{fig:angular_dependence}, both the observed inclination and position angle of a warped system can have significant effects on the observable amount of CGM material. Similarly, as shown in Fig.~\ref{fig:tempest_projected_velocity} and Fig.~\ref{fig:tempest_ideal_pv_diagrams}, the observed position angle of a warped system can have dramatic effects on the observed kinematics, leading to velocity inversions in one orientation \rev{and flat projected velocity curves in the} opposite orientation. \rev{The projected effects of these features can be highly degenerate in various model parameterizations \citep{jozsa07-Tirific,kamphuis15}, particularly at lower inclinations}. If models that are fit to \sRev{the 3-D channel maps} of similar galaxies cannot reliably fit these types of inclination warps, or if we do not have a characteristic signature to know a warp is present, the fits for both rotational and radial velocities may be compromised.

As \HIcm\ observations probe deeper column densities, it should be expected that more signatures of these warps may be found. For instance, \citet{Healy_2024-WarpedMhongooseDisk} shows evidence of a warp in the extended \HI\ disk of NGC 5068 in the MHONGOOSE survey \sRev{using state-of-the-art tilted rings fit to the 3-D datacube}, with the outer regions of the first moment map twisted in an S-shape. The pattern of this velocity field is strikingly similar to the velocity field seen in Tempest at a position angle of 45\textdegree, as seen in the second panel of Fig.~\ref{fig:angular_dependence}. If this warp was rotated in some other orientation with respect to the inclination axis, this signature would not be so easy to point out in a simple moment map and would require advanced modeling techniques to recover.

Similarly, \citet{lin2025-FEASTS-M51} shows a stark velocity inversion between the inner disk and extended, diffuse \HI\ of M51 in the FEASTS survey. Importantly, FEASTS utilizes joint deconvolutions with interferometric data from the VLA \citep{Walter_2008-THINGS} and single dish data from the FAST telescope; however, they do not explicitly claim a warped structure in the extended disk. The velocity field signature is reminiscent of what is seen in Tempest at a position angle of 90\textdegree~(Fig.~\ref{fig:angular_dependence}), however, to a much larger extent. This implies that there is a large warp in the extended, diffuse material of M51, likely related to interactions with the nearby dwarf galaxy. This velocity inversion requires a specific orientation in the sky with respect to the warp. Given the size of the M51's extended disk, it is more likely to be observed than more modest warps.

\subsection{On Spatial Resolution} \label{sec:disc-on_resolution}

Important to the consideration of \HI\ in extended disks is the numerical spatial resolution of the simulations and how it compares to both the spatial resolution of observations as well as the maximum observable scale of an interferometer. The FOGGIE simulations are particularly well suited to this type of study, as their spatial resolution at redshift $z=0$ ($\Delta x= 0.274~ \rm{kpc}$) is of similar resolution to current interferometers (i.e. $\sim$0.262 kpc for MeerKAT at 10 Mpc). Additionally, the forced resolution in the CGM ($\Delta x= 1.10~ \rm{kpc}$) is much smaller than the maximum observable scale ($\sim$60 kpc for the 21-cm line at 10 Mpc). For simulations that do not enforce a fixed spatial resolution in the CGM, caution must be taken when drawing comparisons. If the scale of the resolution element is close to this maximum observable scale, the CGM may not be visible when spatially filtered due to purely numerical resolution effects. 

As discussed previously, the \HI\ emission that is lost due to the missing short baseline problem tends to come from low column density material outside the disk, meaning it may be difficult for classical zoom-in simulations to resolve. Of interest is the mass resolution of the cells in our simulations that end up being filtered out in the spatial filtering step. In this study, the \HI\ weighted average cell mass of these cells is $\sim10^{3} ~M_{\odot}$. This is well below the mass resolution of other state-of-the-art simulations ($\sim10^{4}~M_{\odot}$), implying that FOGGIE is better situated to resolve this missing material. For a more detailed discussion of resolution in the FOGGIE disks and CGM, see Appendix A of \citetalias{trapp25}.

\section{Summary}

This is the second part of a two-paper series in which we investigate the evolution and observable properties of the extended \HI\ disks in Milky Way-mass galaxies using zoom-in simulations from the Figuring Out Gas \& Galaxies in Enzo (FOGGIE) simulation suite. \citetalias{trapp25} focuses on the origin and properties of the extended and misaligned \HI\ disks in these systems. This paper focused on the observable properties of these systems, how they would appear in interferometric radio observations, and how projection effects may alter observed properties. We categorize systems into those with \LessActive\ and \MoreActive\ CGMs following \citetalias{trapp25}. \LessActive\ systems form thin, extended, and coherently rotating disks and have hot inner CGMs near their virial temperature, while \MoreActive\ systems do not. Our conclusions are as follows.

\begin{itemize}

    \item The \HI\ Size-Mass relation for all but one system falls within the 3$\sigma$ scatter from observations. Cyclone is slightly overmassive for its size, likely due to recent satellite interactions leading to a compact disk.
    

    \item To match interferometric studies, we spatially filter our smoothed, synthetic \HIcm\ datacubes to remove spatially diffuse signal. The overall measured flux is largely unaffected by this. The CGM of the \LessActive\ systems with only small-scale clumpy emission is also largely unaffected. The CGM of \MoreActive\ systems, which have more satellite galaxies and relatively more \HI\ outside their central disk, lose up to $\sim15\%$ of their observable signal.

    \item These systems show an over-abundance of low column density ($N_{HI}<10^{20}~\rm{cm}^{-2}$) gas when compared with recent interferometric studies. Spatially filtering these images to remove diffuse signal and mimic the effects of missing short baselines preferentially removes the low column density, low velocity dispersion gas. This brings our simulations into better agreement with recent observations; however, the \MoreActive\ systems still show an overabundance of low column density material.
    
    \item To fully capture the emission from these extended and diffuse \HI\ structures, joint deconvolution with single dish instruments is likely needed \cite[e.g.,][]{eibensteiner2023,2025ApJ...980...25W, koch2025-GBT_VLA_DualConv}

    \item Observed inclination and position angle can alter the amount of detectable material in interferometric studies ($\sim10\%$ based on changing the galaxy's spatial scale on the sky). If significant misaligned gas is present, the observed position angle can have significant effects as well.


\end{itemize}

As we are able to probe deeper and deeper column densities of \HI\ in and around galaxies, we stand to gain new insights into how galaxies grow, evolve, and sustain their star formation. If the diffuse component of the \HI\ emission signal is as prevalent (or more) than what we see in our simulations, current and future interferometry studies may be missing out on a key component of this extended emission, \rev{particularly for interacting systems}. Other simulations may also be missing this same material if their CGM resolution is coarser than what FOGGIE can achieve. In order to constrain this missing component, as well as test and validate simulations, joint deconvolution with large single-dish data is likely required. This, as well as the ability to model complex dynamics of such systems, is highly dependent on the observed viewing angle, implying that simulations will play an important role in interpreting current and future observations and may also need to increase their resolving power to fully track the cycle of baryons in and out of galaxies. 

\begin{acknowledgments}
We thank N. Lehner for helpful comments and suggestions, \rev{and the anonymous reviewer for their insights that greatly improved the manuscript.} CWT was supported for this work in part by NASA via a Theoretical and Computational Astrophysics Networks grant \#80NSSC21K1053 and JWST AR \#5486.
VS was supported for this work in part by NASA via an Astrophysics Theory Program grant \#80NSSC24K0772, HST AR \#17549, and HST GO \#17093.
CWT and VS were additionally supported by HST AR \#16151.
BWO acknowledges support from NSF grants \#1908109 and \#2106575, NASA ATP grants 80NSSC18K1105 and 80NSSC24K0772, and NASA TCAN grant 80NSSC21K1053.
AA acknowledges support from the INAF Large Grant 2022 “Extragalactic Surveys with JWST” (PI Pentericci) and from the European Union – NextGenerationEU RFF M4C2 1.1 PRIN 2022 project 2022ZSL4BL INSIGHT.
RA acknowledges funding from the European Research Council (ERC) under the European Union's Horizon 2020 research and innovation programme (grant agreement 101020943, SPECMAP-CGM).

Computations described in this work were performed using the publicly-available \textsc{Enzo} code (\href{http://enzo-project.org}{http://enzo-project.org}), which is the product of a collaborative effort of many independent scientists from numerous institutions around the world. Their commitment to open science has helped make this work possible. The software package {\sc SoFia-2} \citep{westmeier21-sofia2} was used to extract sources from synthetic images. The python packages {\sc matplotlib} \citep{hunter2007}, {\sc numpy} \citep{walt2011numpy}, {\sc rockstar} \citep{Behroozi2013a}, {\sc tangos} \citep{pontzen2018}, \textsc{scipy} \citep{scipy2020}, {\sc yt} \citep{ytpaper}, and {\sc Astropy} \citep{astropy2013,astropy2018,astropy2022} were all used in parts of this analysis or in products used by this paper. 

Resources supporting this work were provided by the NASA High-End Computing (HEC) Program through the NASA Advanced Supercomputing (NAS) Division at Ames Research Center and were sponsored by NASA's Science Mission Directorate; we are grateful for the superb user-support provided by NAS. Computations described in this work were performed using the publicly-available Enzo code, which is the product of a collaborative effort of many independent scientists from numerous institutions around the world.

\end{acknowledgments}

\begin{contribution}

CWT led the analysis and writing of the paper. MSP, BWO, and JT  are the principal investigators of the FOGGIE collaboration; they obtained funding for, developed, ran the simulations used in this work, and provided valuable feedback and insights throughout the project. ACW characterized satellite evolution and provided feedback on the draft. BDS contributed through discussions and feedback on the draft and developed the initial conditions for the FOGGIE simulations. AA provided insight into synthetic image generation and feedback on the draft. VS assisted in figure design and provided feedback on the draft. RA contributed through discussions during the analysis phase and feedback on the draft.


\end{contribution}

\appendix

\section{Additional Survey Figures} \label{sec:appendix_survey_vis}

\begin{figure}[ht!]
\includegraphics[width=0.49\textwidth]{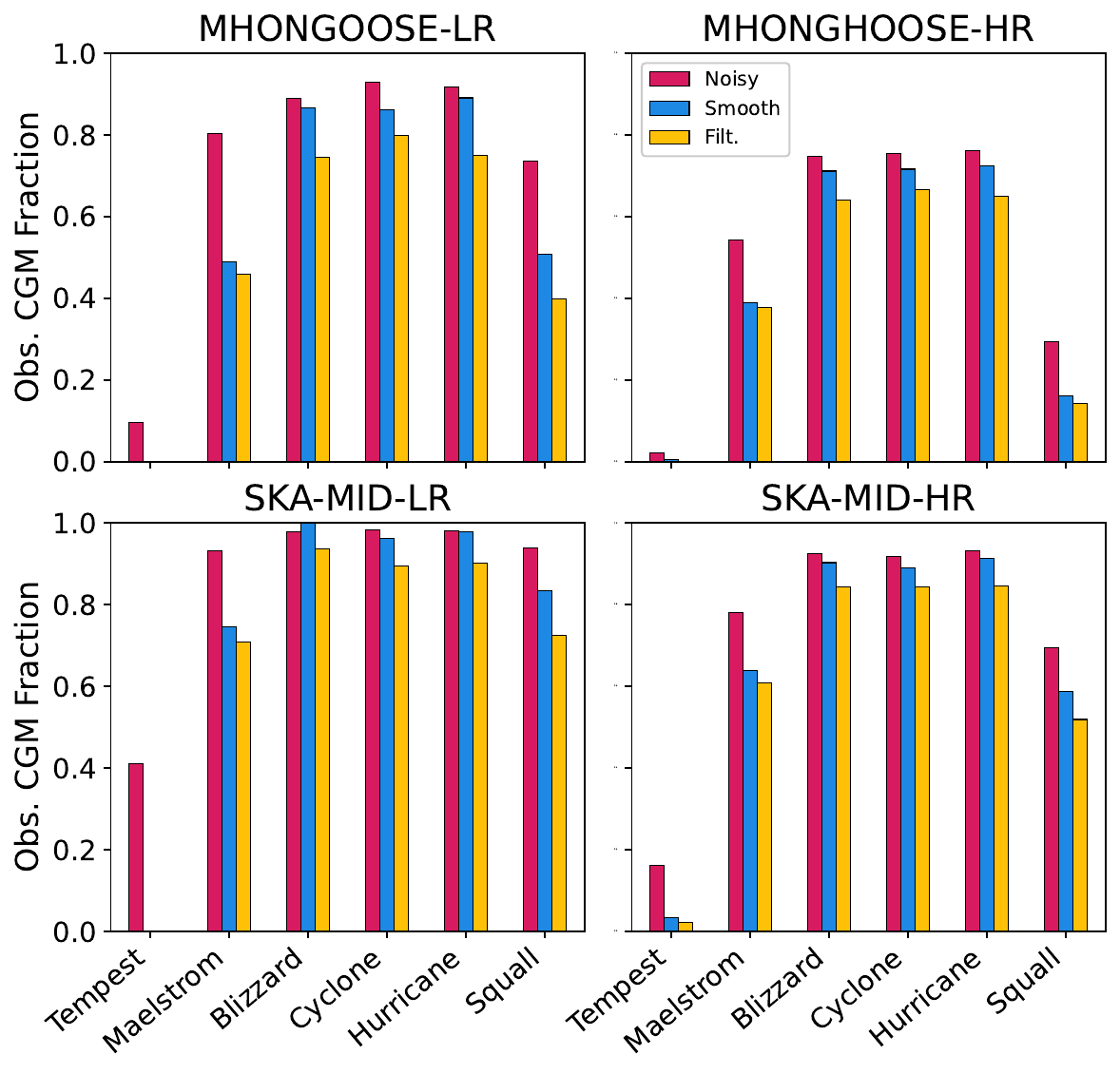}
\caption{
\sRev{Ratio of observable \HI\ to total (ideal) in the CGM at various steps in the synthetic imaging pipeline. The systems with relatively more small-scale structure (Tempest, Maelstrom) lose more signal to smoothing.}
\label{fig:appendix_obs_bar_graph}}
\end{figure}

\sRev{In this section, we briefly present additional quantifications and visualizations for all survey parameters. Fig.~\ref{fig:appendix_obs_bar_graph} shows the numerical effect each step in our imaging pipeline has on the observable \HI\ CGM fraction. The magenta bars (left-most) show the effect of both sensitivity limits and primary beam effects, although we note the latter is relatively minor. These values were calculated by integrating the flux from all spaxels in the ideal datacube above the noise level or within the smoothed {\sc SoFia-2} mask. This enables a more direct comparison, as in some cases {\sc SoFia-2} is able to obtain more flux than a simple sensitivity cut. Note, in the case of SKA-MID-LR, the smoothed image of Blizzard has a slightly higher observable CGM fraction than the prediction from the column density limit, which may be due to noise.}

\sRev{This figure highlights the separate effects of sensitivity limits and spatial smoothing. Tempest and Squall are impacted most severely by sensitivity, as they have relatively low column density \HI\ in their CGM. Similarly, Tempest and Maelstrom are most affected by the smoothing steps, as they have the least amount of diffuse \HI.}

\begin{figure*}[ht!]
\plotone{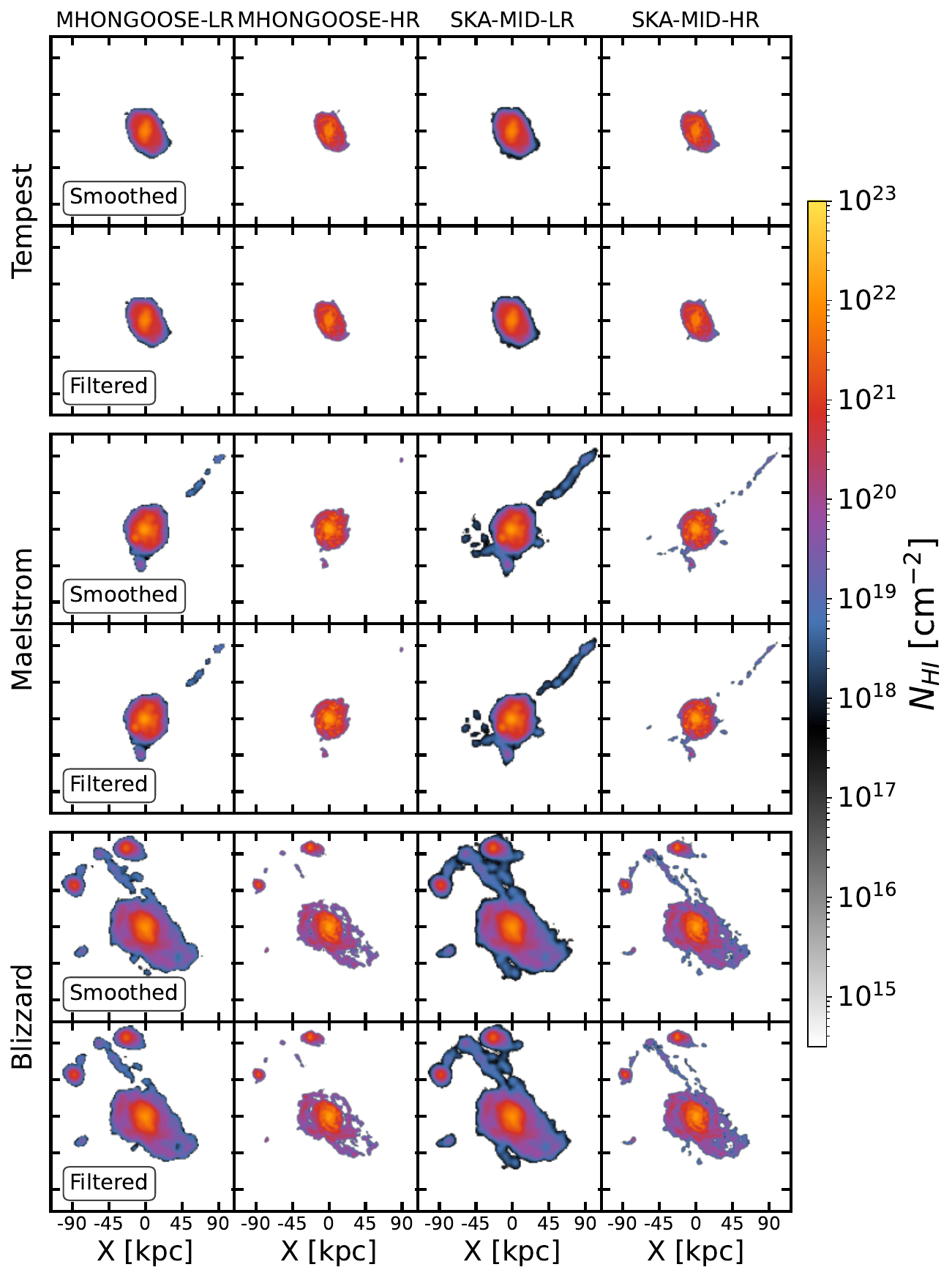}
\caption{
\sRev{Visualizations for observable \HI\ for Tempest, Maelstrom, and Blizzard for all survey parameterizations.}
\label{fig:appendix_lesspop}}
\end{figure*}

\begin{figure*}[ht!]
\plotone{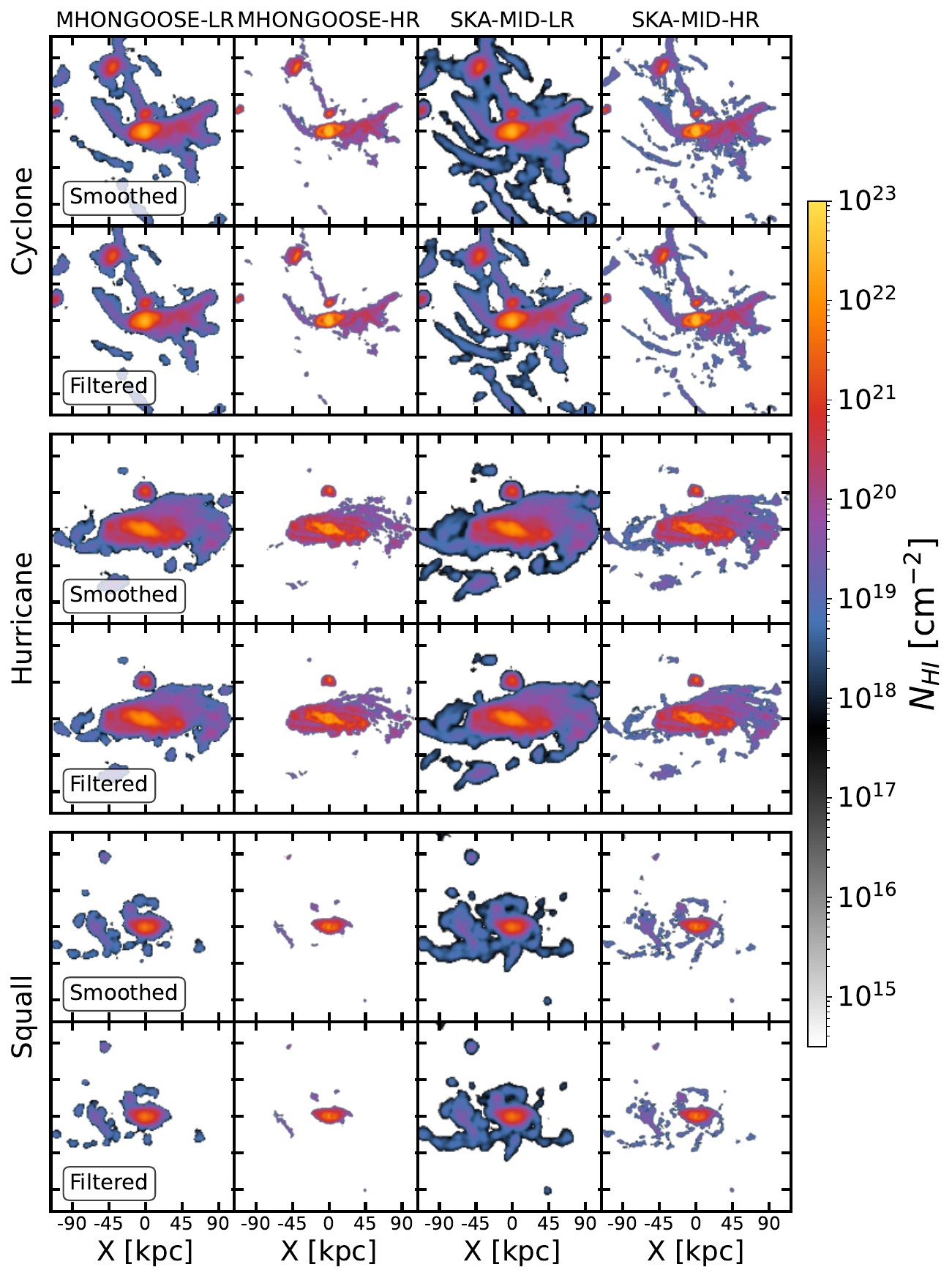}
\caption{
\sRev{Visualizations for observable \HI\ for Cyclone, Hurricane, and Squall for all survey parameterizations.}
\label{fig:appendix_morepop}}
\end{figure*}

\sRev{Fig.~\ref{fig:appendix_lesspop} and Fig.~\ref{fig:appendix_morepop} show visualizations of all survey parameters for the smoothed and filtered datacubes for all halos considered in this study. The projections for MHONGOOSE-LR are identical to what is presented in Fig.~\ref{fig:interferometric_projections} and Fig.~\ref{fig:interferometric_projections_2}. The projections for MHONGOOSE-HR lose much more material due to sensitivity limitations. While the effect of the short-spacing problem is still apparent in Blizzard and Hurricane, there is less overall diffuse material above the sensitivity limit to be filtered out. The projections for SKA-MID-LR are much deeper and are able to retain much more signal. The filtering step in this case removes material from the same areas as in the MHONGOOSE-LR case, but enough signal remains above the noise that {\sc SoFia-2} is able to retain more components. The projections for SKA-MID-HR have a slightly worse sensitivity than MHONGOOSE-LR at the same resolution as MHONGOOSE-HR. The spatial filtering behaves similarly to the MHONGOOSE-LR case, but more small-scale material can be retained.}

\sRev{As evidenced by these figures and Table~\ref{tab:observed_masses}, the overall effect of spatial filtering is slightly lower in the SKA-MID realizations than their MHONGHOOSE counterparts. This has to do with the non-linearity of both the filtering and source extraction. For an \HI\ component near the MHONGOOSE detection limit, the spatial filtering step may remove enough signal to push it below the point where {\sc SoFia-2} can reliably extract it. At SKA-MID sensitivities, however, the component may be able to be extracted, even with reduced flux.}

\bibliography{hi_disks}{}

@ARTICLE{2012ApJ...758...73S,
       author = {{Saintonge}, Am{\'e}lie and {Tacconi}, Linda J. and {Fabello}, Silvia and {Wang}, Jing and {Catinella}, Barbara and {Genzel}, Reinhard and {Graci{\'a}-Carpio}, Javier and {Kramer}, Carsten and {Moran}, Sean and {Heckman}, Timothy M. and {Schiminovich}, David and {Schuster}, Karl and {Wuyts}, Stijn},
        title = "{The Impact of Interactions, Bars, Bulges, and Active Galactic Nuclei on Star Formation Efficiency in Local Massive Galaxies}",
      journal = {\apj},
         year = 2012,
        month = oct,
       volume = {758},
       number = {2},
          eid = {73},
        pages = {73},
          doi = {10.1088/0004-637X/758/2/73},
archivePrefix = {arXiv},
       eprint = {1209.0476},
 primaryClass = {astro-ph.CO},
       adsurl = {https://ui.adsabs.harvard.edu/abs/2012ApJ...758...73S}
}

@ARTICLE{2015ApJ...813...46B,
       author = {{Borthakur}, Sanchayeeta and {Heckman}, Timothy and {Tumlinson}, Jason and {Bordoloi}, Rongmon and {Thom}, Christopher and {Catinella}, Barbara and {Schiminovich}, David and {Dav{\'e}}, Romeel and {Kauffmann}, Guinevere and {Moran}, Sean M. and {Saintonge}, Amelie},
        title = "{Connection between the Circumgalactic Medium and the Interstellar Medium of Galaxies: Results from the COS-GASS Survey}",
      journal = {\apj},
         year = 2015,
        month = nov,
       volume = {813},
       number = {1},
          eid = {46},
        pages = {46},
          doi = {10.1088/0004-637X/813/1/46},
archivePrefix = {arXiv},
       eprint = {1504.01392},
 primaryClass = {astro-ph.GA},
       adsurl = {https://ui.adsabs.harvard.edu/abs/2015ApJ...813...46B}
}

@ARTICLE{wang2024-FEASTS_I,
       author = {{Wang}, Jing and {Lin}, Xuchen and {Yang}, Dong and {Staveley-Smith}, Lister and {Walter}, Fabian and {Wang}, Q. Daniel and {Wang}, Ran and {Battisti}, A.~J. and {Catinella}, Barbara and {Chen}, Hsiao-Wen and {Cortese}, Luca and {Fisher}, D.~B. and {Ho}, Luis C. and {Ji}, Suoqing and {Jiang}, Peng and {Kauffmann}, Guinevere and {Kong}, Xu and {Liu}, Ziming and {Shao}, Li and {Wang}, Jie and {Wang}, Lile and {Wang}, Shun},
        title = "{FEASTS Combined with Interferometry. I. Overall Properties of Diffuse H I and Implications for Gas Accretion in Nearby Galaxies}",
      journal = {\apj},
         year = 2024,
        month = jun,
       volume = {968},
       number = {1},
          eid = {48},
        pages = {48},
          doi = {10.3847/1538-4357/ad3e61},
archivePrefix = {arXiv},
       eprint = {2404.09422},
 primaryClass = {astro-ph.GA},
       adsurl = {https://ui.adsabs.harvard.edu/abs/2024ApJ...968...48W}
}

@ARTICLE{2025ApJ...980...25W,
       author = {{Wang}, Jing and {Yang}, Dong and {Lin}, Xuchen and {Huang}, Qifeng and {Qu}, Zhijie and {Chen}, Hsiao-wen and {Guo}, Hong and {Ho}, Luis C. and {Jiang}, Peng and {Liang}, Zezhong and {P{\'e}roux}, C{\'e}line and {Staveley-Smith}, Lister and {Weng}, Simon},
        title = "{FEASTS: Radial Distribution of H I Surface Densities Down to 0.01 M$_{{\ensuremath{\odot}}}$ pc$^{‑2}$ of 35 Nearby Galaxies}",
      journal = {\apj},
         year = 2025,
        month = feb,
       volume = {980},
       number = {1},
          eid = {25},
        pages = {25},
          doi = {10.3847/1538-4357/ada95a},
archivePrefix = {arXiv},
       eprint = {2501.01289},
 primaryClass = {astro-ph.GA},
       adsurl = {https://ui.adsabs.harvard.edu/abs/2025ApJ...980...25W}
}

@ARTICLE{trapp22,
       author = {{Trapp}, Cameron W. and {Kere{\v{s}}}, Du{\v{s}}an and {Chan}, Tsang Keung and {Escala}, Ivanna and {Hummels}, Cameron and {Hopkins}, Philip F. and {Faucher-Gigu{\`e}re}, Claude-Andr{\'e} and {Murray}, Norman and {Quataert}, Eliot and {Wetzel}, Andrew},
        title = "{Gas infall and radial transport in cosmological simulations of milky way-mass discs}",
      journal = {\mnras},
         year = 2022,
        month = jan,
       volume = {509},
       number = {3},
        pages = {4149-4170},
          doi = {10.1093/mnras/stab3251},
archivePrefix = {arXiv},
       eprint = {2105.11472},
 primaryClass = {astro-ph.GA},
       adsurl = {https://ui.adsabs.harvard.edu/abs/2022MNRAS.509.4149T}
}

@ARTICLE{astropy2013,
       author = {{Astropy Collaboration} and {Robitaille}, Thomas P. and {Tollerud}, Erik J. and {Greenfield}, Perry and {Droettboom}, Michael and {Bray}, Erik and {Aldcroft}, Tom and {Davis}, Matt and {Ginsburg}, Adam and {Price-Whelan}, Adrian M. and {Kerzendorf}, Wolfgang E. and {Conley}, Alexander and {Crighton}, Neil and {Barbary}, Kyle and {Muna}, Demitri and {Ferguson}, Henry and {Grollier}, Fr{\'e}d{\'e}ric and {Parikh}, Madhura M. and {Nair}, Prasanth H. and {Unther}, Hans M. and {Deil}, Christoph and {Woillez}, Julien and {Conseil}, Simon and {Kramer}, Roban and {Turner}, James E.~H. and {Singer}, Leo and {Fox}, Ryan and {Weaver}, Benjamin A. and {Zabalza}, Victor and {Edwards}, Zachary I. and {Azalee Bostroem}, K. and {Burke}, D.~J. and {Casey}, Andrew R. and {Crawford}, Steven M. and {Dencheva}, Nadia and {Ely}, Justin and {Jenness}, Tim and {Labrie}, Kathleen and {Lim}, Pey Lian and {Pierfederici}, Francesco and {Pontzen}, Andrew and {Ptak}, Andy and {Refsdal}, Brian and {Servillat}, Mathieu and {Streicher}, Ole},
        title = "{Astropy: A community Python package for astronomy}",
      journal = {\aap},
         year = 2013,
        month = oct,
       volume = {558},
          eid = {A33},
        pages = {A33},
          doi = {10.1051/0004-6361/201322068},
archivePrefix = {arXiv},
       eprint = {1307.6212},
 primaryClass = {astro-ph.IM},
       adsurl = {https://ui.adsabs.harvard.edu/abs/2013A&A...558A..33A}
}

@ARTICLE{astropy2018,
       author = {{Astropy Collaboration} and {Price-Whelan}, A.~M. and {Sip{\H{o}}cz}, B.~M. and {G{\"u}nther}, H.~M. and {Lim}, P.~L. and {Crawford}, S.~M. and {Conseil}, S. and {Shupe}, D.~L. and {Craig}, M.~W. and {Dencheva}, N. and {Ginsburg}, A. and {VanderPlas}, J.~T. and {Bradley}, L.~D. and {P{\'e}rez-Su{\'a}rez}, D. and {de Val-Borro}, M. and {Aldcroft}, T.~L. and {Cruz}, K.~L. and {Robitaille}, T.~P. and {Tollerud}, E.~J. and {Ardelean}, C. and {Babej}, T. and {Bach}, Y.~P. and {Bachetti}, M. and {Bakanov}, A.~V. and {Bamford}, S.~P. and {Barentsen}, G. and {Barmby}, P. and {Baumbach}, A. and {Berry}, K.~L. and {Biscani}, F. and {Boquien}, M. and {Bostroem}, K.~A. and {Bouma}, L.~G. and {Brammer}, G.~B. and {Bray}, E.~M. and {Breytenbach}, H. and {Buddelmeijer}, H. and {Burke}, D.~J. and {Calderone}, G. and {Cano Rodr{\'\i}guez}, J.~L. and {Cara}, M. and {Cardoso}, J.~V.~M. and {Cheedella}, S. and {Copin}, Y. and {Corrales}, L. and {Crichton}, D. and {D'Avella}, D. and {Deil}, C. and {Depagne}, {\'E}. and {Dietrich}, J.~P. and {Donath}, A. and {Droettboom}, M. and {Earl}, N. and {Erben}, T. and {Fabbro}, S. and {Ferreira}, L.~A. and {Finethy}, T. and {Fox}, R.~T. and {Garrison}, L.~H. and {Gibbons}, S.~L.~J. and {Goldstein}, D.~A. and {Gommers}, R. and {Greco}, J.~P. and {Greenfield}, P. and {Groener}, A.~M. and {Grollier}, F. and {Hagen}, A. and {Hirst}, P. and {Homeier}, D. and {Horton}, A.~J. and {Hosseinzadeh}, G. and {Hu}, L. and {Hunkeler}, J.~S. and {Ivezi{\'c}}, {\v{Z}}. and {Jain}, A. and {Jenness}, T. and {Kanarek}, G. and {Kendrew}, S. and {Kern}, N.~S. and {Kerzendorf}, W.~E. and {Khvalko}, A. and {King}, J. and {Kirkby}, D. and {Kulkarni}, A.~M. and {Kumar}, A. and {Lee}, A. and {Lenz}, D. and {Littlefair}, S.~P. and {Ma}, Z. and {Macleod}, D.~M. and {Mastropietro}, M. and {McCully}, C. and {Montagnac}, S. and {Morris}, B.~M. and {Mueller}, M. and {Mumford}, S.~J. and {Muna}, D. and {Murphy}, N.~A. and {Nelson}, S. and {Nguyen}, G.~H. and {Ninan}, J.~P. and {N{\"o}the}, M. and {Ogaz}, S. and {Oh}, S. and {Parejko}, J.~K. and {Parley}, N. and {Pascual}, S. and {Patil}, R. and {Patil}, A.~A. and {Plunkett}, A.~L. and {Prochaska}, J.~X. and {Rastogi}, T. and {Reddy Janga}, V. and {Sabater}, J. and {Sakurikar}, P. and {Seifert}, M. and {Sherbert}, L.~E. and {Sherwood-Taylor}, H. and {Shih}, A.~Y. and {Sick}, J. and {Silbiger}, M.~T. and {Singanamalla}, S. and {Singer}, L.~P. and {Sladen}, P.~H. and {Sooley}, K.~A. and {Sornarajah}, S. and {Streicher}, O. and {Teuben}, P. and {Thomas}, S.~W. and {Tremblay}, G.~R. and {Turner}, J.~E.~H. and {Terr{\'o}n}, V. and {van Kerkwijk}, M.~H. and {de la Vega}, A. and {Watkins}, L.~L. and {Weaver}, B.~A. and {Whitmore}, J.~B. and {Woillez}, J. and {Zabalza}, V. and {Astropy Contributors}},
        title = "{The Astropy Project: Building an Open-science Project and Status of the v2.0 Core Package}",
      journal = {\aj},
         year = 2018,
        month = sep,
       volume = {156},
       number = {3},
          eid = {123},
        pages = {123},
          doi = {10.3847/1538-3881/aabc4f},
archivePrefix = {arXiv},
       eprint = {1801.02634},
 primaryClass = {astro-ph.IM},
       adsurl = {https://ui.adsabs.harvard.edu/abs/2018AJ....156..123A}
}

@ARTICLE{astropy2022,
       author = {{Astropy Collaboration} and {Price-Whelan}, Adrian M. and {Lim}, Pey Lian and {Earl}, Nicholas and {Starkman}, Nathaniel and {Bradley}, Larry and {Shupe}, David L. and {Patil}, Aarya A. and {Corrales}, Lia and {Brasseur}, C.~E. and {N{\"o}the}, Maximilian and {Donath}, Axel and {Tollerud}, Erik and {Morris}, Brett M. and {Ginsburg}, Adam and {Vaher}, Eero and {Weaver}, Benjamin A. and {Tocknell}, James and {Jamieson}, William and {van Kerkwijk}, Marten H. and {Robitaille}, Thomas P. and {Merry}, Bruce and {Bachetti}, Matteo and {G{\"u}nther}, H. Moritz and {Aldcroft}, Thomas L. and {Alvarado-Montes}, Jaime A. and {Archibald}, Anne M. and {B{\'o}di}, Attila and {Bapat}, Shreyas and {Barentsen}, Geert and {Baz{\'a}n}, Juanjo and {Biswas}, Manish and {Boquien}, M{\'e}d{\'e}ric and {Burke}, D.~J. and {Cara}, Daria and {Cara}, Mihai and {Conroy}, Kyle E. and {Conseil}, Simon and {Craig}, Matthew W. and {Cross}, Robert M. and {Cruz}, Kelle L. and {D'Eugenio}, Francesco and {Dencheva}, Nadia and {Devillepoix}, Hadrien A.~R. and {Dietrich}, J{\"o}rg P. and {Eigenbrot}, Arthur Davis and {Erben}, Thomas and {Ferreira}, Leonardo and {Foreman-Mackey}, Daniel and {Fox}, Ryan and {Freij}, Nabil and {Garg}, Suyog and {Geda}, Robel and {Glattly}, Lauren and {Gondhalekar}, Yash and {Gordon}, Karl D. and {Grant}, David and {Greenfield}, Perry and {Groener}, Austen M. and {Guest}, Steve and {Gurovich}, Sebastian and {Handberg}, Rasmus and {Hart}, Akeem and {Hatfield-Dodds}, Zac and {Homeier}, Derek and {Hosseinzadeh}, Griffin and {Jenness}, Tim and {Jones}, Craig K. and {Joseph}, Prajwel and {Kalmbach}, J. Bryce and {Karamehmetoglu}, Emir and {Ka{\l}uszy{\'n}ski}, Miko{\l}aj and {Kelley}, Michael S.~P. and {Kern}, Nicholas and {Kerzendorf}, Wolfgang E. and {Koch}, Eric W. and {Kulumani}, Shankar and {Lee}, Antony and {Ly}, Chun and {Ma}, Zhiyuan and {MacBride}, Conor and {Maljaars}, Jakob M. and {Muna}, Demitri and {Murphy}, N.~A. and {Norman}, Henrik and {O'Steen}, Richard and {Oman}, Kyle A. and {Pacifici}, Camilla and {Pascual}, Sergio and {Pascual-Granado}, J. and {Patil}, Rohit R. and {Perren}, Gabriel I. and {Pickering}, Timothy E. and {Rastogi}, Tanuj and {Roulston}, Benjamin R. and {Ryan}, Daniel F. and {Rykoff}, Eli S. and {Sabater}, Jose and {Sakurikar}, Parikshit and {Salgado}, Jes{\'u}s and {Sanghi}, Aniket and {Saunders}, Nicholas and {Savchenko}, Volodymyr and {Schwardt}, Ludwig and {Seifert-Eckert}, Michael and {Shih}, Albert Y. and {Jain}, Anany Shrey and {Shukla}, Gyanendra and {Sick}, Jonathan and {Simpson}, Chris and {Singanamalla}, Sudheesh and {Singer}, Leo P. and {Singhal}, Jaladh and {Sinha}, Manodeep and {Sip{\H{o}}cz}, Brigitta M. and {Spitler}, Lee R. and {Stansby}, David and {Streicher}, Ole and {{\v{S}}umak}, Jani and {Swinbank}, John D. and {Taranu}, Dan S. and {Tewary}, Nikita and {Tremblay}, Grant R. and {Val-Borro}, Miguel de and {Van Kooten}, Samuel J. and {Vasovi{\'c}}, Zlatan and {Verma}, Shresth and {de Miranda Cardoso}, Jos{\'e} Vin{\'\i}cius and {Williams}, Peter K.~G. and {Wilson}, Tom J. and {Winkel}, Benjamin and {Wood-Vasey}, W.~M. and {Xue}, Rui and {Yoachim}, Peter and {Zhang}, Chen and {Zonca}, Andrea and {Astropy Project Contributors}},
        title = "{The Astropy Project: Sustaining and Growing a Community-oriented Open-source Project and the Latest Major Release (v5.0) of the Core Package}",
      journal = {\apj},
         year = 2022,
        month = aug,
       volume = {935},
       number = {2},
          eid = {167},
        pages = {167},
          doi = {10.3847/1538-4357/ac7c74},
archivePrefix = {arXiv},
       eprint = {2206.14220},
 primaryClass = {astro-ph.IM},
       adsurl = {https://ui.adsabs.harvard.edu/abs/2022ApJ...935..167A}
}

@ARTICLE{ytpaper,
   author = {{Turk}, M.~J. and {Smith}, B.~D. and {Oishi}, J.~S. and {Skory}, S. and
     {Skillman}, S.~W. and {Abel}, T. and {Norman}, M.~L.},
    title = "{yt: A Multi-code Analysis Toolkit for Astrophysical Simulation Data}",
  journal = {The Astrophysical Journal Supplement Series},
archivePrefix = "arXiv",
   eprint = {1011.3514},
 primaryClass = "astro-ph.IM",
     year = 2011,
    month = jan,
   volume = 192,
      eid = {9},
    pages = {9},
      doi = {10.1088/0067-0049/192/1/9},
   adsurl = {https://ui.adsabs.harvard.edu/abs/2011ApJS..192....9T}
}

@ARTICLE{scipy2020,
       author = {{Virtanen}, Pauli and {Gommers}, Ralf and {Oliphant}, Travis E. and {Haberland}, Matt and {Reddy}, Tyler and {Cournapeau}, David and {Burovski}, Evgeni and {Peterson}, Pearu and {Weckesser}, Warren and {Bright}, Jonathan and {van der Walt}, St{\'e}fan J. and {Brett}, Matthew and {Wilson}, Joshua and {Millman}, K. Jarrod and {Mayorov}, Nikolay and {Nelson}, Andrew R.~J. and {Jones}, Eric and {Kern}, Robert and {Larson}, Eric and {Carey}, C.~J. and {Polat}, {\.I}lhan and {Feng}, Yu and {Moore}, Eric W. and {VanderPlas}, Jake and {Laxalde}, Denis and {Perktold}, Josef and {Cimrman}, Robert and {Henriksen}, Ian and {Quintero}, E.~A. and {Harris}, Charles R. and {Archibald}, Anne M. and {Ribeiro}, Ant{\^o}nio H. and {Pedregosa}, Fabian and {van Mulbregt}, Paul and {SciPy 1. 0 Contributors}},
        title = "{SciPy 1.0: fundamental algorithms for scientific computing in Python}",
      journal = {Nature Methods},
         year = 2020,
        month = feb,
       volume = {17},
        pages = {261-272},
          doi = {10.1038/s41592-019-0686-2},
archivePrefix = {arXiv},
       eprint = {1907.10121},
 primaryClass = {cs.MS},
       adsurl = {https://ui.adsabs.harvard.edu/abs/2020NatMe..17..261V}
}

@ARTICLE{hunter2007,
       author = {{Hunter}, John D.},
        title = "{Matplotlib: A 2D Graphics Environment}",
      journal = {Computing in Science and Engineering},
         year = 2007,
        month = may,
       volume = {9},
       number = {3},
        pages = {90-95},
          doi = {10.1109/MCSE.2007.55},
       adsurl = {https://ui.adsabs.harvard.edu/abs/2007CSE.....9...90H}
}

@article{walt2011numpy,
  title={The NumPy array: a structure for efficient numerical computation},
  author={Walt, St{\'e}fan van der and Colbert, S Chris and Varoquaux, Gael},
  journal={Computing in Science \& Engineering},
  volume={13},
  number={2},
  pages={22--30},
  year={2011},
  publisher={IEEE}
}

@ARTICLE{Behroozi2013a,
       author = {{Behroozi}, Peter S. and {Wechsler}, Risa H. and {Wu}, Hao-Yi},
        title = "{The ROCKSTAR Phase-space Temporal Halo Finder and the Velocity Offsets of Cluster Cores}",
      journal = {\apj},
         year = 2013,
        month = jan,
       volume = {762},
       number = {2},
          eid = {109},
        pages = {109},
          doi = {10.1088/0004-637X/762/2/109},
archivePrefix = {arXiv},
       eprint = {1110.4372},
 primaryClass = {astro-ph.CO},
       adsurl = {https://ui.adsabs.harvard.edu/abs/2013ApJ...762..109B}
}

@ARTICLE{pontzen2018,
       author = {{Pontzen}, Andrew and {Tremmel}, Michael},
        title = "{TANGOS: The Agile Numerical Galaxy Organization System}",
      journal = {\apjs},
         year = 2018,
        month = aug,
       volume = {237},
       number = {2},
          eid = {23},
        pages = {23},
          doi = {10.3847/1538-4365/aac832},
archivePrefix = {arXiv},
       eprint = {1803.00010},
 primaryClass = {astro-ph.IM},
       adsurl = {https://ui.adsabs.harvard.edu/abs/2018ApJS..237...23P}
}

@ARTICLE{werk14,
       author = {{Werk}, Jessica K. and {Prochaska}, J. Xavier and {Tumlinson}, Jason and {Peeples}, Molly S. and {Tripp}, Todd M. and {Fox}, Andrew J. and {Lehner}, Nicolas and {Thom}, Christopher and {O'Meara}, John M. and {Ford}, Amanda Brady and {Bordoloi}, Rongmon and {Katz}, Neal and {Tejos}, Nicolas and {Oppenheimer}, Benjamin D. and {Dav{\'e}}, Romeel and {Weinberg}, David H.},
        title = "{The COS-Halos Survey: Physical Conditions and Baryonic Mass in the Low-redshift Circumgalactic Medium}",
      journal = {\apj},
         year = 2014,
        month = sep,
       volume = {792},
       number = {1},
          eid = {8},
        pages = {8},
          doi = {10.1088/0004-637X/792/1/8},
archivePrefix = {arXiv},
       eprint = {1403.0947},
 primaryClass = {astro-ph.CO},
       adsurl = {https://ui.adsabs.harvard.edu/abs/2014ApJ...792....8W}
}

@ARTICLE{brummel-smith19,
       author = {{Brummel-Smith}, Corey and {Bryan}, Greg and {Butsky}, Iryna and {Corlies}, Lauren and {Emerick}, Andrew and {Forbes}, John and {Fujimoto}, Yusuke and {Goldbaum}, Nathan and {Grete}, Philipp and {Hummels}, Cameron and {Kim}, Ji-hoon and {Koh}, Daegene and {Li}, Miao and {Li}, Yuan and {Li}, Xinyu and {OShea}, Brian and {Peeples}, Molly and {Regan}, John and {Salem}, Munier and {Schmidt}, Wolfram and {Simpson}, Christine and {Smith}, Britton and {Tumlinson}, Jason and {Turk}, Matthew and {Wise}, John and {Abel}, Tom and {Bordner}, James and {Cen}, Renyue and {Collins}, David and {Crosby}, Brian and {Edelmann}, Philipp and {Hahn}, Oliver and {Harkness}, Robert and {Harper-Clark}, Elizabeth and {Kong}, Shuo and {Kritsuk}, Alexei and {Kuhlen}, Michael and {Larrue}, James and {Lee}, Eve and {Meece}, Greg and {Norman}, Michael and {Oishi}, Jeffrey and {Paschos}, Pascal and {Peruta}, Carolyn and {Razoumov}, Alex and {Reynolds}, Daniel and {Silvia}, Devin and {Skillman}, Samuel and {Skory}, Stephen and {So}, Geoffrey and {Tasker}, Elizabeth and {Wagner}, Rick and {Wang}, Peng and {Xu}, Hao and {Zhao}, Fen},
        title = "{ENZO: An Adaptive Mesh Refinement Code for Astrophysics (Version 2.6)}",
      journal = {The Journal of Open Source Software},
         year = 2019,
        month = oct,
       volume = {4},
       number = {42},
          eid = {1636},
        pages = {1636},
          doi = {10.21105/joss.01636},
       adsurl = {https://ui.adsabs.harvard.edu/abs/2019JOSS....4.1636B}
}

@ARTICLE{bryan14,
       author = {{Bryan}, Greg L. and {Norman}, Michael L. and {O'Shea}, Brian W. and {Abel}, Tom and {Wise}, John H. and {Turk}, Matthew J. and {Reynolds}, Daniel R. and {Collins}, David C. and {Wang}, Peng and {Skillman}, Samuel W. and {Smith}, Britton and {Harkness}, Robert P. and {Bordner}, James and {Kim}, Ji-hoon and {Kuhlen}, Michael and {Xu}, Hao and {Goldbaum}, Nathan and {Hummels}, Cameron and {Kritsuk}, Alexei G. and {Tasker}, Elizabeth and {Skory}, Stephen and {Simpson}, Christine M. and {Hahn}, Oliver and {Oishi}, Jeffrey S. and {So}, Geoffrey C. and {Zhao}, Fen and {Cen}, Renyue and {Li}, Yuan and {Enzo Collaboration}},
        title = "{ENZO: An Adaptive Mesh Refinement Code for Astrophysics}",
      journal = {\apjs},
         year = 2014,
        month = apr,
       volume = {211},
       number = {2},
          eid = {19},
        pages = {19},
          doi = {10.1088/0067-0049/211/2/19},
archivePrefix = {arXiv},
       eprint = {1307.2265},
 primaryClass = {astro-ph.IM},
       adsurl = {https://ui.adsabs.harvard.edu/abs/2014ApJS..211...19B}
}

@ARTICLE{lochhaas23,
       author = {{Lochhaas}, Cassandra and {Tumlinson}, Jason and {Peeples}, Molly S. and {O'Shea}, Brian W. and {Werk}, Jessica K. and {Simons}, Raymond C. and {Juno}, James and {Kopenhafer}, Claire and {Augustin}, Ramona and {Wright}, Anna C. and {Acharyya}, Ayan and {Smith}, Britton D.},
        title = "{Figuring Out Gas \& Galaxies in Enzo (FOGGIE). VI. The Circumgalactic Medium of L $^{{\ensuremath{*}}}$ Galaxies Is Supported in an Emergent, Nonhydrostatic Equilibrium}",
      journal = {\apj},
         year = 2023,
        month = may,
       volume = {948},
       number = {1},
          eid = {43},
        pages = {43},
          doi = {10.3847/1538-4357/acbb06},
archivePrefix = {arXiv},
       eprint = {2206.09925},
 primaryClass = {astro-ph.GA},
       adsurl = {https://ui.adsabs.harvard.edu/abs/2023ApJ...948...43L}
}

@ARTICLE{lochhaas21,
       author = {{Lochhaas}, Cassandra and {Tumlinson}, Jason and {O'Shea}, Brian W. and {Peeples}, Molly S. and {Smith}, Britton D. and {Werk}, Jessica K. and {Augustin}, Ramona and {Simons}, Raymond C.},
        title = "{Figuring Out Gas \& Galaxies In Enzo (FOGGIE). V. The Virial Temperature Does Not Describe Gas in a Virialized Galaxy Halo}",
      journal = {\apj},
         year = 2021,
        month = dec,
       volume = {922},
       number = {2},
          eid = {121},
        pages = {121},
          doi = {10.3847/1538-4357/ac2496},
archivePrefix = {arXiv},
       eprint = {2102.08393},
 primaryClass = {astro-ph.GA},
       adsurl = {https://ui.adsabs.harvard.edu/abs/2021ApJ...922..121L}
}

@ARTICLE{lochhaas25,
       author = {{Lochhaas}, Cassandra and {Peeples}, Molly S. and {O'Shea}, Brian W. and {Tumlinson}, Jason and {Corlies}, Lauren and {Saeedzadeh}, Vida and {Lehner}, Nicolas and {Wright}, Anna C. and {Werk}, Jessica K. and {Trapp}, Cameron W. and {Augustin}, Ramona and {Acharyya}, Ayan and {Smith}, Britton D.},
        title = "{Figuring Out Gas \& Galaxies In Enzo (FOGGIE) XI: Circumgalactic O VI Emission Traces Clumpy Inflowing Recycled Gas}",
      journal = {arXiv e-prints},
         year = 2025,
        month = oct,
          eid = {arXiv:2510.25844},
        pages = {arXiv:2510.25844},
archivePrefix = {arXiv},
       eprint = {2510.25844},
 primaryClass = {astro-ph.GA},
       adsurl = {https://ui.adsabs.harvard.edu/abs/2025arXiv251025844L}
}

@ARTICLE{simons20,
       author = {{Simons}, Raymond C. and {Peeples}, Molly S. and {Tumlinson}, Jason and {O'Shea}, Brian W. and {Smith}, Britton D. and {Corlies}, Lauren and {Lochhaas}, Cassandra and {Zheng}, Yong and {Augustin}, Ramona and {Prasad}, Deovrat and {Snyder}, Gregory F. and {Tollerud}, Erik},
        title = "{Figuring Out Gas \& Galaxies in Enzo (FOGGIE). IV. The Stochasticity of Ram Pressure Stripping in Galactic Halos}",
      journal = {\apj},
         year = 2020,
        month = dec,
       volume = {905},
       number = {2},
          eid = {167},
        pages = {167},
          doi = {10.3847/1538-4357/abc5b8},
archivePrefix = {arXiv},
       eprint = {2004.14394},
 primaryClass = {astro-ph.GA},
       adsurl = {https://ui.adsabs.harvard.edu/abs/2020ApJ...905..167S}
}

@ARTICLE{wright24,
       IDs={wright23},
       author = {{Wright}, Anna C. and {Tumlinson}, Jason and {Peeples}, Molly S. and {O'Shea}, Brian W. and {Lochhaas}, Cassandra and {Corlies}, Lauren and {Smith}, Britton D. and {Binh}, Nguyen and {Augustin}, Ramona and {Simons}, Raymond C.},
        title = "{Figuring Out Gas and Galaxies in Enzo (FOGGIE). VII. The (Dis)assembly of Stellar Halos}",
      journal = {\apj},
         year = 2024,
        month = jul,
       volume = {970},
       number = {1},
          eid = {70},
        pages = {70},
          doi = {10.3847/1538-4357/ad49a3},
       adsurl = {https://ui.adsabs.harvard.edu/abs/2024ApJ...970...70W}
}

@ARTICLE{zheng20,
       author = {{Zheng}, Yong and {Peeples}, Molly S. and {O'Shea}, Brian W. and {Simons}, Raymond C. and {Lochhaas}, Cassandra and {Corlies}, Lauren and {Tumlinson}, Jason and {Smith}, Britton D. and {Augustin}, Ramona},
        title = "{Figuring Out Gas \& Galaxies in Enzo (FOGGIE). III. The Mocky Way: Investigating Biases in Observing the Milky Way's Circumgalactic Medium}",
      journal = {\apj},
         year = 2020,
        month = jun,
       volume = {896},
       number = {2},
          eid = {143},
        pages = {143},
          doi = {10.3847/1538-4357/ab960a},
archivePrefix = {arXiv},
       eprint = {2001.07736},
 primaryClass = {astro-ph.GA},
       adsurl = {https://ui.adsabs.harvard.edu/abs/2020ApJ...896..143Z}
}

@ARTICLE{cen_ostriker_06,
       author = {{Cen}, Renyue and {Ostriker}, Jeremiah P.},
        title = "{Where Are the Baryons? II. Feedback Effects}",
      journal = {\apj},
         year = 2006,
        month = oct,
       volume = {650},
       number = {2},
        pages = {560-572},
          doi = {10.1086/506505},
archivePrefix = {arXiv},
       eprint = {astro-ph/0601008},
 primaryClass = {astro-ph},
       adsurl = {https://ui.adsabs.harvard.edu/abs/2006ApJ...650..560C}
}

@ARTICLE{corlies20,
       author = {{Corlies}, Lauren and {Peeples}, Molly S. and {Tumlinson}, Jason and {O'Shea}, Brian W. and {Lehner}, Nicolas and {Howk}, J. Christopher and {O'Meara}, John M. and {Smith}, Britton D.},
        title = "{Figuring Out Gas \& Galaxies in Enzo (FOGGIE). II. Emission from the z = 3 Circumgalactic Medium}",
      journal = {\apj},
         year = 2020,
        month = jun,
       volume = {896},
       number = {2},
          eid = {125},
        pages = {125},
          doi = {10.3847/1538-4357/ab9310},
archivePrefix = {arXiv},
       eprint = {1811.05060},
 primaryClass = {astro-ph.GA},
       adsurl = {https://ui.adsabs.harvard.edu/abs/2020ApJ...896..125C}
}

@ARTICLE{Acharyya24,
       author = {{Acharyya}, Ayan and {Peeples}, Molly S. and {Tumlinson}, Jason and {O'Shea}, Brian W. and {Lochhaas}, Cassandra and {Wright}, Anna C. and {Simons}, Raymond C. and {Augustin}, Ramona and {Smith}, Britton D. and {Lee}, Eugene Hyeonmin},
        title = "{Figuring Out Gas and Galaxies In Enzo (FOGGIE). VIII. Complex and Stochastic Metallicity Gradients at z > 2}",
      journal = {\apj},
         year = 2025,
        month = feb,
       volume = {979},
       number = {2},
          eid = {129},
        pages = {129},
          doi = {10.3847/1538-4357/ad9dd8},
archivePrefix = {arXiv},
       eprint = {2404.06613},
 primaryClass = {astro-ph.GA},
       adsurl = {https://ui.adsabs.harvard.edu/abs/2025ApJ...979..129A}
}

@ARTICLE{peeples19,
        IDs = {peeples19, Peeples19},
       author = {{Peeples}, Molly S. and {Corlies}, Lauren and {Tumlinson}, Jason and {O'Shea}, Brian W. and {Lehner}, Nicolas and {O'Meara}, John M. and {Howk}, J. Christopher and {Earl}, Nicholas and {Smith}, Britton D. and {Wise}, John H. and {Hummels}, Cameron B.},
        title = "{Figuring Out Gas \& Galaxies in Enzo (FOGGIE). I. Resolving Simulated Circumgalactic Absorption at 2 {\ensuremath{\leq}} z {\ensuremath{\leq}} 2.5}",
      journal = {\apj},
         year = 2019,
        month = mar,
       volume = {873},
       number = {2},
          eid = {129},
        pages = {129},
          doi = {10.3847/1538-4357/ab0654},
archivePrefix = {arXiv},
       eprint = {1810.06566},
 primaryClass = {astro-ph.GA},
       adsurl = {https://ui.adsabs.harvard.edu/abs/2019ApJ...873..129P}
}

@article{lin2025-FEASTS-M51,
       author = {{Lin}, Xuchen and {Wang}, Jing and {Staveley-Smith}, Lister and {Ji}, Suoqing and {Yang}, Dong and {Chen}, Xinkai and {Walter}, Fabian and {Chen}, Hsiao-Wen and {Ho}, Luis C. and {Jiang}, Peng and {Mandelker}, Nir and {Oh}, Se-Heon and {Peng}, Bo and {P{\'e}roux}, C{\'e}line and {Qu}, Zhijie and {Wang}, Q. Daniel},
        title = "{FEASTS Combined with Interferometry. III. The Low Column Density H I Around M51 and Possibility of Turbulent-mixing Gas Accretion}",
      journal = {\apj},
         year = 2025,
        month = apr,
       volume = {982},
       number = {2},
          eid = {151},
        pages = {151},
          doi = {10.3847/1538-4357/adb718},
archivePrefix = {arXiv},
       eprint = {2502.10672},
 primaryClass = {astro-ph.GA},
       adsurl = {https://ui.adsabs.harvard.edu/abs/2025ApJ...982..151L}
}

@ARTICLE{mcGaugh00-TullyFisher,
       author = {{McGaugh}, S.~S. and {Schombert}, J.~M. and {Bothun}, G.~D. and {de Blok}, W.~J.~G.},
        title = "{The Baryonic Tully-Fisher Relation}",
      journal = {\apjl},
         year = 2000,
        month = apr,
       volume = {533},
       number = {2},
        pages = {L99-L102},
          doi = {10.1086/312628},
archivePrefix = {arXiv},
       eprint = {astro-ph/0003001},
 primaryClass = {astro-ph},
       adsurl = {https://ui.adsabs.harvard.edu/abs/2000ApJ...533L..99M}
}

@ARTICLE{broeils97-HISizeMass,
       author = {{Broeils}, A.~H. and {Rhee}, M. -H.},
        title = "{Short 21-cm WSRT observations of spiral and irregular galaxies. HI properties.}",
      journal = {\aap},
         year = 1997,
        month = aug,
       volume = {324},
        pages = {877-887},
       adsurl = {https://ui.adsabs.harvard.edu/abs/1997A&A...324..877B}
}

@ARTICLE{wang16-HISizeMass,
       author = {{Wang}, Jing and {Koribalski}, B{\"a}rbel S. and {Serra}, Paolo and {van der Hulst}, Thijs and {Roychowdhury}, Sambit and {Kamphuis}, Peter and {Chengalur}, Jayaram N.},
        title = "{New lessons from the H I size-mass relation of galaxies}",
      journal = {\mnras},
         year = 2016,
        month = aug,
       volume = {460},
       number = {2},
        pages = {2143-2151},
          doi = {10.1093/mnras/stw1099},
archivePrefix = {arXiv},
       eprint = {1605.01489},
 primaryClass = {astro-ph.GA},
       adsurl = {https://ui.adsabs.harvard.edu/abs/2016MNRAS.460.2143W}
}

@ARTICLE{hogbom74-CLEAN,
       author = {{H{\"o}gbom}, J.~A.},
        title = "{Aperture Synthesis with a Non-Regular Distribution of Interferometer Baselines}",
      journal = {\aaps},
         year = 1974,
        month = jun,
       volume = {15},
        pages = {417},
       adsurl = {https://ui.adsabs.harvard.edu/abs/1974A&AS...15..417H}
}

@article{deBlok24-Mhongoose,
   title={MHONGOOSE: A MeerKAT nearby galaxy H I survey},
   volume={688},
   ISSN={1432-0746},
   url={http://dx.doi.org/10.1051/0004-6361/202348297},
   DOI={10.1051/0004-6361/202348297},
   journal={Astronomy \& Astrophysics},
   publisher={EDP Sciences},
   author={de Blok, W. J. G. and Healy, J. and Maccagni, F. M. and Pisano, D. J. and Bosma, A. and English, J. and Jarrett, T. and Marasco, A. and Meurer, G. R. and Veronese, S. and Bigiel, F. and Chemin, L. and Fraternali, F. and Holwerda, B. W. and Kamphuis, P. and Klöckner, H. R. and Kleiner, D. and Leroy, A. K. and Mogotsi, M. and Oman, K. A. and Schinnerer, E. and Verdes-Montenegro, L. and Westmeier, T. and Wong, O. I. and Zabel, N. and Amram, P. and Carignan, C. and Combes, F. and Brinks, E. and Dettmar, R. J. and Gibson, B. K. and Jozsa, G. I. G. and Koribalski, B. S. and McGaugh, S. S. and Oosterloo, T. A. and Spekkens, K. and Schröder, A. C. and Adams, E. A. K. and Athanassoula, E. and Bershady, M. A. and Beswick, R. J. and Blyth, S. and Elson, E. C. and Frank, B. S. and Heald, G. and Henning, P. A. and Kurapati, S. and Loubser, S. I. and Lucero, D. and Meyer, M. and Namumba, B. and Oh, S.-H. and Sardone, A. and Sheth, K. and Smith, M. W. L. and Sorgho, A. and Walter, F. and Williams, T. and Woudt, P. A. and Zijlstra, A.},
   year={2024},
   month=aug, pages={A109} }

@ARTICLE{westmeier21-sofia2,
       author = {{Westmeier}, T. and {Kitaeff}, S. and {Pallot}, D. and {Serra}, P. and {van der Hulst}, J.~M. and {Jurek}, R.~J. and {Elagali}, A. and {For}, B. -Q. and {Kleiner}, D. and {Koribalski}, B.~S. and {Lee-Waddell}, K. and {Mould}, J.~R. and {Reynolds}, T.~N. and {Rhee}, J. and {Staveley-Smith}, L.},
        title = "{SOFIA 2 - an automated, parallel H I source finding pipeline for the WALLABY survey}",
      journal = {\mnras},
         year = 2021,
        month = sep,
       volume = {506},
       number = {3},
        pages = {3962-3976},
          doi = {10.1093/mnras/stab1881},
archivePrefix = {arXiv},
       eprint = {2106.15789},
 primaryClass = {astro-ph.IM},
       adsurl = {https://ui.adsabs.harvard.edu/abs/2021MNRAS.506.3962W}
}

@article{marasco25:mhongoose-sim-comp,
   title={HI within and around observed and simulated galaxy discs: Comparing MeerKAT observations with mock data from TNG50 and FIRE-2},
   volume={697},
   ISSN={1432-0746},
   url={http://dx.doi.org/10.1051/0004-6361/202453172},
   DOI={10.1051/0004-6361/202453172},
   journal={Astronomy \& Astrophysics},
   publisher={EDP Sciences},
   author={Marasco, A. and de Blok, W. J. G. and Maccagni, F. M. and Fraternali, F. and Oman, K. A. and Oosterloo, T. and Combes, F. and McGaugh, S. S. and Kamphuis, P. and Spekkens, K. and Kleiner, D. and Veronese, S. and Amram, P. and Chemin, L. and Brinks, E.},
   year={2025},
   month=may, pages={A86} }

@ARTICLE{bluebird20:chilesVI,
       author = {{Blue Bird}, J. and {Davis}, J. and {Luber}, N. and {van Gorkom}, J.~H. and {Wilcots}, E. and {Pisano}, D.~J. and {Gim}, H.~B. and {Momjian}, E. and {Fernandez}, X. and {Hess}, K.~M. and {Lucero}, D. and {Dodson}, R. and {Vinsen}, K. and {Popping}, A. and {Chung}, A. and {Kreckel}, K. and {van der Hulst}, J.~M. and {Yun}, M.},
        title = "{CHILES VI: H I and H {\ensuremath{\alpha}} observations for z < 0.1 galaxies; probing H I spin alignment with filaments in the cosmic web}",
      journal = {\mnras},
         year = 2020,
        month = feb,
       volume = {492},
       number = {1},
        pages = {153-176},
          doi = {10.1093/mnras/stz3357},
archivePrefix = {arXiv},
       eprint = {1912.01062},
 primaryClass = {astro-ph.GA},
       adsurl = {https://ui.adsabs.harvard.edu/abs/2020MNRAS.492..153B}
}

@article{Healy_2024-WarpedMhongooseDisk,
   title={Possible origins of anomalous HI gas around MHONGOOSE galaxy, NGC 5068},
   volume={687},
   ISSN={1432-0746},
   url={http://dx.doi.org/10.1051/0004-6361/202347475},
   DOI={10.1051/0004-6361/202347475},
   journal={Astronomy \& Astrophysics},
   publisher={EDP Sciences},
   author={Healy, J. and de Blok, W. J. G. and Maccagni, F. M. and Amram, P. and Chemin, L. and Combes, F. and Holwerda, B. W. and Kamphuis, P. and Pisano, D. J. and Schinnerer, E. and Spekkens, K. and Verdes-Montenegro, L. and Walter, F. and Adams, E. A. K. and Gibson, B. K. and Kleiner, D. and Veronese, S. and Zabel, N. and English, J. and Carignan, C.},
   year={2024},
   month=jul, pages={A254} }

@article{Walter_2008-THINGS,
   title={THINGS: THE H I NEARBY GALAXY SURVEY},
   volume={136},
   ISSN={1538-3881},
   url={http://dx.doi.org/10.1088/0004-6256/136/6/2563},
   DOI={10.1088/0004-6256/136/6/2563},
   number={6},
   journal={The Astronomical Journal},
   publisher={American Astronomical Society},
   author={Walter, Fabian and Brinks, Elias and de Blok, W. J. G. and Bigiel, Frank and Kennicutt, Robert C. and Thornley, Michele D. and Leroy, Adam},
   year={2008},
   month=nov, pages={2563–2647} }

@ARTICLE{kennicutt98,
       author = {{Kennicutt}, Jr., Robert C.},
        title = "{The Global Schmidt Law in Star-forming Galaxies}",
      journal = {\apj},
         year = 1998,
        month = may,
       volume = {498},
       number = {2},
        pages = {541-552},
          doi = {10.1086/305588},
archivePrefix = {arXiv},
       eprint = {astro-ph/9712213},
 primaryClass = {astro-ph},
       adsurl = {https://ui.adsabs.harvard.edu/abs/1998ApJ...498..541K}
}

@ARTICLE{koch2025-GBT_VLA_DualConv,
       author = {{Koch}, Eric W. and {Leroy}, Adam K. and {Rosolowsky}, Erik W. and {Chomiuk}, Laura and {Dalcanton}, Julianne J. and {Pingel}, Nickolas M. and {Sarbadhicary}, Sumit K. and {Stanimirovi{\'c}}, Sne{\v{z}}ana and {Walter}, Fabian and {Archer}, Haylee N. and {Bolatto}, Alberto D. and {Busch}, Michael P. and {Chen}, Hongxing and {Chown}, Ryan and {Corbould}, Harrisen and {Cronin}, Serena A. and {Darling}, Jeremy and {Do}, Thomas and {Meyer}, Jennifer Donovan and {Eibensteiner}, Cosima and {Hunter}, Deidre and {Indebetouw}, R{\'e}my and {Jagannathan}, Preshanth and {Kepley}, Amanda A. and {Kim}, Chang-Goo and {Kim}, Shin-Jeong and {Kovacs}, Timea O. and {Marvil}, Joshua and {Murphy}, Eric J. and {Murray}, Claire E. and {Ott}, J{\"u}rgen and {Pisano}, D.~J. and {Putman}, Mary and {Rybarczyk}, Daniel R. and {Roman-Duval}, Julia and {Sandstrom}, Karin and {Schinnerer}, Eva and {Skillman}, Evan D. and {Smercina}, Adam and {Stelea}, Ioana and {Strader}, Jay and {Sun}, Jiayi and {Tallapaneni}, Devisree and {Tarantino}, Elizabeth and {Villanueva}, Vicente and {Weisz}, Daniel R. and {Williams}, Thomas G. and {Wong}, Tony},
        title = "{The Karl G. Jansky Very Large Array Local Group L-Band Survey (LGLBS)}",
      journal = {\apjs},
         year = 2025,
        month = aug,
       volume = {279},
       number = {2},
          eid = {35},
        pages = {35},
          doi = {10.3847/1538-4365/ade0ad},
archivePrefix = {arXiv},
       eprint = {2506.11792},
 primaryClass = {astro-ph.GA},
       adsurl = {https://ui.adsabs.harvard.edu/abs/2025ApJS..279...35K}
}

@misc{trapp25,
       author = {{Trapp}, Cameron W. and {Peeples}, Molly S. and {Tumlinson}, Jason and {O'Shea}, Brian W. and {Lochhaas}, Cassandra and {Wright}, Anna C. and {Smith}, Britton D. and {Saeedzadeh}, Vida and {Acharyya}, Ayan and {Augustin}, Ramona and {Simons}, Raymond C.},
        title = "{FOGGIE: Figuring Out Gas and Galaxies In Enzo. XII. The Formation and Evolution of Extended H I Galactic Disks and Warps with a Dynamic Circumgalactic Medium}",
      journal = {\apj},
         year = 2026,
        month = jun,
       volume = {1003},
       number = {2},
          eid = {186},
        pages = {186},
          doi = {10.3847/1538-4357/ae6503},
archivePrefix = {arXiv},
       eprint = {2511.00158},
 primaryClass = {astro-ph.GA},
       adsurl = {https://ui.adsabs.harvard.edu/abs/2026ApJ..1003..186T}
}

@ARTICLE{pisano2014:GBT-NGC2997,
       author = {{Pisano}, D.~J.},
        title = "{Green Bank Telescope Observations of Low Column Density H I around NGC 2997 and NGC 6946}",
      journal = {\aj},
         year = 2014,
        month = mar,
       volume = {147},
       number = {3},
          eid = {48},
        pages = {48},
          doi = {10.1088/0004-6256/147/3/48},
archivePrefix = {arXiv},
       eprint = {1312.3953},
 primaryClass = {astro-ph.GA},
       adsurl = {https://ui.adsabs.harvard.edu/abs/2014AJ....147...48P}
}

@ARTICLE{deBlok2014:GBT-NGC2403,
       author = {{de Blok}, W.~J.~G. and {Keating}, K.~M. and {Pisano}, D.~J. and {Fraternali}, F. and {Walter}, F. and {Oosterloo}, T. and {Brinks}, E. and {Bigiel}, F. and {Leroy}, A.},
        title = "{A low H I column density filament in NGC 2403: signature of interaction or accretion}",
      journal = {\aap},
         year = 2014,
        month = sep,
       volume = {569},
          eid = {A68},
        pages = {A68},
          doi = {10.1051/0004-6361/201423880},
archivePrefix = {arXiv},
       eprint = {1407.3648},
 primaryClass = {astro-ph.GA},
       adsurl = {https://ui.adsabs.harvard.edu/abs/2014A&A...569A..68D}
}

@ARTICLE{Qian2025:FAST-NGC2683,
       author = {{Jiao}, Qian and {Zhu}, Ming and {Vollmer}, Bernd and {Ai}, Mei and {Yu}, Haiyang and {Tan}, Qinghua and {Cheng}, Cheng and {Gao}, Yang},
        title = "{FAST Reveals the Extended H I Halo and Accretion Signatures of NGC 2683}",
      journal = {\apj},
         year = 2025,
        month = jun,
       volume = {986},
       number = {1},
          eid = {46},
        pages = {46},
          doi = {10.3847/1538-4357/add0ba},
archivePrefix = {arXiv},
       eprint = {2505.03158},
 primaryClass = {astro-ph.GA},
       adsurl = {https://ui.adsabs.harvard.edu/abs/2025ApJ...986...46J}
}

@ARTICLE{heald2011:HALOGAS,
       author = {{Heald}, G. and {J{\'o}zsa}, G. and {Serra}, P. and {Zschaechner}, L. and {Rand}, R. and {Fraternali}, F. and {Oosterloo}, T. and {Walterbos}, R. and {J{\"u}tte}, E. and {Gentile}, G.},
        title = "{The Westerbork Hydrogen Accretion in LOcal GAlaxieS (HALOGAS) survey. I. Survey description and pilot observations}",
      journal = {\aap},
         year = 2011,
        month = feb,
       volume = {526},
          eid = {A118},
        pages = {A118},
          doi = {10.1051/0004-6361/201015938},
archivePrefix = {arXiv},
       eprint = {1012.0816},
 primaryClass = {astro-ph.CO},
       adsurl = {https://ui.adsabs.harvard.edu/abs/2011A&A...526A.118H}
}

@ARTICLE{zwaan1997:optically-thin-justification,
       author = {{Zwaan}, Martin A. and {Briggs}, Frank H. and {Sprayberry}, David and {Sorar}, Ertu},
        title = "{The H I Mass Function of Galaxies from a Deep Survey in the 21 Centimeter Line}",
      journal = {\apj},
         year = 1997,
        month = nov,
       volume = {490},
       number = {1},
        pages = {173-186},
          doi = {10.1086/304872},
archivePrefix = {arXiv},
       eprint = {astro-ph/9707109},
 primaryClass = {astro-ph},
       adsurl = {https://ui.adsabs.harvard.edu/abs/1997ApJ...490..173Z}
}

@ARTICLE{veronese2025,
       author = {{Veronese}, S. and {de Blok}, W.~J.~G. and {Healy}, J. and {Kleiner}, D. and {Marasco}, A. and {Maccagni}, F.~M. and {Kamphuis}, P. and {Brinks}, E. and {Holwerda}, B.~W. and {Zabel}, N. and {Chemin}, L. and {Adams}, E.~A.~K. and {Kurapati}, S. and {Sorgho}, A. and {Spekkens}, K. and {Combes}, F. and {Pisano}, D.~J. and {Walter}, F. and {Amram}, P. and {Bigiel}, F. and {Wong}, O.~I. and {Athanassoula}, E.},
        title = "{Searching for HI around MHONGOOSE galaxies via spectral stacking}",
      journal = {\aap},
         year = 2025,
        month = jan,
       volume = {693},
          eid = {A97},
        pages = {A97},
          doi = {10.1051/0004-6361/202452085},
archivePrefix = {arXiv},
       eprint = {2411.11584},
 primaryClass = {astro-ph.GA},
       adsurl = {https://ui.adsabs.harvard.edu/abs/2025A&A...693A..97V}
}

@ARTICLE{hemler21:TNG50-Zgradients,
       author = {{Hemler}, Z.~S. and {Torrey}, Paul and {Qi}, Jia and {Hernquist}, Lars and {Vogelsberger}, Mark and {Ma}, Xiangcheng and {Kewley}, Lisa J. and {Nelson}, Dylan and {Pillepich}, Annalisa and {Pakmor}, R{\"u}diger and {Marinacci}, Federico},
        title = "{Gas-phase metallicity gradients of TNG50 star-forming galaxies}",
      journal = {\mnras},
         year = 2021,
        month = sep,
       volume = {506},
       number = {2},
        pages = {3024-3048},
          doi = {10.1093/mnras/stab1803},
archivePrefix = {arXiv},
       eprint = {2007.10993},
 primaryClass = {astro-ph.GA},
       adsurl = {https://ui.adsabs.harvard.edu/abs/2021MNRAS.506.3024H}
}

@ARTICLE{graf24:FIRE-MetallicityGradients,
       author = {{Graf}, Russell L. and {Wetzel}, Andrew and {Bailin}, Jeremy and {Orr}, Matthew E.},
        title = "{Inside-out versus Upside-down: The Origin and Evolution of Metallicity Radial Gradients in FIRE Simulations of Milky Way-mass Galaxies and the Essential Role of Gas Mixing}",
      journal = {\apj},
         year = 2025,
        month = oct,
       volume = {991},
       number = {2},
          eid = {139},
        pages = {139},
          doi = {10.3847/1538-4357/adfa07},
archivePrefix = {arXiv},
       eprint = {2410.21377},
 primaryClass = {astro-ph.GA},
       adsurl = {https://ui.adsabs.harvard.edu/abs/2025ApJ...991..139G}
}

@ARTICLE{hahn2022,
       author = {{Hahn}, ChangHoon and {Eickenberg}, Michael and {Ho}, Shirley and {Hou}, Jiamin and {Lemos}, Pablo and {Massara}, Elena and {Modi}, Chirag and {Moradinezhad Dizgah}, Azadeh and {R{\'e}galdo-Saint Blancard}, Bruno and {Abidi}, Muntazir M.},
        title = "{${\rm S{\scriptsize IM}BIG}$: A Forward Modeling Approach To Analyzing Galaxy Clustering}",
      journal = {arXiv e-prints},
         year = 2022,
        month = nov,
          eid = {arXiv:2211.00723},
        pages = {arXiv:2211.00723},
          doi = {10.48550/arXiv.2211.00723},
archivePrefix = {arXiv},
       eprint = {2211.00723},
 primaryClass = {astro-ph.CO},
       adsurl = {https://ui.adsabs.harvard.edu/abs/2022arXiv221100723H}
}

@ARTICLE{berg2019,
       author = {{Berg}, Michelle A. and {Howk}, J. Christopher and {Lehner}, Nicolas and {Wotta}, Christopher B. and {O'Meara}, John M. and {Bowen}, David V. and {Burchett}, Joseph N. and {Peeples}, Molly S. and {Tejos}, Nicolas},
        title = "{The Red Dead Redemption Survey of Circumgalactic Gas about Massive Galaxies. I. Mass and Metallicity of the Cool Phase}",
      journal = {\apj},
         year = 2019,
        month = sep,
       volume = {883},
       number = {1},
          eid = {5},
        pages = {5},
          doi = {10.3847/1538-4357/ab378e},
archivePrefix = {arXiv},
       eprint = {1811.10717},
 primaryClass = {astro-ph.GA},
       adsurl = {https://ui.adsabs.harvard.edu/abs/2019ApJ...883....5B}
}

@ARTICLE{eibensteiner2023,
       author = {{Eibensteiner}, C. and {Bigiel}, F. and {Leroy}, A.~K. and {Koch}, E.~W. and {Rosolowsky}, E. and {Schinnerer}, E. and {Sardone}, A. and {Meidt}, S. and {de Blok}, W.~J.~G. and {Thilker}, D. and {Pisano}, D.~J. and {Ott}, J. and {Barnes}, A. and {Querejeta}, M. and {Emsellem}, E. and {Puschnig}, J. and {Utomo}, D. and {Be{\v{s}}li{\'c}}, I. and {den Brok}, J. and {Faridani}, S. and {Glover}, S.~C.~O. and {Grasha}, K. and {Hassani}, H. and {Henshaw}, J.~D. and {Jim{\'e}nez-Donaire}, M.~J. and {Kerp}, J. and {Dale}, D.~A. and {Kruijssen}, J.~M.~D. and {Laudage}, S. and {Sanchez-Blazquez}, P. and {Smith}, R. and {Stuber}, S. and {Pessa}, I. and {Watkins}, E.~J. and {Williams}, T.~G. and {Winkel}, B.},
        title = "{Kinematic analysis of the super-extended H I disk of the nearby spiral galaxy M 83}",
      journal = {\aap},
         year = 2023,
        month = jul,
       volume = {675},
          eid = {A37},
        pages = {A37},
          doi = {10.1051/0004-6361/202245290},
archivePrefix = {arXiv},
       eprint = {2304.02037},
 primaryClass = {astro-ph.GA},
       adsurl = {https://ui.adsabs.harvard.edu/abs/2023A&A...675A..37E}
}

@ARTICLE{rahul23-TNG50_CGM,
       author = {{Ramesh}, Rahul and {Nelson}, Dylan and {Pillepich}, Annalisa},
        title = "{The circumgalactic medium of Milky Way-like galaxies in the TNG50 simulation - I: halo gas properties and the role of SMBH feedback}",
      journal = {\mnras},
         year = 2023,
        month = jan,
       volume = {518},
       number = {4},
        pages = {5754-5777},
          doi = {10.1093/mnras/stac3524},
archivePrefix = {arXiv},
       eprint = {2211.00020},
 primaryClass = {astro-ph.GA},
       adsurl = {https://ui.adsabs.harvard.edu/abs/2023MNRAS.518.5754R}
}

@ARTICLE{mina21,
       author = {{Mina}, Mattia and {Shen}, Sijing and {Keller}, Benjamin Walter and {Mayer}, Lucio and {Madau}, Piero and {Wadsley}, James},
        title = "{The baryon cycle of Seven Dwarfs with superbubble feedback}",
      journal = {\aap},
         year = 2021,
        month = nov,
       volume = {655},
          eid = {A22},
        pages = {A22},
          doi = {10.1051/0004-6361/202039420},
archivePrefix = {arXiv},
       eprint = {2009.06646},
 primaryClass = {astro-ph.GA},
       adsurl = {https://ui.adsabs.harvard.edu/abs/2021A&A...655A..22M}
}

@ARTICLE{piacitelli25,
       author = {{Piacitelli}, Daniel R. and {Brooks}, Alyson M. and {Christensen}, Charlotte and {Sanchez}, N. Nicole and {Faerman}, Yakov and {Shen}, Sijing and {Cruz}, Akaxia and {Keller}, Ben and {Quinn}, Thomas R. and {Wadsley}, James},
        title = "{Marvelous Metals: Surveying the Circumgalactic Medium of Simulated Dwarf Galaxies}",
      journal = {\apj},
         year = 2025,
        month = nov,
       volume = {993},
       number = {2},
          eid = {230},
        pages = {230},
          doi = {10.3847/1538-4357/ae06a0},
archivePrefix = {arXiv},
       eprint = {2505.08861},
 primaryClass = {astro-ph.GA},
       adsurl = {https://ui.adsabs.harvard.edu/abs/2025ApJ...993..230P}
}

@ARTICLE{maccio16,
       author = {{Macci{\`o}}, Andrea V. and {Udrescu}, Silviu M. and {Dutton}, Aaron A. and {Obreja}, Aura and {Wang}, Liang and {Stinson}, Greg R. and {Kang}, Xi},
        title = "{NIHAO X: reconciling the local galaxy velocity function with cold dark matter via mock H I observations}",
      journal = {\mnras},
         year = 2016,
        month = nov,
       volume = {463},
       number = {1},
        pages = {L69-L73},
          doi = {10.1093/mnrasl/slw147},
       adsurl = {https://ui.adsabs.harvard.edu/abs/2016MNRAS.463L..69M}
}

@ARTICLE{brooks17,
       author = {{Brooks}, Alyson M. and {Papastergis}, Emmanouil and {Christensen}, Charlotte R. and {Governato}, Fabio and {Stilp}, Adrienne and {Quinn}, Thomas R. and {Wadsley}, James},
        title = "{How to Reconcile the Observed Velocity Function of Galaxies with Theory}",
      journal = {\apj},
         year = 2017,
        month = nov,
       volume = {850},
       number = {1},
          eid = {97},
        pages = {97},
          doi = {10.3847/1538-4357/aa9576},
archivePrefix = {arXiv},
       eprint = {1701.07835},
 primaryClass = {astro-ph.GA},
       adsurl = {https://ui.adsabs.harvard.edu/abs/2017ApJ...850...97B}
}

@ARTICLE{ruan25,
       author = {{Ruan}, Dilys and {Brooks}, Alyson M. and {Cruz}, Akaxia and {Peter}, Annika H.~G. and {Keller}, Benjamin W. and {Quinn}, Thomas and {Wadsley}, James and {Adams}, Elizabeth A.~K.},
        title = "{Predictions for detecting a turndown in the baryonic Tully{\textendash}Fisher relation}",
      journal = {\mnras},
         year = 2025,
        month = aug,
       volume = {541},
       number = {3},
        pages = {2180-2196},
          doi = {10.1093/mnras/staf1099},
archivePrefix = {arXiv},
       eprint = {2503.16607},
 primaryClass = {astro-ph.GA},
       adsurl = {https://ui.adsabs.harvard.edu/abs/2025MNRAS.541.2180R}
}

@ARTICLE{lehner25,
       author = {{Lehner}, Nicolas and {Howk}, J. Christopher and {Collins}, Lucy and {Sameer} and {Wakker}, Bart P. and {Augustin}, Ramona and {Barger}, Kathleen A. and {Berg}, Michelle A. and {Bordoloi}, Rongmon and {Brown}, Thomas M. and {Cashman}, Frances H. and {Faucher-Gigu{\`e}re}, Claude-Andr{\'e} and {Fox}, Andrew J. and {French}, David M. and {Gilbert}, Karoline M. and {Guhathakurta}, Puragra and {O'Meara}, John M. and {O'Shea}, Brian W. and {Peeples}, Molly S. and {Pisano}, D.~J. and {Prochaska}, J. Xavier and {Stern}, Jonathan and {Tumlinson}, Jason and {Werk}, Jessica K. and {Williams}, Benjamin F.},
        title = "{Project AMIGA: The Inner Circumgalactic Medium of Andromeda from Thick Disk to Halo}",
      journal = {\apj},
         year = 2026,
        month = feb,
       volume = {997},
       number = {2},
          eid = {183},
        pages = {183},
          doi = {10.3847/1538-4357/ae1f13},
archivePrefix = {arXiv},
       eprint = {2506.16573},
 primaryClass = {astro-ph.GA},
       adsurl = {https://ui.adsabs.harvard.edu/abs/2026ApJ...997..183L}
}

@ARTICLE{lehner22,
       author = {{Lehner}, Nicolas and {Kopenhafer}, Claire and {O'Meara}, John M. and {Howk}, J. Christopher and {Fumagalli}, Michele and {Prochaska}, J. Xavier and {Acharyya}, Ayan and {O'Shea}, Brian W. and {Peeples}, Molly S. and {Tumlinson}, Jason and {Hummels}, Cameron B.},
        title = "{KODIAQ-Z: Metals and Baryons in the Cool Intergalactic and Circumgalactic Gas at 2.2 {\ensuremath{\lesssim}} z {\ensuremath{\lesssim}} 3.6}",
      journal = {\apj},
         year = 2022,
        month = sep,
       volume = {936},
       number = {2},
          eid = {156},
        pages = {156},
          doi = {10.3847/1538-4357/ac7400},
archivePrefix = {arXiv},
       eprint = {2112.03304},
 primaryClass = {astro-ph.GA},
       adsurl = {https://ui.adsabs.harvard.edu/abs/2022ApJ...936..156L}
}

@ARTICLE{chen12,
       author = {{Chen}, Hsiao-Wen},
        title = "{The unchanging circumgalactic medium over the past 11 billion years}",
      journal = {\mnras},
         year = 2012,
        month = dec,
       volume = {427},
       number = {2},
        pages = {1238-1244},
          doi = {10.1111/j.1365-2966.2012.22053.x},
archivePrefix = {arXiv},
       eprint = {1209.1094},
 primaryClass = {astro-ph.CO},
       adsurl = {https://ui.adsabs.harvard.edu/abs/2012MNRAS.427.1238C}
}

@ARTICLE{peroux19,
       author = {{P{\'e}roux}, C{\'e}line and {Zwaan}, Martin A. and {Klitsch}, Anne and {Augustin}, Ramona and {Hamanowicz}, Aleksandra and {Rahmani}, Hadi and {Pettini}, Max and {Kulkarni}, Varsha and {Straka}, Lorrie A. and {Biggs}, Andy D. and {York}, Donald G. and {Milliard}, Bruno},
        title = "{Multiphase circumgalactic medium probed with MUSE and ALMA}",
      journal = {\mnras},
         year = 2019,
        month = may,
       volume = {485},
       number = {2},
        pages = {1595-1613},
          doi = {10.1093/mnras/stz202},
archivePrefix = {arXiv},
       eprint = {1901.05217},
 primaryClass = {astro-ph.GA},
       adsurl = {https://ui.adsabs.harvard.edu/abs/2019MNRAS.485.1595P}
}

@ARTICLE{nielsen13,
       author = {{Nielsen}, Nikole M. and {Churchill}, Christopher W. and {Kacprzak}, Glenn G.},
        title = "{MAGIICAT II. General Characteristics of the Mg II Absorbing Circumgalactic Medium}",
      journal = {\apj},
         year = 2013,
        month = oct,
       volume = {776},
       number = {2},
          eid = {115},
        pages = {115},
          doi = {10.1088/0004-637X/776/2/115},
archivePrefix = {arXiv},
       eprint = {1211.1380},
 primaryClass = {astro-ph.CO},
       adsurl = {https://ui.adsabs.harvard.edu/abs/2013ApJ...776..115N}
}

@ARTICLE{zahedy19,
       author = {{Zahedy}, Fakhri S. and {Chen}, Hsiao-Wen and {Johnson}, Sean D. and {Pierce}, Rebecca M. and {Rauch}, Michael and {Huang}, Yun-Hsin and {Weiner}, Benjamin J. and {Gauthier}, Jean-Ren{\'e}},
        title = "{Characterizing circumgalactic gas around massive ellipticals at z {\ensuremath{\sim}} 0.4 - II. Physical properties and elemental abundances}",
      journal = {\mnras},
         year = 2019,
        month = apr,
       volume = {484},
       number = {2},
        pages = {2257-2280},
          doi = {10.1093/mnras/sty3482},
archivePrefix = {arXiv},
       eprint = {1809.05115},
 primaryClass = {astro-ph.GA},
       adsurl = {https://ui.adsabs.harvard.edu/abs/2019MNRAS.484.2257Z}
}

@ARTICLE{kacprzak15,
       author = {{Kacprzak}, Glenn G. and {Muzahid}, Sowgat and {Churchill}, Christopher W. and {Nielsen}, Nikole M. and {Charlton}, Jane C.},
        title = "{The Azimuthal Dependence of Outflows and Accretion Detected Using O VI Absorption}",
      journal = {\apj},
         year = 2015,
        month = dec,
       volume = {815},
       number = {1},
          eid = {22},
        pages = {22},
          doi = {10.1088/0004-637X/815/1/22},
archivePrefix = {arXiv},
       eprint = {1511.03275},
 primaryClass = {astro-ph.GA},
       adsurl = {https://ui.adsabs.harvard.edu/abs/2015ApJ...815...22K}
}

@ARTICLE{fraternali02,
       author = {{Fraternali}, Filippo and {van Moorsel}, Gustaaf and {Sancisi}, Renzo and {Oosterloo}, Tom},
        title = "{Deep H I Survey of the Spiral Galaxy NGC 2403}",
      journal = {\aj},
         year = 2002,
        month = jun,
       volume = {123},
       number = {6},
        pages = {3124-3140},
          doi = {10.1086/340358},
archivePrefix = {arXiv},
       eprint = {astro-ph/0203405},
 primaryClass = {astro-ph},
       adsurl = {https://ui.adsabs.harvard.edu/abs/2002AJ....123.3124F}
}

@ARTICLE{thilker04,
       author = {{Thilker}, David A. and {Braun}, Robert and {Walterbos}, Ren{\'e} A.~M. and {Corbelli}, Edvige and {Lockman}, Felix J. and {Murphy}, Edward and {Maddalena}, Ronald},
        title = "{On the Continuing Formation of the Andromeda Galaxy: Detection of H I Clouds in the M31 Halo}",
      journal = {\apjl},
         year = 2004,
        month = jan,
       volume = {601},
       number = {1},
        pages = {L39-L42},
          doi = {10.1086/381703},
archivePrefix = {arXiv},
       eprint = {astro-ph/0311571},
 primaryClass = {astro-ph},
       adsurl = {https://ui.adsabs.harvard.edu/abs/2004ApJ...601L..39T}
}

@ARTICLE{1977A&A....54..661T,
       author = {{Tully}, R.~B. and {Fisher}, J.~R.},
        title = "{A new method of determining distances to galaxies.}",
      journal = {\aap},
         year = 1977,
        month = feb,
       volume = {54},
        pages = {661-673},
       adsurl = {https://ui.adsabs.harvard.edu/abs/1977A&A....54..661T}
}

@ARTICLE{lin2025-FeastsCompWithSimulations,
       author = {{Lin}, Xuchen and {Wang}, Jing and {Kauffmann}, Guinevere and {Springel}, Volker and {Pakmor}, R{\"u}diger},
        title = "{FEASTS Compared with Simulations: Abnormally Irregular and Extended H I Morphologies at a Column Density of {}10$^{18}$ cm$^{{\ensuremath{-}}2}$ in TNG50 and Auriga}",
      journal = {\apj},
         year = 2026,
        month = feb,
       volume = {998},
       number = {1},
          eid = {56},
        pages = {56},
          doi = {10.3847/1538-4357/ae29b7},
archivePrefix = {arXiv},
       eprint = {2512.07223},
 primaryClass = {astro-ph.GA},
       adsurl = {https://ui.adsabs.harvard.edu/abs/2026ApJ...998...56L}
}

@ARTICLE{pingel18,
       author = {{Pingel}, N.~M. and {Pisano}, D.~J. and {Heald}, G. and {Jarrett}, T.~H. and {de Blok}, W.~J.~G. and {J{\'o}zsa}, G.~I.~G. and {J{\"u}tte}, E. and {Rand}, R.~J. and {Oosterloo}, T. and {Winkel}, B.},
        title = "{A GBT Survey of the HALOGAS Galaxies and Their Environments. I. Revealing the Full Extent of H I around NGC 891, NGC 925, NGC 4414, and NGC 4565}",
      journal = {\apj},
         year = 2018,
        month = sep,
       volume = {865},
       number = {1},
          eid = {36},
        pages = {36},
          doi = {10.3847/1538-4357/aad816},
archivePrefix = {arXiv},
       eprint = {1808.02041},
 primaryClass = {astro-ph.GA},
       adsurl = {https://ui.adsabs.harvard.edu/abs/2018ApJ...865...36P}
}

@ARTICLE{haardt12,
       author = {{Haardt}, Francesco and {Madau}, Piero},
        title = "{Radiative Transfer in a Clumpy Universe. IV. New Synthesis Models of the Cosmic UV/X-Ray Background}",
      journal = {\apj},
         year = 2012,
        month = feb,
       volume = {746},
       number = {2},
          eid = {125},
        pages = {125},
          doi = {10.1088/0004-637X/746/2/125},
       adsurl = {https://ui.adsabs.harvard.edu/abs/2012ApJ...746..125H}
}

@ARTICLE{smith17,
       author = {{Smith}, Britton D. and {Bryan}, Greg L. and {Glover}, Simon C.~O. and {Goldbaum}, Nathan J. and {Turk}, Matthew J. and {Regan}, John and {Wise}, John H. and {Schive}, Hsi-Yu and {Abel}, Tom and {Emerick}, Andrew and {O'Shea}, Brian W. and {Anninos}, Peter and {Hummels}, Cameron B. and {Khochfar}, Sadegh},
        title = "{GRACKLE: a chemistry and cooling library for astrophysics}",
      journal = {\mnras},
         year = 2017,
        month = apr,
       volume = {466},
       number = {2},
        pages = {2217-2234},
          doi = {10.1093/mnras/stw3291},
archivePrefix = {arXiv},
       eprint = {1610.09591},
 primaryClass = {astro-ph.CO},
       adsurl = {https://ui.adsabs.harvard.edu/abs/2017MNRAS.466.2217S}
}

@ARTICLE{emerick19,
       author = {{Emerick}, Andrew and {Bryan}, Greg L. and {Mac Low}, Mordecai-Mark},
        title = "{Simulating an isolated dwarf galaxy with multichannel feedback and chemical yields from individual stars}",
      journal = {\mnras},
         year = 2019,
        month = jan,
       volume = {482},
       number = {1},
        pages = {1304-1329},
          doi = {10.1093/mnras/sty2689},
archivePrefix = {arXiv},
       eprint = {1807.07182},
 primaryClass = {astro-ph.GA},
       adsurl = {https://ui.adsabs.harvard.edu/abs/2019MNRAS.482.1304E}
}

@ARTICLE{saeedzadeh25,
       author = {{Saeedzadeh}, Vida and {Tumlinson}, Jason and {Peeples}, Molly S. and {O'Shea}, Brian W. and {Lochhaas}, Cassandra and {Corlies}, Lauren and {Trapp}, Cameron W. and {Smith}, Britton D. and {Werk}, Jessica K. and {Acharyya}, Ayan and {Augustin}, Ramona and {Fox}, Andrew J. and {Lehner}, Nicolas and {Wright}, Anna C.},
        title = "{Figuring Out Gas \& Galaxies In Enzo (FOGGIE). XIV. The Observability of Emission from Accretion and Feedback in the Circumgalactic Medium with Current and Future Instruments}",
      journal = {\apj},
         year = 2026,
        month = jun,
       volume = {1004},
       number = {1},
          eid = {9},
        pages = {9},
          doi = {10.3847/1538-4357/ae658a},
archivePrefix = {arXiv},
       eprint = {2511.05644},
 primaryClass = {astro-ph.GA},
       adsurl = {https://ui.adsabs.harvard.edu/abs/2026ApJ..1004....9S}
}

@ARTICLE{braun85,
       author = {{Braun}, R. and {Walterbos}, R.~A.~M.},
        title = "{A solution to the short spacing problem in radio interferometry.}",
      journal = {\aap},
         year = 1985,
        month = feb,
       volume = {143},
        pages = {307-312},
       adsurl = {https://ui.adsabs.harvard.edu/abs/1985A&A...143..307B}
}

@ARTICLE{reynolds22-Wallaby, 
       author = {{Reynolds}, T.~N. and {Catinella}, B. and {Cortese}, L. and {Westmeier}, T. and {Meurer}, G.~R. and {Shao}, L. and {Obreschkow}, D. and {Rom{\'a}n}, J. and {Verdes-Montenegro}, L. and {Deg}, N. and {D{\'e}nes}, H. and {For}, B.-Q. and {Kleiner}, D. and {Koribalski}, B.~S. and {Lee-Waddell}, K. and {Murugeshan}, C. and {Oh}, S.-H. and {Rhee}, J. and {Spekkens}, K. and {Staveley-Smith}, L. and {Stevens}, A.~R.~H. and {van der Hulst}, J.~M. and {Wang}, J. and {Wong}, O.~I. and {Holwerda}, B.~W. and {Bosma}, A. and {Madrid}, J.~P. and {Bekki}, K.},
        title = "{WALLABY pilot survey: H I gas disc truncation and star formation of galaxies falling into the Hydra I cluster}",
      journal = {\mnras},
         year = 2022,
        month = feb,
       volume = {510},
       number = {2},
        pages = {1716-1732},
          doi = {10.1093/mnras/stab3522},
archivePrefix = {arXiv},
       eprint = {2112.00231},
 primaryClass = {astro-ph.GA},
       adsurl = {https://ui.adsabs.harvard.edu/abs/2022MNRAS.510.1716R}
}

@ARTICLE{faridani18-JointDecovolution,
       author = {{Faridani}, S. and {Bigiel}, F. and {Fl{\"o}er}, L. and {Kerp}, J. and {Stanimirovi{\'c}}, S.},
        title = "{A new approach for short-spacing correction of radio interferometric datasets}",
      journal = {Astronomische Nachrichten},
         year = 2018,
        month = jan,
       volume = {339},
       number = {1},
        pages = {87-100},
          doi = {10.1002/asna.201713381},
archivePrefix = {arXiv},
       eprint = {1709.09365},
 primaryClass = {astro-ph.IM},
       adsurl = {https://ui.adsabs.harvard.edu/abs/2018AN....339...87F}
}

@ARTICLE{plunkett23,
       author = {{Plunkett}, Adele and {Hacar}, Alvaro and {Moser-Fischer}, Lydia and {Petry}, Dirk and {Teuben}, Peter and {Pingel}, Nickolas and {Kunneriath}, Devaky and {Takagi}, Toshinobu and {Miyamoto}, Yusuke and {Moravec}, Emily and {Suri}, S{\"u}meyye and {Hess}, Kelley M. and {Hoffman}, Melissa and {Mason}, Brian},
        title = "{Data Combination: Interferometry and Single-dish Imaging in Radio Astronomy}",
      journal = {\pasp},
         year = 2023,
        month = mar,
       volume = {135},
       number = {1045},
          eid = {034501},
        pages = {034501},
          doi = {10.1088/1538-3873/acb9bd},
archivePrefix = {arXiv},
       eprint = {2303.02177},
 primaryClass = {astro-ph.IM},
       adsurl = {https://ui.adsabs.harvard.edu/abs/2023PASP..135c4501P}
}

@ARTICLE{deul87,
       author = {{Deul}, E.~R. and {van der Hulst}, J.~M.},
        title = "{A survey of the neutral atomic hydrogen in M 33.}",
      journal = {\aaps},
         year = 1987,
        month = feb,
       volume = {67},
        pages = {509-539},
       adsurl = {https://ui.adsabs.harvard.edu/abs/1987A&AS...67..509D}
}

@ARTICLE{stanimirovic99,
       author = {{Stanimirovic}, S. and {Staveley-Smith}, L. and {Dickey}, J.~M. and {Sault}, R.~J. and {Snowden}, S.~L.},
        title = "{The large-scale HI structure of the Small Magellanic Cloud}",
      journal = {\mnras},
         year = 1999,
        month = jan,
       volume = {302},
       number = {3},
        pages = {417-436},
          doi = {10.1046/j.1365-8711.1999.02013.x},
       adsurl = {https://ui.adsabs.harvard.edu/abs/1999MNRAS.302..417S}
}

@INPROCEEDINGS{clark99,
       author = {{Clark}, B.~G.},
        title = "{Coherence in Radio Astronomy}",
    booktitle = {Synthesis Imaging in Radio Astronomy II},
         year = 1999,
       editor = {{Taylor}, G.~B. and {Carilli}, C.~L. and {Perley}, R.~A.},
       series = {Astronomical Society of the Pacific Conference Series},
       volume = {180},
        month = jan,
        pages = {1},
       adsurl = {https://ui.adsabs.harvard.edu/abs/1999ASPC..180....1C}
}

@INPROCEEDINGS{thompson99,
       author = {{Thompson}, A. Richard},
        title = "{Fundamentals of Radio Interferometry}",
    booktitle = {Synthesis Imaging in Radio Astronomy II},
         year = 1999,
       editor = {{Taylor}, G.~B. and {Carilli}, C.~L. and {Perley}, R.~A.},
       series = {Astronomical Society of the Pacific Conference Series},
       volume = {180},
        month = jan,
        pages = {11},
       adsurl = {https://ui.adsabs.harvard.edu/abs/1999ASPC..180...11T}
}

@ARTICLE{rogstad74,
       author = {{Rogstad}, D.~H. and {Lockhart}, I.~A. and {Wright}, M.~C.~H.},
        title = "{Aperture-synthesis observations of H I in the galaxy M83.}",
      journal = {\apj},
         year = 1974,
        month = oct,
       volume = {193},
        pages = {309-319},
          doi = {10.1086/153164},
       adsurl = {https://ui.adsabs.harvard.edu/abs/1974ApJ...193..309R}
}

@ARTICLE{kamphuis15,
       author = {{Kamphuis}, P. and {J{\'o}zsa}, G.~I.~G. and {Oh}, S.-. H. and {Spekkens}, K. and {Urbancic}, N. and {Serra}, P. and {Koribalski}, B.~S. and {Dettmar}, R.-J.},
        title = "{Automated kinematic modelling of warped galaxy discs in large H I surveys: 3D tilted-ring fitting of H I emission cubes}",
      journal = {\mnras},
         year = 2015,
        month = sep,
       volume = {452},
       number = {3},
        pages = {3139-3158},
          doi = {10.1093/mnras/stv1480},
archivePrefix = {arXiv},
       eprint = {1507.00413},
 primaryClass = {astro-ph.GA},
       adsurl = {https://ui.adsabs.harvard.edu/abs/2015MNRAS.452.3139K}
}

@ARTICLE{jozsa07-Tirific,
       author = {{J{\'o}zsa}, G.~I.~G. and {Kenn}, F. and {Klein}, U. and {Oosterloo}, T.~A.},
        title = "{Kinematic modelling of disk galaxies. I. A new method to fit tilted rings to data cubes}",
      journal = {\aap},
         year = 2007,
        month = jun,
       volume = {468},
       number = {2},
        pages = {731-774},
          doi = {10.1051/0004-6361:20066164},
archivePrefix = {arXiv},
       eprint = {astro-ph/0703207},
 primaryClass = {astro-ph},
       adsurl = {https://ui.adsabs.harvard.edu/abs/2007A&A...468..731J}
}

@ARTICLE{shachar26-RotCurve,
       author = {{Nestor Shachar}, A. and {Sternberg}, A. and {Price}, S.~H. and {F{\"o}rster Schreiber}, N.~M. and {Genzel}, R. and {Tacconi}, L.~J. and {{\"U}bler}, H. and {Barfety}, C. and {Burkert}, A. and {Chen}, J. and {Davies}, R. and {Eisenhauer}, F. and {Espejo Salcedo}, J.~M. and {Herrera-Camus}, R. and {Jolly}, J.~B. and {Lee}, L.~L. and {Naab}, T. and {Pastras}, S. and {Pulsoni}, C. and {Shimizu}, T.~T. and {Tozzi}, G.},
        title = "{RotCurves: a PYTHON package for efficient modelling and fitting of galactic rotation curves at high-z}",
      journal = {\mnras},
         year = 2026,
        month = mar,
       volume = {546},
       number = {4},
          eid = {stag134},
        pages = {stag134},
          doi = {10.1093/mnras/stag134},
archivePrefix = {arXiv},
       eprint = {2601.08348},
 primaryClass = {astro-ph.GA},
       adsurl = {https://ui.adsabs.harvard.edu/abs/2026MNRAS.546ag134N}
}

@ARTICLE{Bouche15-Galpak3d,
       author = {{Bouch{\'e}}, N. and {Carfantan}, H. and {Schroetter}, I. and {Michel-Dansac}, L. and {Contini}, T.},
        title = "{GalPak$^{3D}$: A Bayesian Parametric Tool for Extracting Morphokinematics of Galaxies from 3D Data}",
      journal = {\aj},
         year = 2015,
        month = sep,
       volume = {150},
       number = {3},
          eid = {92},
        pages = {92},
          doi = {10.1088/0004-6256/150/3/92},
archivePrefix = {arXiv},
       eprint = {1501.06586},
 primaryClass = {astro-ph.IM},
       adsurl = {https://ui.adsabs.harvard.edu/abs/2015AJ....150...92B}
}

@ARTICLE{price21-DysmalPy,
       author = {{Price}, S.~H. and {Shimizu}, T.~T. and {Genzel}, R. and {{\"U}bler}, H. and {F{\"o}rster Schreiber}, N.~M. and {Tacconi}, L.~J. and {Davies}, R.~I. and {Coogan}, R.~T. and {Lutz}, D. and {Wuyts}, S. and {Wisnioski}, E. and {Nestor}, A. and {Sternberg}, A. and {Burkert}, A. and {Bender}, R. and {Contursi}, A. and {Davies}, R.~L. and {Herrera-Camus}, R. and {Lee}, M.-J. and {Naab}, T. and {Neri}, R. and {Renzini}, A. and {Saglia}, R. and {Schruba}, A. and {Schuster}, K.},
        title = "{Rotation Curves in z   1-2 Star-forming Disks: Comparison of Dark Matter Fractions and Disk Properties for Different Fitting Methods}",
      journal = {\apj},
         year = 2021,
        month = dec,
       volume = {922},
       number = {2},
          eid = {143},
        pages = {143},
          doi = {10.3847/1538-4357/ac22ad},
archivePrefix = {arXiv},
       eprint = {2109.02659},
 primaryClass = {astro-ph.GA},
       adsurl = {https://ui.adsabs.harvard.edu/abs/2021ApJ...922..143P}
}

@ARTICLE{oh18-2DBAT,
       author = {{Oh}, Se-Heon and {Staveley-Smith}, Lister and {Spekkens}, Kristine and {Kamphuis}, Peter and {Koribalski}, B{\"a}rbel S.},
        title = "{2D Bayesian automated tilted-ring fitting of disc galaxies in large H I galaxy surveys: 2DBAT}",
      journal = {\mnras},
         year = 2018,
        month = jan,
       volume = {473},
       number = {3},
        pages = {3256-3298},
          doi = {10.1093/mnras/stx2304},
archivePrefix = {arXiv},
       eprint = {1709.02049},
 primaryClass = {astro-ph.GA},
       adsurl = {https://ui.adsabs.harvard.edu/abs/2018MNRAS.473.3256O}
}

@ARTICLE{sellwood15-Diskfit,
       author = {{Sellwood}, J.~A. and {Spekkens}, Kristine},
        title = "{DiskFit: a code to fit simple non-axisymmetric galaxy models either to photometric images or to kinematic maps}",
      journal = {arXiv e-prints},
         year = 2015,
        month = sep,
          eid = {arXiv:1509.07120},
        pages = {arXiv:1509.07120},
          doi = {10.48550/arXiv.1509.07120},
archivePrefix = {arXiv},
       eprint = {1509.07120},
 primaryClass = {astro-ph.GA},
       adsurl = {https://ui.adsabs.harvard.edu/abs/2015arXiv150907120S}
}

@ARTICLE{teodoro15-3dbarolo,
       author = {{Di Teodoro}, E.~M. and {Fraternali}, F.},
        title = "{$^{3D}$ BAROLO: a new 3D algorithm to derive rotation curves of galaxies}",
      journal = {\mnras},
         year = 2015,
        month = aug,
       volume = {451},
       number = {3},
        pages = {3021-3033},
          doi = {10.1093/mnras/stv1213},
archivePrefix = {arXiv},
       eprint = {1505.07834},
 primaryClass = {astro-ph.GA},
       adsurl = {https://ui.adsabs.harvard.edu/abs/2015MNRAS.451.3021D}
}

@ARTICLE{burstein91,
       author = {{Burstein}, David and {Haynes}, Martha P. and {Faber}, M.},
        title = "{Dependence of galaxy properties on viewing angle.}",
      journal = {\nat},
         year = 1991,
        month = oct,
       volume = {353},
        pages = {515-521},
          doi = {10.1038/353515a0},
       adsurl = {https://ui.adsabs.harvard.edu/abs/1991Natur.353..515B}
}

@ARTICLE{wong04,
   author = {{Wong}, T. and {Blitz}, L. and {Bosma}, A.},
    title = "{A Search for Kinematic Evidence of Radial Gas Flows in Spiral Galaxies}",
  journal = {\apj},
   eprint = {astro-ph/0401187},
     year = 2004,
    month = apr,
   volume = 605,
    pages = {183-204},
      doi = {10.1086/382215},
   adsurl = {http://adsabs.harvard.edu/abs/2004ApJ...605..183W}
}

@ARTICLE{lucchini26,
       author = {{Lucchini}, Scott and {Abramson}, Cecilia and {Hummels}, Cameron and {Conroy}, Charlie and {Hernquist}, Lars and {Smith}, Aaron},
        title = "{ENhanced Galactic Atmospheres With Arepo: Resolving the CGM at 200 pc with the ENGAWA Simulations}",
      journal = {arXiv e-prints},
         year = 2026,
        month = mar,
          eid = {arXiv:2603.05584},
        pages = {arXiv:2603.05584},
          doi = {10.48550/arXiv.2603.05584},
archivePrefix = {arXiv},
       eprint = {2603.05584},
 primaryClass = {astro-ph.GA},
       adsurl = {https://ui.adsabs.harvard.edu/abs/2026arXiv260305584L}
}

@ARTICLE{wang26-FeastsAndMhongoose,
       author = {{Wang}, Jing and {Lin}, Xuchen and {Liang}, Ze-Zhong and {De Blok}, W.~J.~G. and {Guo}, Hong and {Qu}, Zhijie and {P{\'e}roux}, C{\'e}line and {Nagamine}, Kentaro and {Ho}, Luis C. and {Yang}, Dong and {Weng}, Simon and {Lagos}, Claudia Del P. and {Chen}, Xinkai and {Heald}, George and {Healy}, J. and {Huang}, Qifeng and {Kamphuis}, Peter and {Kleiner}, D. and {Li}, Di and {Liu}, Siqi and {Maccagni}, F.~M. and {Staveley-Smith}, Lister and {Su}, Zherong and {Van De Voort}, Freeke and {Walter}, Fabian and {Zhong}, Fangxiong and {Zou}, Siwei},
        title = "{FEASTS and MHONGOOSE: HI Column Density Distribution at $z=0$ for $N_\mathrm{HI}>10^{17.8}\, \mathrm{cm}^{-2}$}",
      journal = {arXiv e-prints},
         year = 2026,
        month = mar,
          eid = {arXiv:2603.02670},
        pages = {arXiv:2603.02670},
          doi = {10.48550/arXiv.2603.02670},
archivePrefix = {arXiv},
       eprint = {2603.02670},
 primaryClass = {astro-ph.GA},
       adsurl = {https://ui.adsabs.harvard.edu/abs/2026arXiv260302670W}
}
\bibliographystyle{aasjournalv7}



\end{document}